\documentclass[superscriptaddress,nofootinbib,aps,prd,10pt]{revtex4-1}

\usepackage{graphicx}
\usepackage{dcolumn}
\usepackage{bm}

\usepackage{amssymb}

\def\Tr{{\rm Tr}}
\def\spose#1{\hbox to 0pt{#1\hss}}
\def\ltapprox{\mathrel{\spose{\lower 3pt\hbox{$\mathchar"218$}}
 \raise 2.0pt\hbox{$\mathchar"13C$}}}
\def\gtapprox{\mathrel{\spose{\lower 3pt\hbox{$\mathchar"218$}}
 \raise 2.0pt\hbox{$\mathchar"13E$}}}
\catcode`~=11 
\newcommand{\urltilde}{\kern -.15em\lower .7ex\hbox{~}\kern .04em}
\catcode`~=13 

\begin{document}

\title{{\Large {\bf
Ghost sector and geometry in minimal Landau gauge: \\[1mm]
further constraining the infinite-volume limit \\[1mm]
      }}}

\author{Attilio~Cucchieri}
\email{attilio@ifsc.usp.br}
\affiliation{Instituto de F\'\i sica de S\~ao Carlos, Universidade de S\~ao Paulo,
             Caixa Postal 369, \\
             13560-970 S\~ao Carlos, SP, Brazil}
\author{Tereza~Mendes}
\email{mendes@ifsc.usp.br}
\affiliation{Instituto de F\'\i sica de S\~ao Carlos, Universidade de S\~ao Paulo,
             Caixa Postal 369, \\
             13560-970 S\~ao Carlos, SP, Brazil}

\date{\today}

\begin{abstract}
We present improved upper and lower bounds for the momentum-space
ghost propagator of Yang-Mills theories in terms of the
two smallest nonzero eigenvalues (and their corresponding
eigenvectors) of the Faddeev-Popov matrix. These results are
verified using data from four-dimensional numerical simulations of
SU(2) lattice gauge theory in minimal Landau gauge at $\beta = 2.2$,
for lattice sides $N = 16$, $32$, $48$ and $64$. Gribov-copy effects
are discussed by considering four different sets of numerical
minima. We then present a lower bound for the smallest
nonzero eigenvalue of the Faddeev-Popov matrix in terms of
the smallest nonzero momentum on the lattice and of a parameter
characterizing the geometry of the first Gribov region $\Omega$.
This allows a simple and intuitive description of the infinite-volume
limit in the ghost sector. In particular, we show how nonperturbative
effects may be quantified by the rate at which typical thermalized and
gauge-fixed configurations approach the boundary of $\Omega$,
known as the first Gribov horizon. As a result, a simple
and concrete explanation emerges for why lattice studies do not
observe an enhanced ghost propagator in the deep infrared limit.
Most of the simulations have been performed on the Blue
Gene/P--IBM supercomputer shared by Rice University and S\~ao
Paulo University.
\end{abstract}


\maketitle


\section{Introduction}
\label{sec:intro}

It is well known that the infinite-volume extrapolation of
numerical data is a key step to obtain continuum results for
minimal-Landau-gauge-fixed Green's functions in Yang-Mills theories
(see e.g.\ \cite{Cucchieri:2008yp,Cucchieri:2008mv,Cucchieri:2010xr}
and references therein).
Extracting the relevant infinite-volume information from lattice
simulations is, however, 
a difficult computational task, since analytic results that might
guide the necessary extrapolation are still limited.
In particular, the main --- and widely accepted --- assumption governing 
the infinite-volume limit in minimal Landau gauge is that

\vskip 4mm
\begin{center}
\parbox{4in}{
\noindent
{\em At very large volumes, the functional integration gets
     concentrated on the boundary $\partial\Omega$ of the first 
     Gribov region $\Omega$ [defined by transverse gauge configurations with 
     all nonnegative eigenvalues 
     of the Faddeev-Popov (FP) matrix ${\cal M}(b,x;c,y)$]. 
}}
\end{center}

\vskip 1mm 
\noindent
This is explained by considering the interplay among the
volume of configuration space, the Boltzmann weight associated to the gauge
configurations and the
step function used to constrain the functional integration to the region
$\Omega$. Indeed, since the configuration space has very
large dimensionality, we expect that, for large volumes, entropy
will favor configurations near the boundary $\partial \Omega$ of
$\Omega$ \cite{Zwanziger:1991ac, Zwanziger:1992qr}. 
(For a pictorial visualization of this competition
see Figure 2 of Ref.\ \cite{Zwanziger:1991ac} and Figure 3.5
of Ref.\ \cite{Vandersickel:2011zc} or, equivalently, Figure 2.5 of
Ref.\ \cite{Vandersickel:2012tz}.) The above statement is supported
by numerical data: the average value of the smallest nonzero
eigenvalue $\lambda_1$ of ${\cal M}(b,x;c,y)$ goes to zero as the 
lattice volume increases
in 2d \cite{Maas:2007uv}, 3d \cite{Cucchieri:2006tf} and 4d
\cite{Sternbeck:2005vs} minimal Landau gauge. 

These results give
rise to a simple picture. For very small physical volumes one
expects the measure to be concentrated around $A=0$, i.e.\ the
lattice configuration should be purely perturbative (see for
example \cite{Damm:1998pd,Cucchieri:1999ky} for numerical
simulations in the perturbative regime). Conversely, for
sufficiently large physical volumes, nonperturbative effects (and
in particular the confinement mechanism) should be at work.
Such effects should be encoded in configurations on and around
the boundary of $\Omega$. Thus, if one accepts the above assumption,
the functional integration should be strongly dominated at very 
large volumes by configurations belonging to a thin
layer close to $\partial \Omega$. This indeed
seems to be the case if one looks at the plots reported on the left
column of Figure 1 $\mbox{in}$ Ref.\ \cite{Sternbeck:2005vs} (see
also Figures 1 and 2 $\mbox{in}$ Ref.\ \cite{Sternbeck:2012mf}).

At the same time, the study of Green's functions of Yang-Mills theories 
in minimal Landau gauge is complicated by the existence of Gribov copies 
\cite{Gribov:1977wm,Dokshitzer:2004ie,Sobreiro:2005ec}, corresponding to 
different local minima\footnote{With respect to gauge transformations
$\left\{ g(x) \right\}$.} of the minimizing functional
\begin{equation}
{\cal E}[A] \, = \, \int \, d^dx \; \Tr
  \left[ \, A_{\mu}^{(g)}(x) \, A_{\mu}^{(g)}(x) \right] \; ,
\label{eq:min}
\end{equation}
used to define this gauge (in the continuum). [Here,
$A_{\mu}^{(g)}(x)$ is the gauge-transformed gauge field.]
Indeed, at finite volume, different sets of local minima
have been shown to yield different results for the numerical data
\cite{Cucchieri:1997dx,Cucchieri:1997ns,Sternbeck:2005vs,
Sternbeck:2012mf,Bakeev:2003rr,Silva:2004bv,Nakajima:2004vc,
Sternbeck:2005tk,Lokhov:2005ra,Bogolubsky:2005wf,Bogolubsky:2007bw,
Maas:2008ri,Bornyakov:2008yx,Maas:2009se,Maas:2009ph,
Bornyakov:2009ug}. Thus, in the infinite-volume limit,
systematic effects due to Gribov copies should be taken
into account. With respect to these effects, two different
possibilities have been discussed: 
\begin{itemize}
\item The influence of Gribov copies becomes smaller as the
volume increases \cite{Bogolubsky:2005wf,Lokhov:2005ra,
Bogolubsky:2007bw}. These results are supported by the
analysis carried out in Ref.\ \cite{Zwanziger:2003cf},
which showed that the normalized probability distributions
over the first Gribov region $\Omega$ and over the fundamental
modular region $\Lambda$ --- defined \cite{Zwanziger:1991ac,
Zwanziger:1993dh} by the absolute minima of ${\cal E}[A]$ --- 
are equal (in the sense that their moments of finite order are equal). 
\item Gribov copies
should be related to different possible infrared behaviors of
minimal-Landau-gauge Green's functions, i.e.\ different
sets of numerical minima --- such as those considered in
Refs.\ \cite{Sternbeck:2012mf,Maas:2009se,Maas:2013vd} ---
would yield qualitatively different results, and these
differences would survive the infinite-volume limit.
Very recently, in Ref.\ \cite{Bornyakov:2013ysa}, 
different infinite-volume results were indeed obtained, for the gluon
propagator in the $3d$ SU(2) case, depending on whether
the averages were taken in $\Omega$ or restricted to $\Lambda$.
\end{itemize}

\vskip 3mm
In Refs.\ \cite{Cucchieri:2007rg,Cucchieri:2008fc} we
introduced upper and lower bounds for the gluon\footnote{Interesting
bounds in the gluon sector have also been recently proven
in Refs.\ \cite{Zwanziger:2012xg,Maas:2013me}.} and the ghost
propagators using a magnetization-like quantity in the former
case and the smallest nonzero eigenvalue $\lambda_1$ of
the FP matrix (and the corresponding eigenvector) in the latter.
These bounds allow a better control of the extrapolation
of the data to infinite volume \cite{Cucchieri:2007rg,
Cucchieri:2008fc,Cucchieri:2008yp,Cucchieri:2008mv,
Cucchieri:2010xr}. They also provide a better intuition of the
relevant aspects of the theory as very large lattice sizes
$N$ are considered. In particular, in the ghost case, one can
show that the inverse of $\lambda_1$ is an upper bound for
the ghost propagator $G(p)$ [see Eq.\ (\ref{eq:ineq}) in the
next section]. As explained above, for very large $N$,
one expects the relevant configurations to be very
close to $\partial \Omega$, i.e.\ these configurations should be
characterized by very small values for $\lambda_1$. Thus,
in order to describe the extrapolation $N \to + \infty$ in
the ghost sector, we can re-formulate the above assumption
and say that

\vskip 4mm
\begin{center}
\parbox{4in}{
\noindent
{\em The key point seems to be the rate at which $\lambda_1$
     goes to zero, which, in turn, should be related to the
     rate at which a thermalized and gauge-fixed configuration
     approaches $\partial \Omega$.}}
\end{center}

\vskip 1mm
\noindent
This observation has partially prompted the present work. Indeed,
we present here (see Section \ref{sec:lambdamin}) a lower bound
for $\lambda_1$ that allows us to give a simple mathematical
realization of the above statement.

A relevant aspect of the bounds proven in
\cite{Cucchieri:2008fc}, as well as of
the new bounds presented here, is that they apply
to any configuration of $\Omega$. As a consequence, they can
be used for any Gribov copy of the minimal Landau gauge. Of
course, Gribov-copy effects may (or may not) be present also
in the quantities entering the formulae of the
bounds.\footnote{For example, we already know that this is
the case for the smallest nonzero eigenvalue $\lambda_1$
\cite{Sternbeck:2005vs,Cucchieri:1997ns}.} On the other hand,
the possibility of comparing Gribov-copy effects for several,
related, quantities could be useful to understand what triggers
these effects and when they are relevant.

\vskip 3mm
The present work is organized as follows. In the next section
we review the proof presented in Ref.\ \cite{Cucchieri:2008fc}
and we show how the previous bounds can be systematically improved, 
by considering
a larger set of eigenvalues (and eigenvectors) of the FP
matrix. These results are numerically verified in Section
\ref{sec:num}. Let us recall that a complete verification of
the bounds was lacking in Ref.\ \cite{Cucchieri:2008fc},
since in most of our old simulations we did not collect data
for the smallest eigenvalues $\lambda_1$ or for the
corresponding eigenvector. This omission is now rectified,
considering SU(2) data at $\beta = 2.2$, for lattice volumes
up to $64^4$. In Section \ref{sec:num} we also discuss Gribov-copy
effects by considering four different sets of numerical minima.
A lower bound for $\lambda_1$ is proven in Section
\ref{sec:lambdamin} and numerically verified in Section
\ref{sec:num2}, by an unconventional approach. 
This bound, which is written in terms of the
smallest nonzero lattice momentum $p_{min}$ and of a parameter
characterizing the geometry of the first Gribov region $\Omega$,
implies i) a simple upper bound for the ghost propagator
[see Eq.\ (\ref{eq:inequpp}) below], ii) a stronger version of
the so-called no-pole condition [see Eq.\ (\ref{eq:sigma})]
and iii) a new bound for the so-called horizon function [see Eq.\
(\ref{eq:hbound})]. It also allows a better understanding of 
the infinite-volume limit and a simple explanation of some of 
the results published in the literature (see Section \ref{sec:limit}). 
Finally, in Section \ref{sec:con}, we present our conclusions.


\section{Bounds for $G(p_{min})$ (old and new)}
\label{sec:ineq}

In Ref.\ \cite{Cucchieri:2008fc} (see also \cite{Cucchieri:2008yp,
Cucchieri:2008mv,Cucchieri:2010xr}) we have proven that, in the
SU($N_c$) case, for any nonzero momentum $p$ and for any gauge-fixed
configuration that belongs to the interior of the first Gribov region
$\Omega$, the minimal-Landau-gauge ghost propagator $G(p)$ satisfies
the bounds
\begin{equation}
\frac{1}{N_c^2 - 1} \, \frac{1}{\lambda_1} \, \sum_b \,
  | {\widetilde \psi_1(b,p)} |^2
   \, \leq \, G(p) \, \leq \, \frac{1}{\lambda_1} \; .
\label{eq:ineq}
\end{equation}
Here, $b = 1, 2, \ldots, N_c^2 - 1$ is a color index running over
the $N_c^2 - 1$ generators of SU($N_c$) and $p(k)$ is the lattice
momentum, whose components are $p_{\mu}(k) = 2 \sin{(\pi k_{\mu} /N)}$
with $k_{\mu} = 0, 1, \ldots, N-1$, where $N$ is the lattice size.
We indicate with $\lambda_1$ the smallest nonzero eigenvalue
of the FP matrix ${\cal M}(b,x;c,y)$, with $\, \psi_1(b,x) \,$
the corresponding eigenvector and with ${\widetilde \psi_1(b,p)}$
its Fourier transform, for which we use the definition
\begin{equation}
{\widetilde \psi_i(b,p)} \, = \, \frac{1}{\sqrt{V}}
     \sum_x \psi_i(b,x) \, e^{- 2 \pi i k \cdot x} \; ,
\label{eq:psifourier}
\end{equation}
where $V = N^d$ is the lattice volume (in the $d$-dimensional case)
and we indicate with $\psi_i(b,x)$ a generic eigenvector of
${\cal M}(b,x;c,y)$. These bounds were obtained by considering the
equations
\begin{equation}
G(p) \, = \, \frac{1}{N_c^2 - 1} \sum_{x\mbox{,}\, y\mbox{,}\, b}
                 \frac{e^{- 2 \pi i \, k \cdot (x - y)}}{V}\,
        {\cal M}^{- 1}(b,x;b,y) 
     \, = \, \frac{1}{N_c^2 - 1} \, \sum_{i: \lambda_i \neq 0} \,
                      \sum_b \, \lambda_i^{-1} \,
                    | {\widetilde \psi_i(b,p)} |^2 \; ,
\label{eq:Gp}
\end{equation}
valid in the space orthogonal to the kernel of the FP matrix. Then,
using the inequalities\footnote{Note that, at finite volume, the
spectrum is discrete. For simplicity, we assume that the 
nonzero eigenvalues are non-degenerate. This is usually the case
for nontrivial gauge configurations $A\neq 0$.}
\begin{equation}
0 \, < \, \lambda_1 \, < \, \lambda_2 \, < \, \lambda_3 \, \ldots \; ,
\label{eq:ineqlambdas}
\end{equation}
we can write\footnote{Clearly, due to the relations (\ref{eq:ineqlambdas}),
all the inequalities considered in this section usually hold strictly.
However, we write all of them as non-strict inequalities, since we are
mainly interested in estimates for the lower and upper bounds of
the ghost propagator.}
\begin{equation}
\frac{1}{N_c^2 - 1} \, \frac{1}{\lambda_1} \, \sum_b \,
  | {\widetilde \psi_1(b,p)} |^2 \, \leq \, G(p) 
\end{equation}
and
\begin{equation}
G(p) \, \leq \, \frac{1}{N_c^2 - 1} \, \frac{1}{\lambda_1}
          \, \sum_{i: \lambda_i \neq 0} \, \sum_b \,
             | {\widetilde \psi_i(b,p)} \, |^2 \; .
\label{eq:ineq2ini}
\end{equation}
Thus, after summing and subtracting in Eq.\ (\ref{eq:ineq2ini}) the
contributions from the eigenvectors spanning the kernel of
${\cal M}(b,x;c,y)$ and using the completeness relation
\begin{equation}
\sum_{i} \, \psi_i(b,x) \, \psi_i^*(c,y)
         \, = \, \delta_{bc} \, \delta_{xy} \; ,
\label{eq:complet}
\end{equation}
where $^*$ indicates complex conjugation, we find
\begin{equation}
G(p) \, \leq \, \frac{1}{\lambda_1}
     \, \left[ \, 1 \,-\, \frac{1}{N_c^2 - 1}
            \sum_{j: \lambda_j = 0} \, \sum_b \,
                | {\widetilde \psi_j(b,p)} \, |^2 \, \right] \; .
\label{eq:ineq2}
\end{equation}
In Landau gauge the eigenvectors corresponding to the null
eigenvalue are constant modes and, for any nonzero momentum $p$,
we immediately obtain the upper bound in Eq.\ (\ref{eq:ineq}).
One should note that the above upper bound becomes an equality
if and only if the eigenvalues of the FP matrix are all degenerate 
and equal to $\lambda_1$.

\vskip 3mm
The above result can be (systematically) improved, by
considering also the eigenvalues $\lambda_2, \lambda_3, \ldots$
and the inequalities (\ref{eq:ineqlambdas}). For example, if
$\lambda_2$ and ${\widetilde \psi_2(b,p)}$ are also known, we
have
\begin{equation}
\label{eq:ineq2low}
\frac{1}{N_c^2 - 1} \, \sum_b \, \left[ \,
    \frac{1}{\lambda_1} \, | {\widetilde \psi_1(b,p)} |^2
      \, + \, \frac{1}{\lambda_2} \, | {\widetilde \psi_2(b,p)} |^2
       \, \right] \, \leq \, G(p)
\end{equation}
and
\begin{equation}
G(p) \, \leq \, \frac{1}{N_c^2 - 1} \, \sum_b \, \left[ \,
   \frac{1}{\lambda_1} \, | {\widetilde \psi_1(b,p)} |^2
      \, + \, \frac{1}{\lambda_2} \,
        \sum_{i: \lambda_i \neq 0, \lambda_1} \,
            | {\widetilde \psi_i(b,p)} \, |^2 \right] \; .
\label{eq:ineq2ininew}
\end{equation}
After adding and subtracting the quantity
\begin{equation}
\frac{1}{N_c^2 - 1} \, \frac{1}{\lambda_2} \, \sum_b \,
      \sum_{j: \lambda_j = 0, \lambda_1} \,
           | {\widetilde \psi_j(b,p)} \, |^2
\end{equation}
and using again the completeness relation (\ref{eq:complet}),
the above upper bound can also be written as
\begin{equation}
G(p) \, \leq \, \frac{1}{N_c^2 - 1} \, \sum_b
   \, \Biggl\{ \, \frac{1}{\lambda_1} \,
       | {\widetilde \psi_1(b,p)} |^2 \,
     + \, \frac{1}{\lambda_2} \, \Bigl[ 1 - \,
          | {\widetilde \psi_1(b,p)} |^2 \, - \,
     \sum_{j: \lambda_j = 0} \,
        | {\widetilde \psi_j(b,p)} \, |^2 \Bigr] \, \Biggr\}
\end{equation}
and, for nonzero momenta $p$, we are left with
\begin{eqnarray}
G(p) & \leq & \frac{1}{N_c^2 - 1} \, \sum_b \, \left\{ \,
    \frac{1}{\lambda_1} \,
         | {\widetilde \psi_1(b,p)} |^2 \,
     + \, \frac{1}{\lambda_2} \, \left[ \, 1 \, - \,
        | {\widetilde \psi_1(b,p)} |^2 \, \right] \, \right\}
\nonumber \\[2mm]
 & = & \left( \frac{1}{\lambda_1} - \frac{1}{\lambda_2} \right)
     \, \left( \, \frac{1}{N_c^2 - 1} \, \sum_b \, 
          | {\widetilde \psi_1(b,p)} |^2 \, \right) \,
                  + \, \frac{1}{\lambda_2} \; .
\label{eq:newupper}
\end{eqnarray}
One can easily check that the new upper bound (\ref{eq:newupper})
is in general an improved bound compared to the original one [see Eq.\
(\ref{eq:ineq}) above]. Indeed, since the FP matrix is obtained
from a second-order expansion, it can always be written
in a symmetric form and its eigenvectors $\psi_i(b,x)$ can be
assumed to be orthogonal to each other and normalized as
$ \sum_{b, x} \, | \psi_i(b,x) |^2 = 1$. Then, we have that\footnote{Using
the normalization condition for the eigenvalues $\psi_i(b,x)$
and considering the plane-wave functions $\phi_p(b,x) =
\delta^{bc} \exp{(2 \pi i k \cdot x)} / \sqrt{V}$, with $c= 1, 2, \ldots,
N_c^2 - 1$, the inequality (\ref{eq:ineqpsi}) is just a
consequence of the Cauchy-Schwarz-Bunyakovsky inequality and
of the relation $\sum_{b, x} \, | \phi_p(b,x) |^2 = N_c^2 - 1$.}
\begin{equation}
\frac{1}{N_c^2 - 1} \, \sum_b \,
 | {\widetilde \psi_1(b,p)} |^2 \leq 1
\label{eq:ineqpsi}
\end{equation}
and the r.h.s.\ of the inequality (\ref{eq:newupper}) becomes
\begin{equation}
\left( \frac{1}{\lambda_1} - \frac{1}{\lambda_2} \right)
     \, \left( \frac{1}{N_c^2 - 1} \, \sum_b \, 
          | {\widetilde \psi_1(b,p)} |^2 \right)
                  + \frac{1}{\lambda_2}
 \leq \frac{1}{N_c^2 - 1} \, \sum_b \, \left\{ \,
    \left[ \frac{1}{\lambda_1} - \frac{1}{\lambda_2} \right]
     + \, \frac{1}{\lambda_2} \, \right\} \, = \,
                 \frac{1}{\lambda_1} \; .
\end{equation}
For intermediate values of the lattice size $N$, one expects
$\lambda_1 < \lambda_2$ and, at the same time, the upper
bound (\ref{eq:newupper}) should be (numerically) strictly
smaller than the previous upper bound $1/\lambda_1$. On the
other hand, in the infinite-volume limit the spectrum of the
FP matrix becomes continuous.\footnote{For a numerical
verification of this statement see, for example, Ref.\
\cite{Greensite:2004ur} (Coulomb gauge) and Ref.\
\cite{Sternbeck:2005vs} (Landau gauge).} Thus, for
very large values of $N$, the two smallest eigenvalues
$\lambda_1$ and $\lambda_2$ should be almost equal,
i.e.\ $\lambda_1 \ltapprox \lambda_2$, and the two upper
bounds will also be (almost) numerically equivalent. Therefore,
the main conclusion of Ref.\ \cite{Cucchieri:2008fc} is not
modified: if $\lambda_1$ behaves as $\, N^{-2-\alpha} \,$
in the infinite-volume limit, we have that $\alpha > 0$ is a
necessary condition to obtain a ghost propagator $G(p)$ enhanced
in the infrared region (compared to the tree-level behavior $1/p^2$).
Note that the bounds discussed here are valid for all momenta $p$,
but we will generally be interested in the behavior of $G(p)$
for the smallest nonzero momentum $p_{min}$.


\section{Bounds for $G(p_{min})$: numerical results}
\label{sec:num}

\begin{figure}
\centering
\includegraphics[trim=80 0 40 0, clip, scale=1.00]{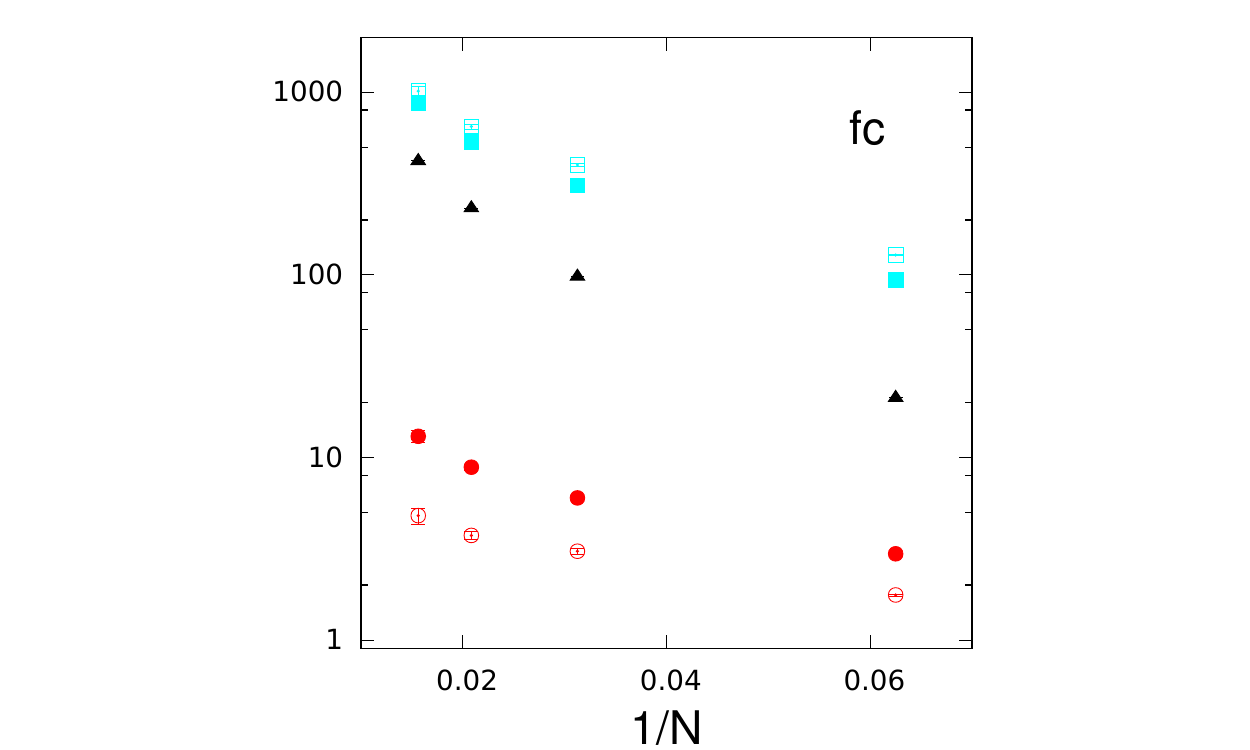}
\hspace{-5mm}
\includegraphics[trim=80 0 40 0, clip, scale=1.00]{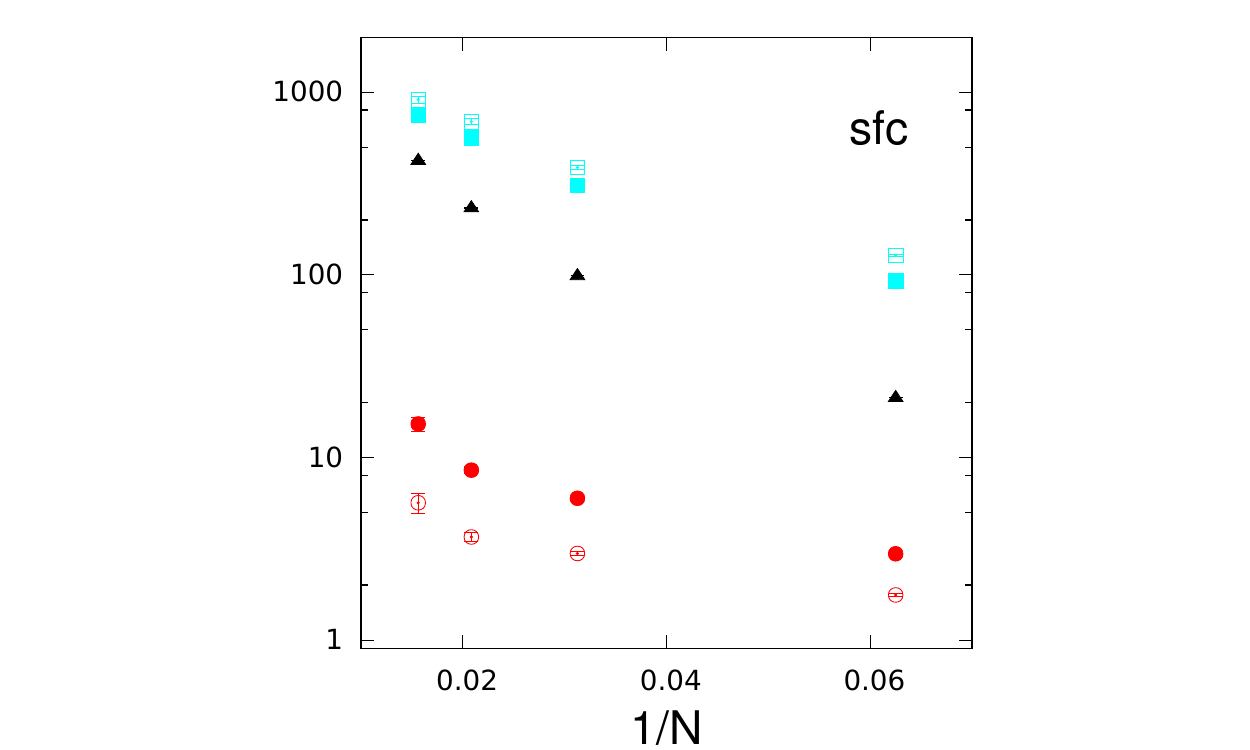}
\\
\vspace{5mm}
\includegraphics[trim=80 0 40 0, clip, scale=1.00]{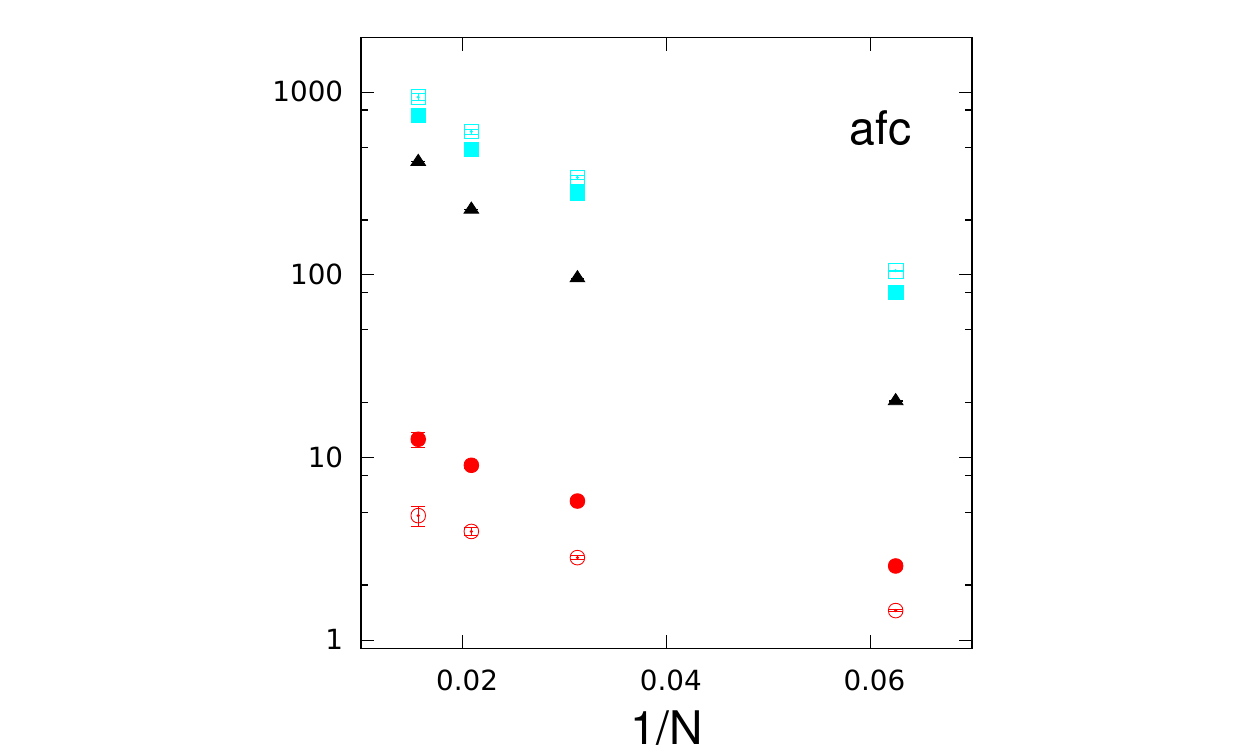}
\hspace{-5mm}
\includegraphics[trim=80 0 40 0, clip, scale=1.00]{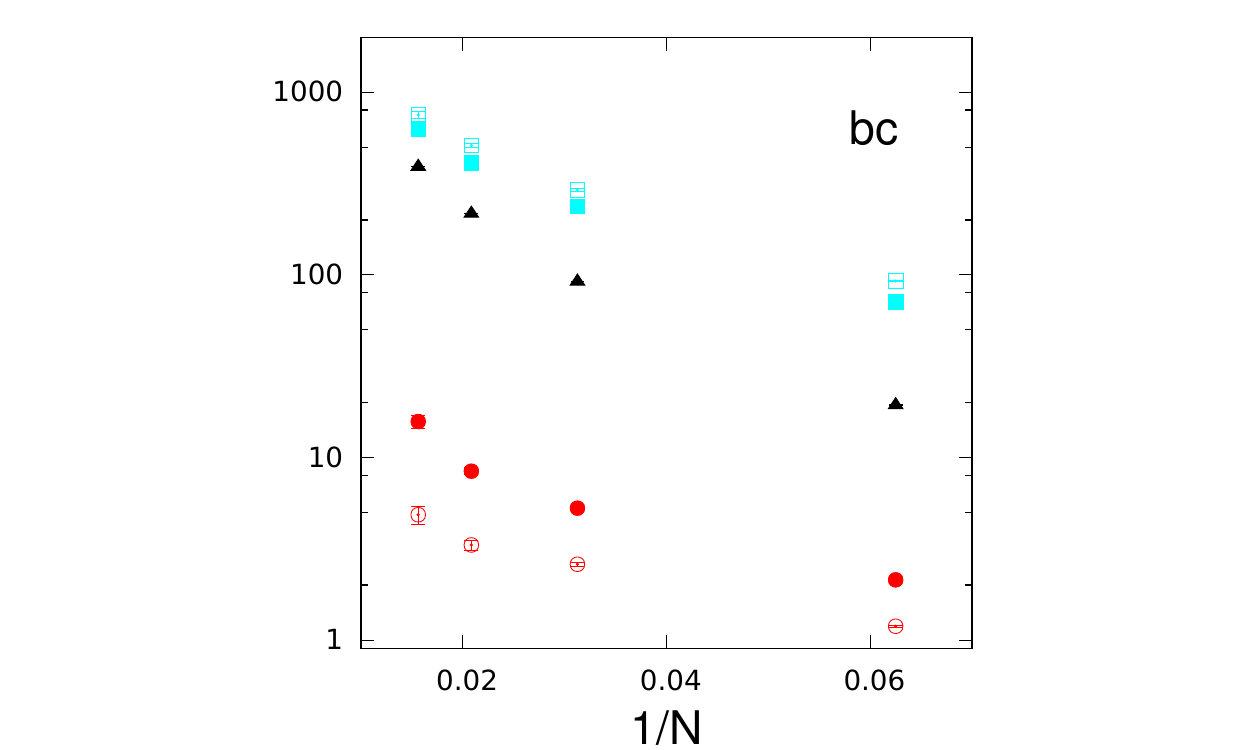}
\vspace{2mm}
\caption{\label{fig:bounds} The ghost propagator $G(p_{min})$
(full triangles), the lower bounds in Eqs.\ (\ref{eq:ineq})
and (\ref{eq:ineq2low}) (respectively empty and full circles)
and the upper bounds in Eqs.\ (\ref{eq:ineq}) and
(\ref{eq:newupper}) (respectively empty and full squares) as a
function of the inverse lattice size $1/N$. All quantities
are in lattice units. Four types of gauge-fixing prescription are 
considered (see Section \ref{sec:num}): {\em fc} (upper left plot),
{\em sfc} (upper right plot), {\em afc} (lower left plot)
and {\em bc} (lower right plot). Note the logarithmic scale
on the $y$ axis. 
The data points represent averages over gauge configurations, 
error bars correspond to one standard deviation.
(We consider the statistical error only.)
}
\end{figure}

\begin{figure}
\centering
\includegraphics[trim=58 0 50 0, clip, scale=1.00]{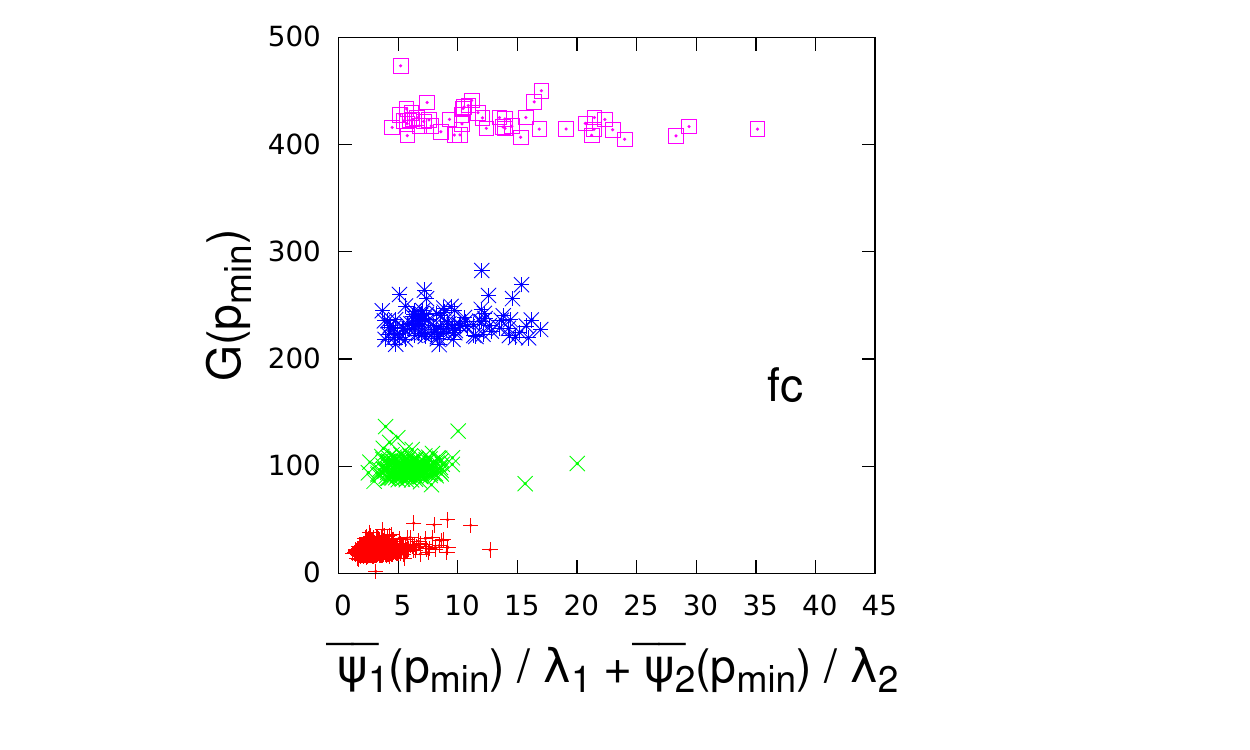}
\hspace{-12mm}
\includegraphics[trim=58 0 50 0, clip, scale=1.00]{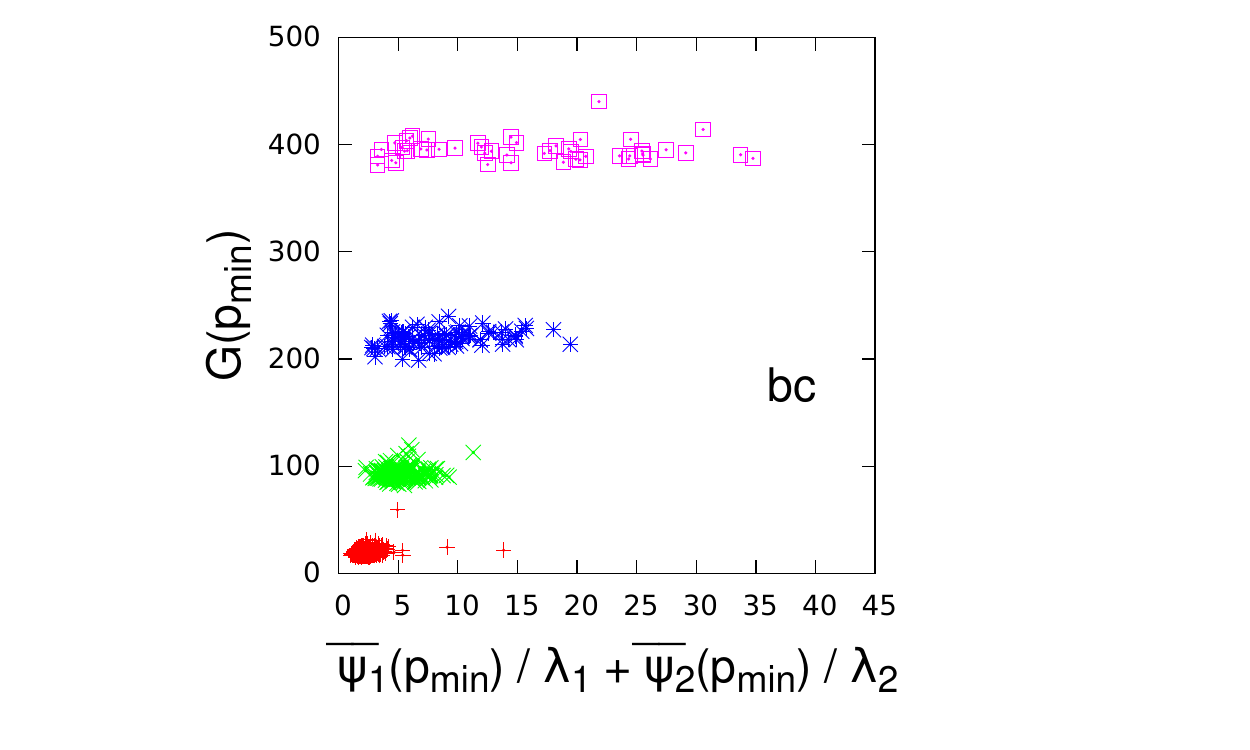}
\vspace{2mm}
\caption{\label{fig:proj-gmin} The ghost propagator
$G(p_{min})$ vs.\ the contribution to its value
from the two smallest nonzero eigenvalues of the
FP matrix [first two terms in Eq.\ (\ref{eq:Gp})]
for the lattice volumes $V = 16^4$ (red $+$), $32^4$
(green $\times$), $48^4$ (blue $*$) and $64^4$
(magenta $\square$) and for all the configurations
used in our analysis. The quantities are in
lattice units. Two types of Gribov copies are
considered (see Section \ref{sec:num}): {\em fc}
(left plot) and {\em bc} (right plot). Similar
results have been obtained for the sets of
Gribov copies indicated with {\em sfc} and {\em afc}.
}
\end{figure}

\begin{figure}
\centering
\includegraphics[trim=68 0 40 0, clip, scale=1.00]{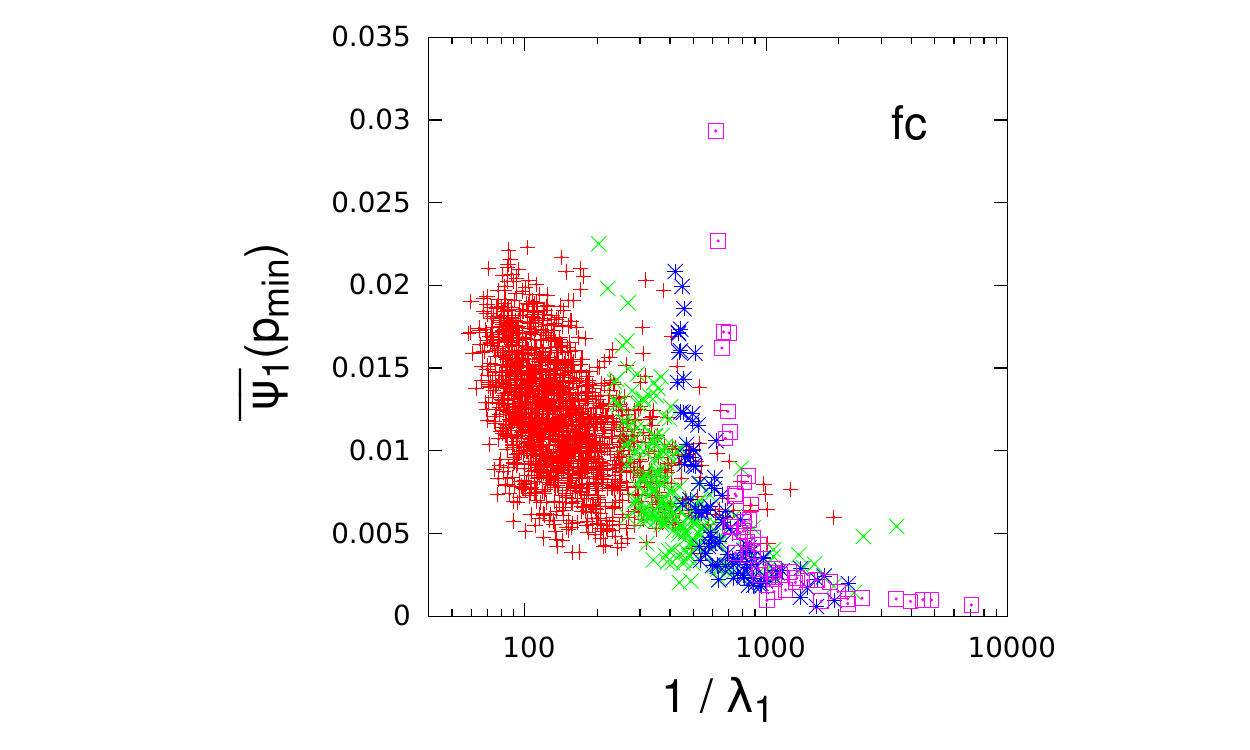}
\hspace{-2mm}
\includegraphics[trim=68 0 40 0, clip, scale=1.00]{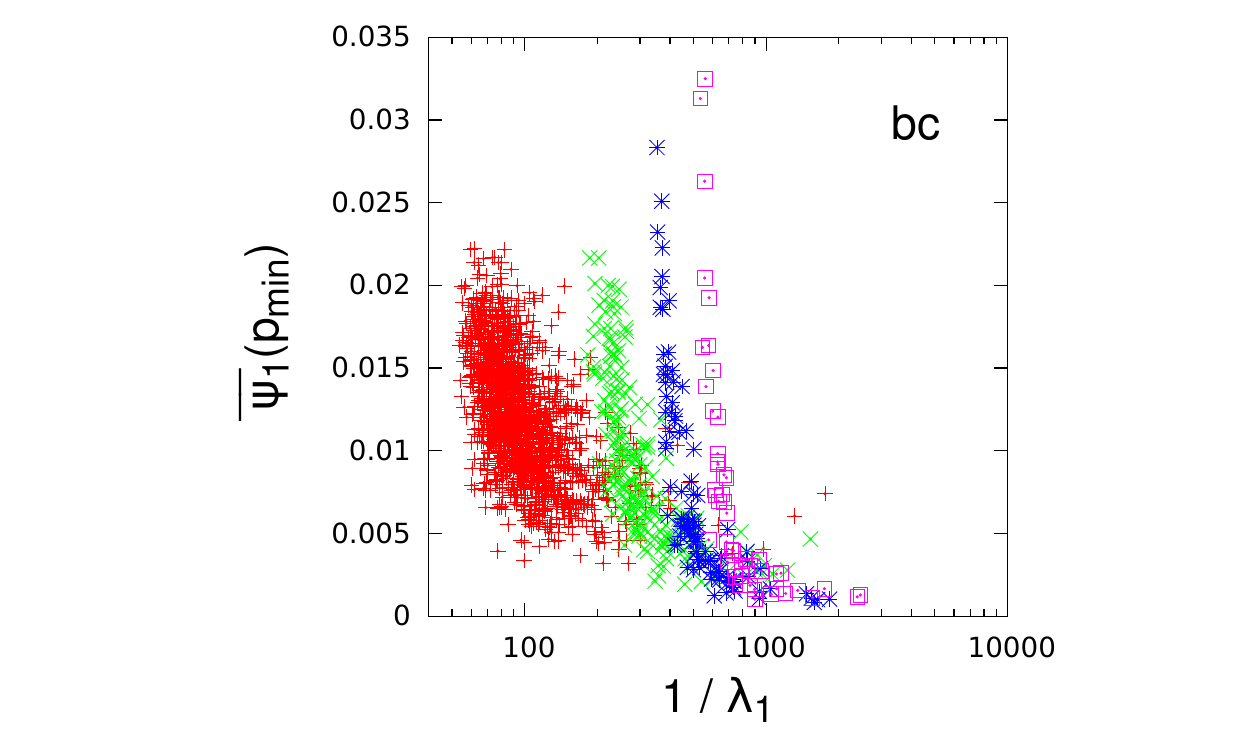}
\vspace{2mm}
\caption{\label{fig:lambdamin-proj} Plot of the
average projection $\overline{\psi}_1(p_{min})$ vs.\ the
inverse of the smallest nonzero eigenvalue $\lambda_1$
for the lattice volumes $V = 16^4$ (red $+$), $32^4$
(green $\times$), $48^4$ (blue $*$) and $64^4$
(magenta $\square$) and for all the configurations
used in our analysis. The quantities are in
lattice units. Two types of Gribov copies are
considered (see Section \ref{sec:num}): {\em fc}
(left plot) and {\em bc} (right plot). Similar
results have been obtained for the sets of
Gribov copies indicated with {\em sfc} and {\em afc}.
Note the logarithmic scale on the $x$ axis.
}
\end{figure}

\begin{figure}
\centering
\includegraphics[trim=68 0 40 0, clip, scale=1.00]{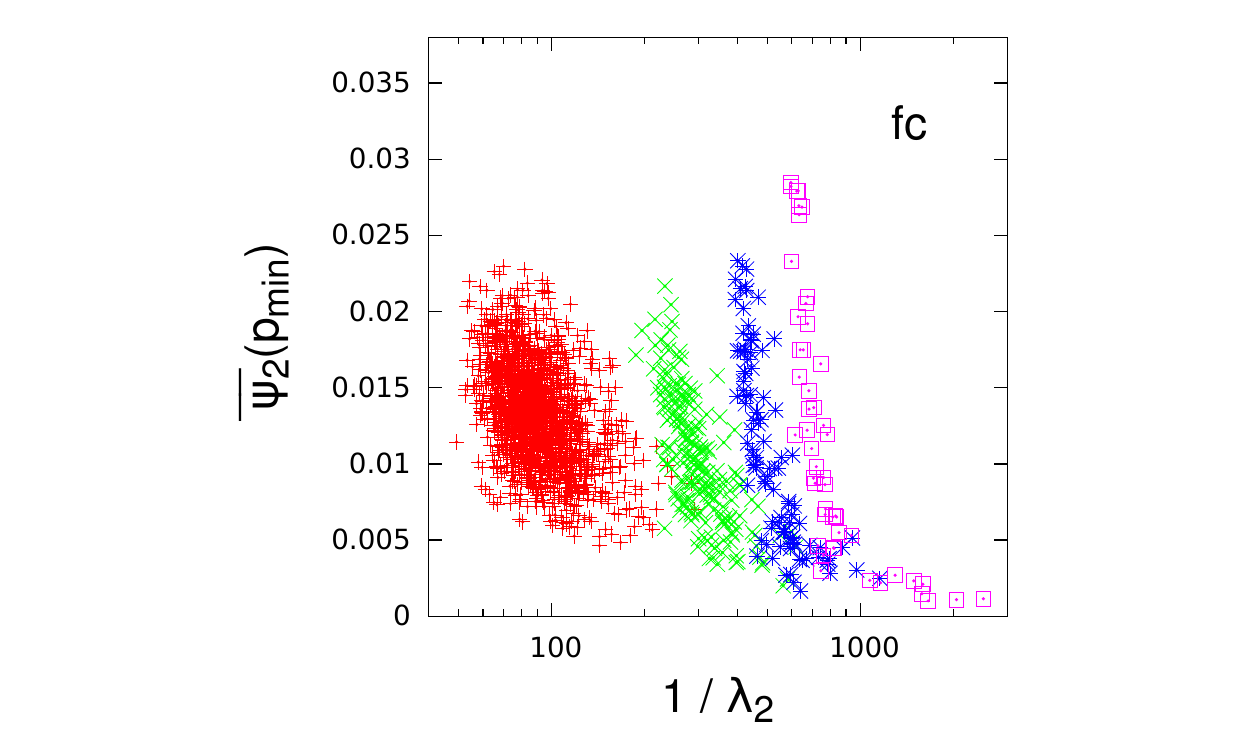}
\hspace{-2mm}
\includegraphics[trim=68 0 40 0, clip, scale=1.00]{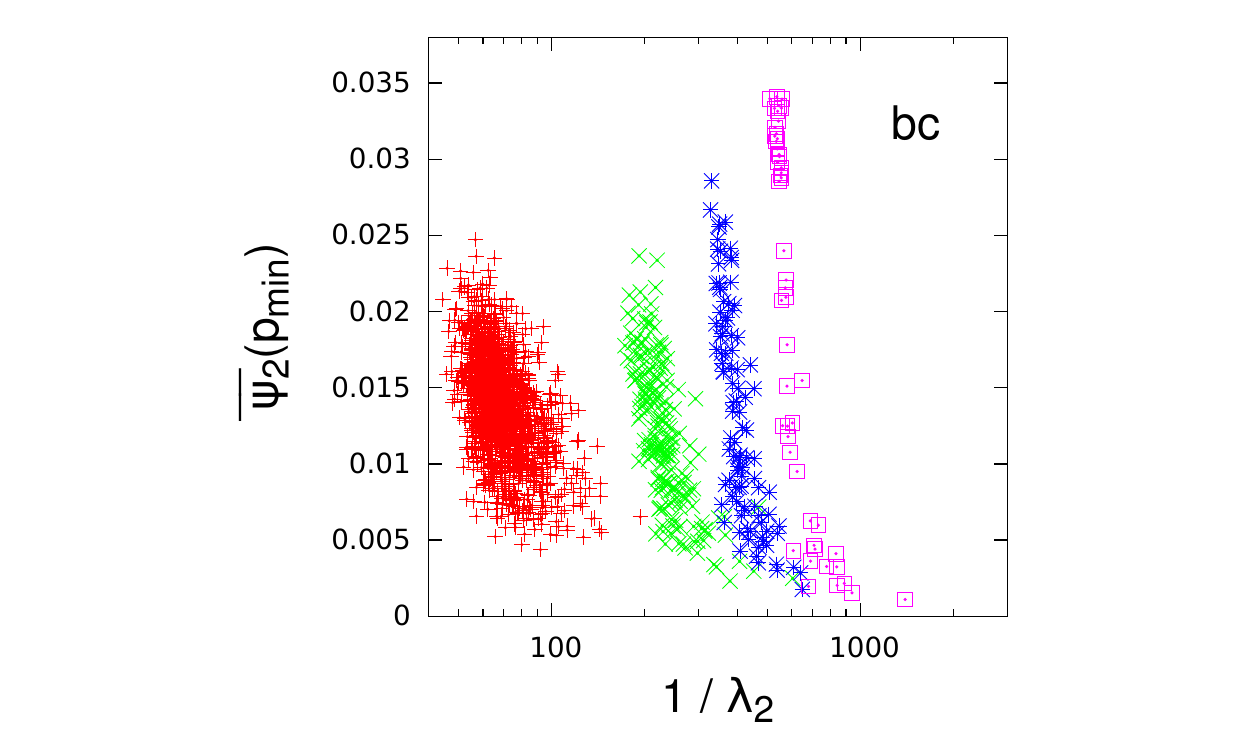}
\vspace{2mm}
\caption{\label{fig:lambda2-proj2} Plot of the
average projection $\overline{\psi}_2(p_{min})$ vs.\ the
inverse of the second smallest nonzero eigenvalue
$\lambda_2$ for the lattice volumes $V = 16^4$ (red $+$),
$32^4$ (green $\times$), $48^4$ (blue $*$) and $64^4$
(magenta $\square$) and for all the configurations
used in our analysis. Two types of Gribov copies are
considered (see Section \ref{sec:num}): {\em fc}
(left plot) and {\em bc} (right plot). Similar
results have been obtained for the sets of
Gribov copies indicated with {\em sfc} and {\em afc}.
Note the logarithmic scale on the $x$ axis.
}
\end{figure}

\begin{figure}
\centering
\includegraphics[trim=68 0 40 0, clip, scale=1.00]{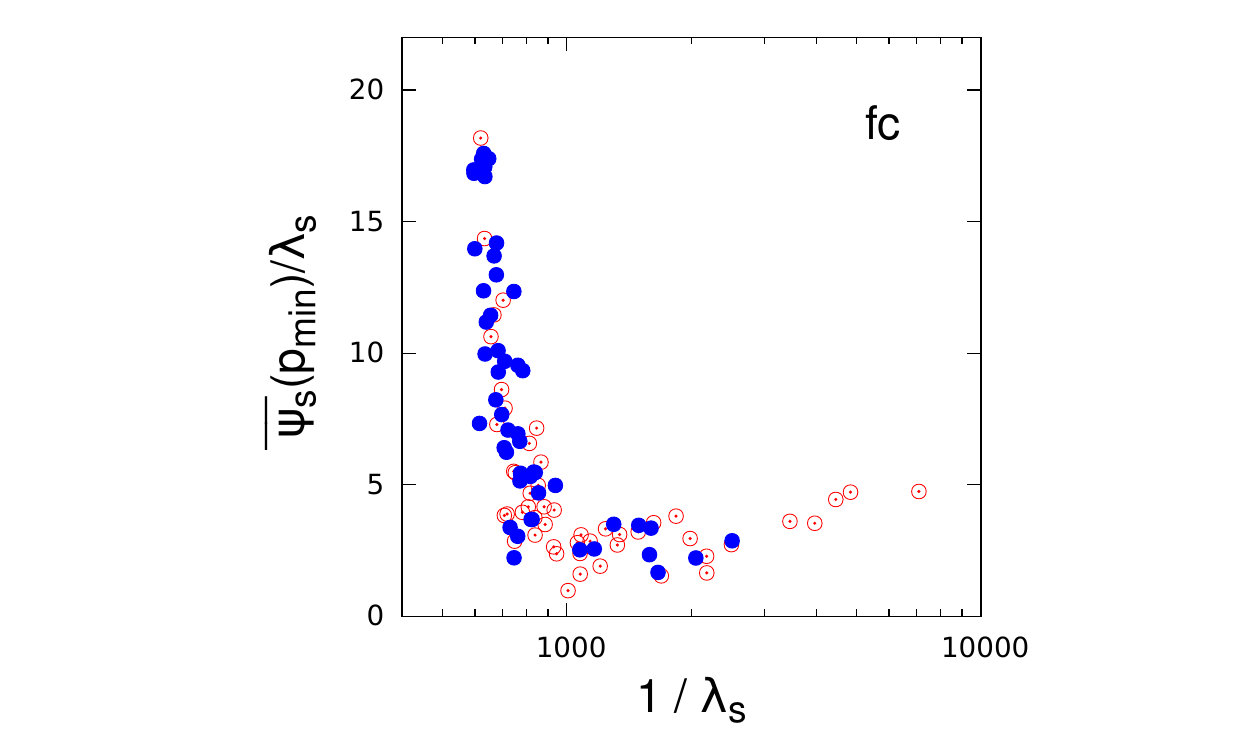}
\hspace{-2mm}
\includegraphics[trim=68 0 40 0, clip, scale=1.00]{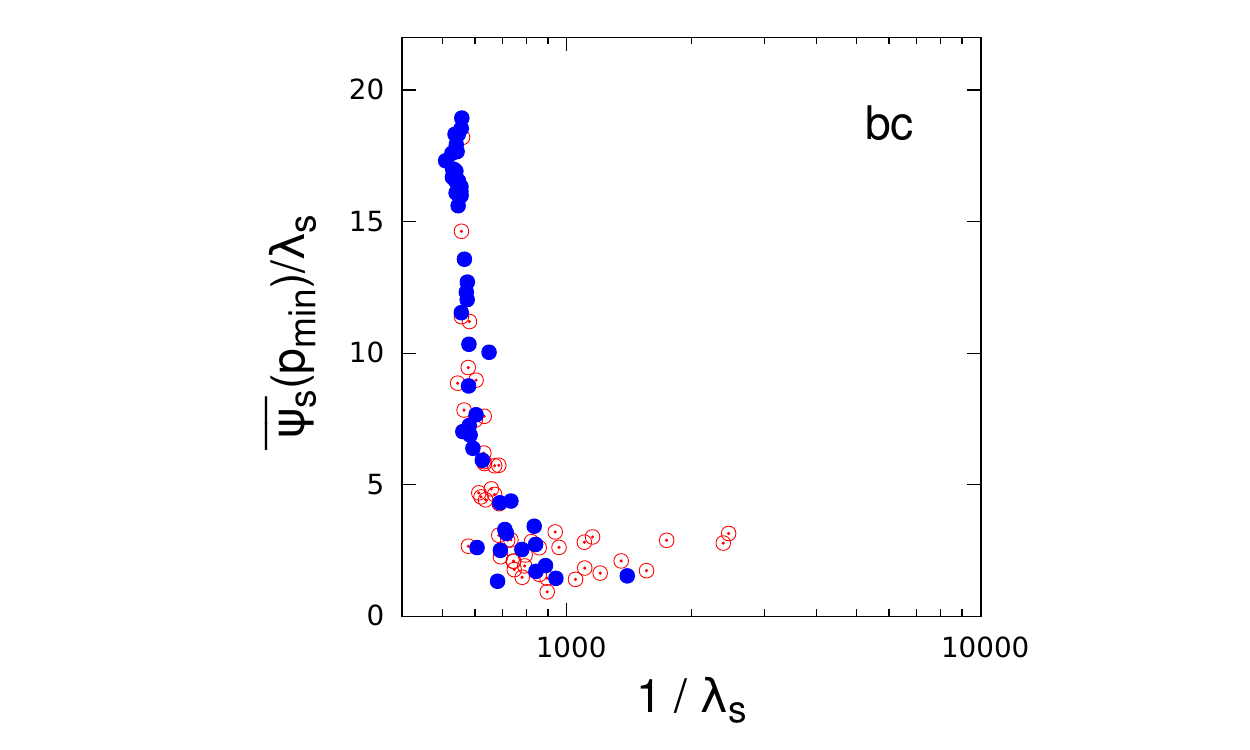}
\vspace{2mm}
\caption{\label{fig:distri-i} Plot of the
average projections $\overline{\psi}_s(p_{min}) / \lambda_s$ vs.\ the
inverse eigenvalue $1/\lambda_s$ for the lattice volume $V = 64^4$
and $s=1$ (empty circles) or $s=2$ (full circles), for all
the configurations used in our analysis. The quantities are
in lattice units. Two types of Gribov copies are
considered (see Section \ref{sec:num}): {\em fc}
(left plot) and {\em bc} (right plot). Similar
results have been obtained for the sets of
Gribov copies indicated with {\em sfc} and {\em afc}.
Note the logarithmic scale on the $x$ axis.
}
\end{figure}

\begin{figure}
\centering
\includegraphics[trim=68 0 40 0, clip, scale=1.00]{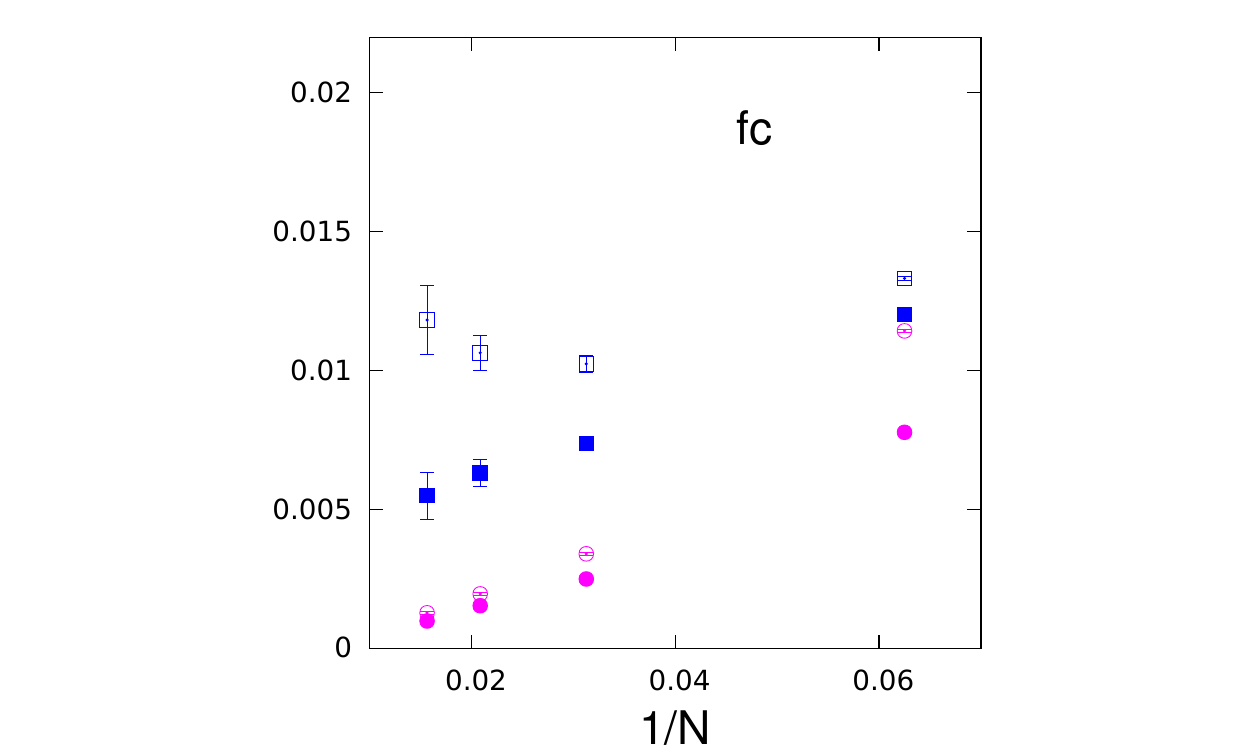}
\hspace{-2mm}
\includegraphics[trim=68 0 40 0, clip, scale=1.00]{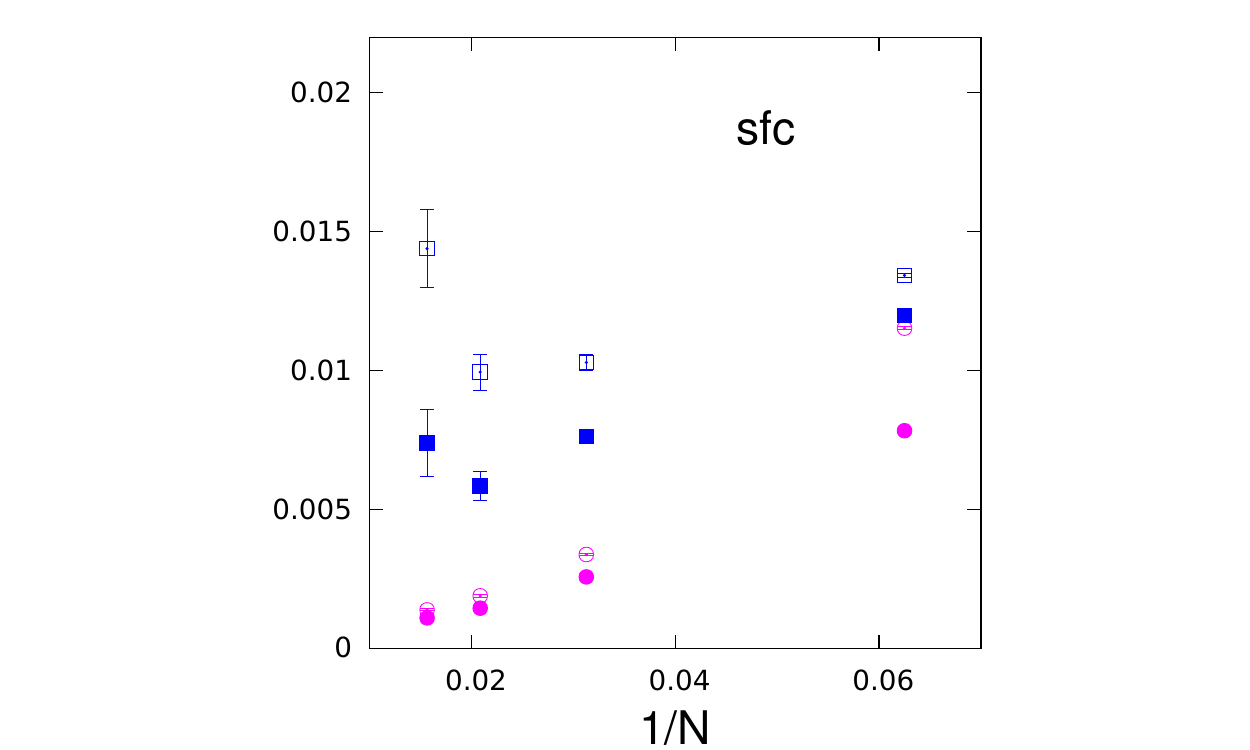}
\\
\vspace{5mm}
\includegraphics[trim=68 0 40 0, clip, scale=1.00]{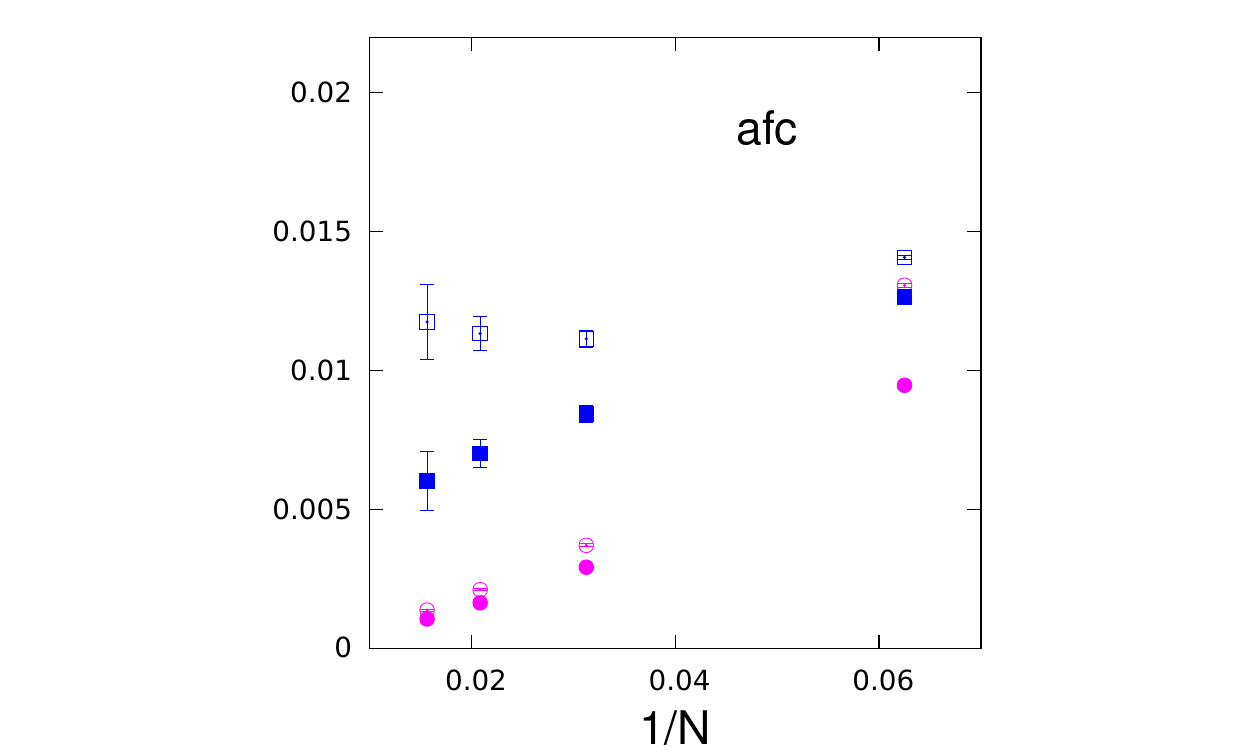}
\hspace{-2mm}
\includegraphics[trim=68 0 40 0, clip, scale=1.00]{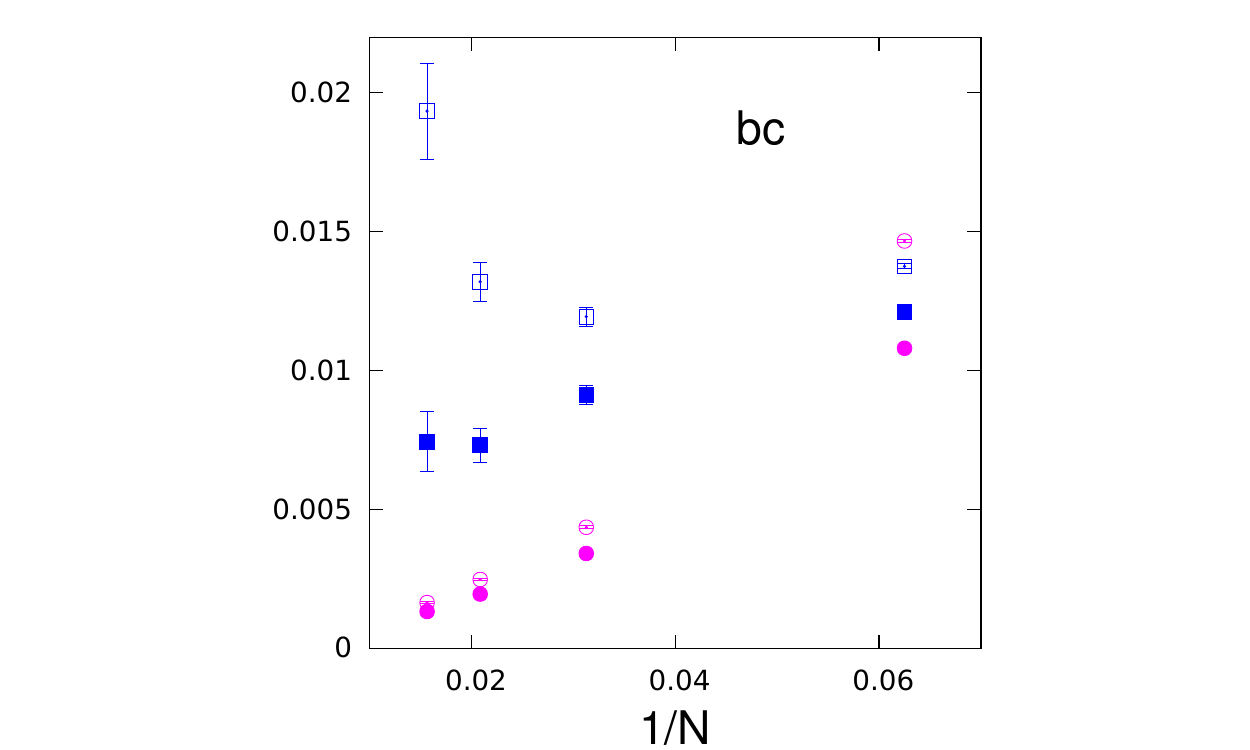}
\vspace{2mm}
\caption{\label{fig:lambda} The two smallest nonzero eigenvalues
$\lambda_1$ (full circles) and $\lambda_2$ (empty circles)
and the two average projections $\overline{\psi}_1(p_{min})$
(full squares) and $\overline{\psi}_2(p_{min})$ (empty squares)
[see Eq.\ (\ref{eq:proj})], as a function of the inverse
lattice size $1/N$. All quantities are in lattice units.
Four types of gauge-fixing prescription are considered (see 
Section \ref{sec:num}):
{\em fc} (upper left plot), {\em sfc} (upper right plot), 
{\em afc} (lower left plot) and {\em bc} (lower right plot).
The data points represent averages over gauge configurations, 
error bars correspond to one standard deviation.
(We consider the statistical error only.)
}
\end{figure}

\begin{figure}
\centering
\includegraphics[trim=68 0 40 0, clip, scale=1.00]{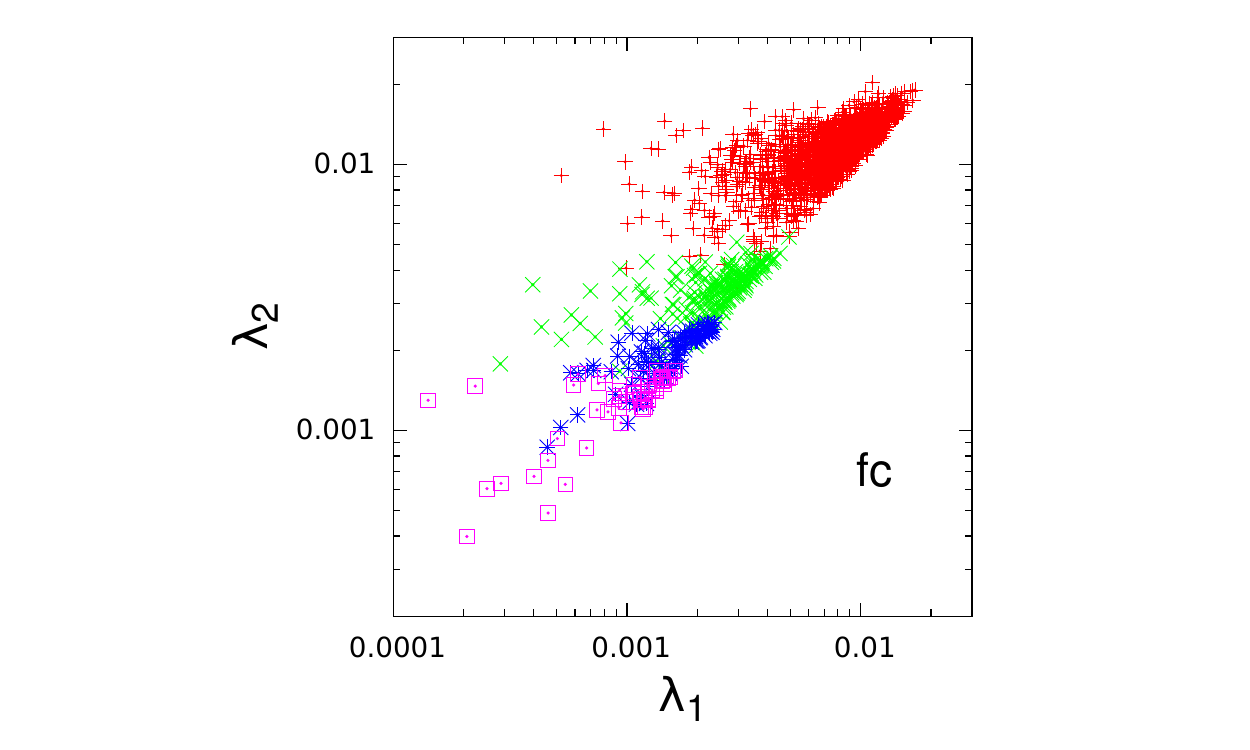}
\hspace{-2mm}
\includegraphics[trim=68 0 40 0, clip, scale=1.00]{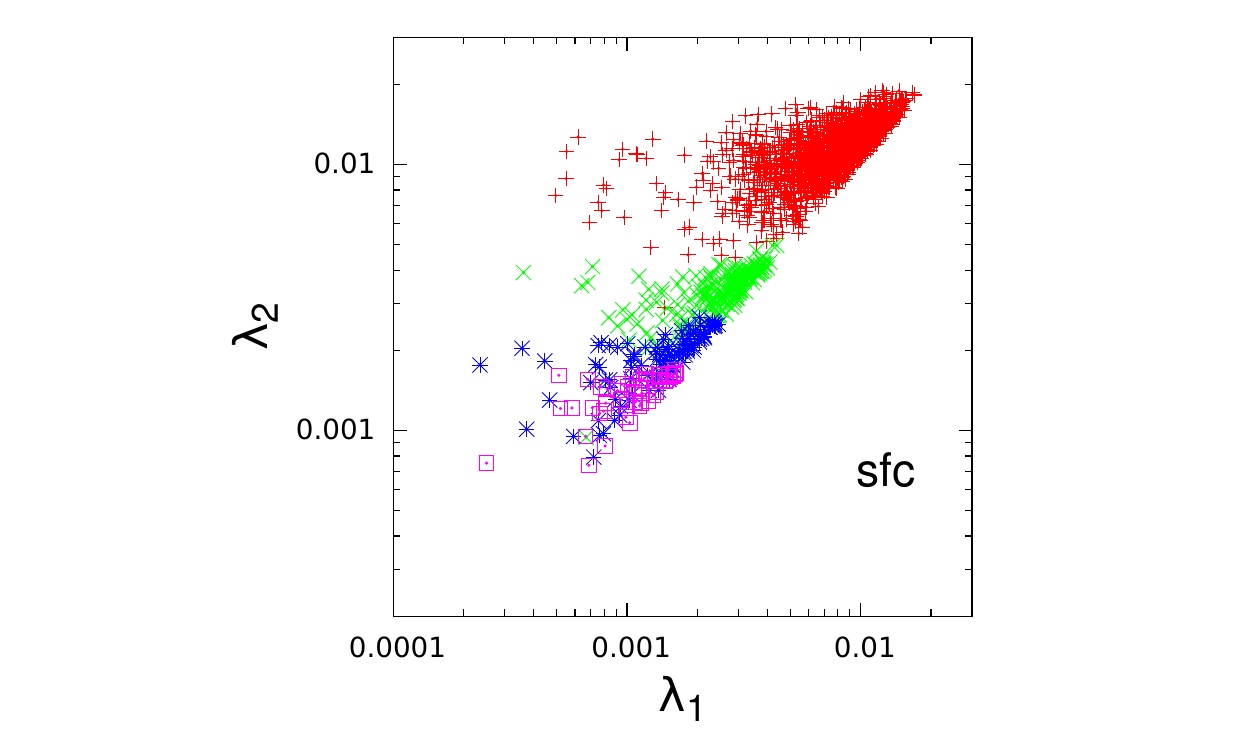}
\\
\vspace{5mm}
\includegraphics[trim=68 0 40 0, clip, scale=1.00]{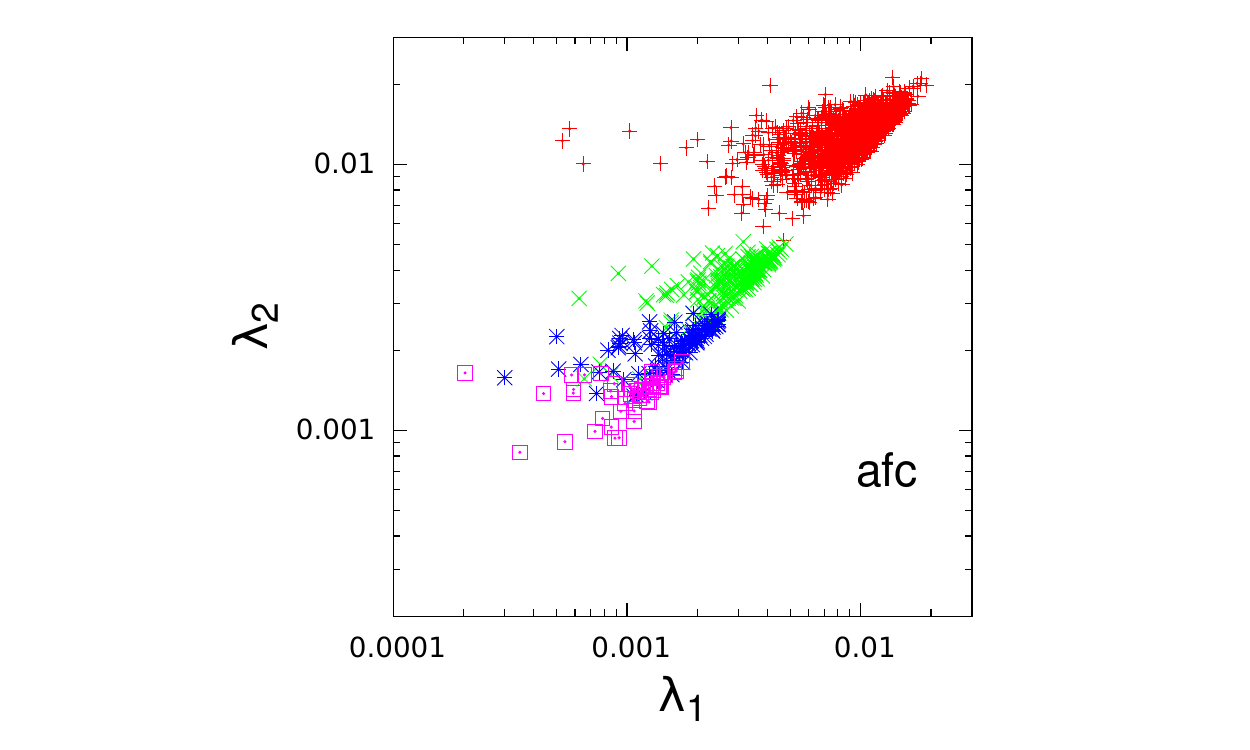}
\hspace{-2mm}
\includegraphics[trim=68 0 40 0, clip, scale=1.00]{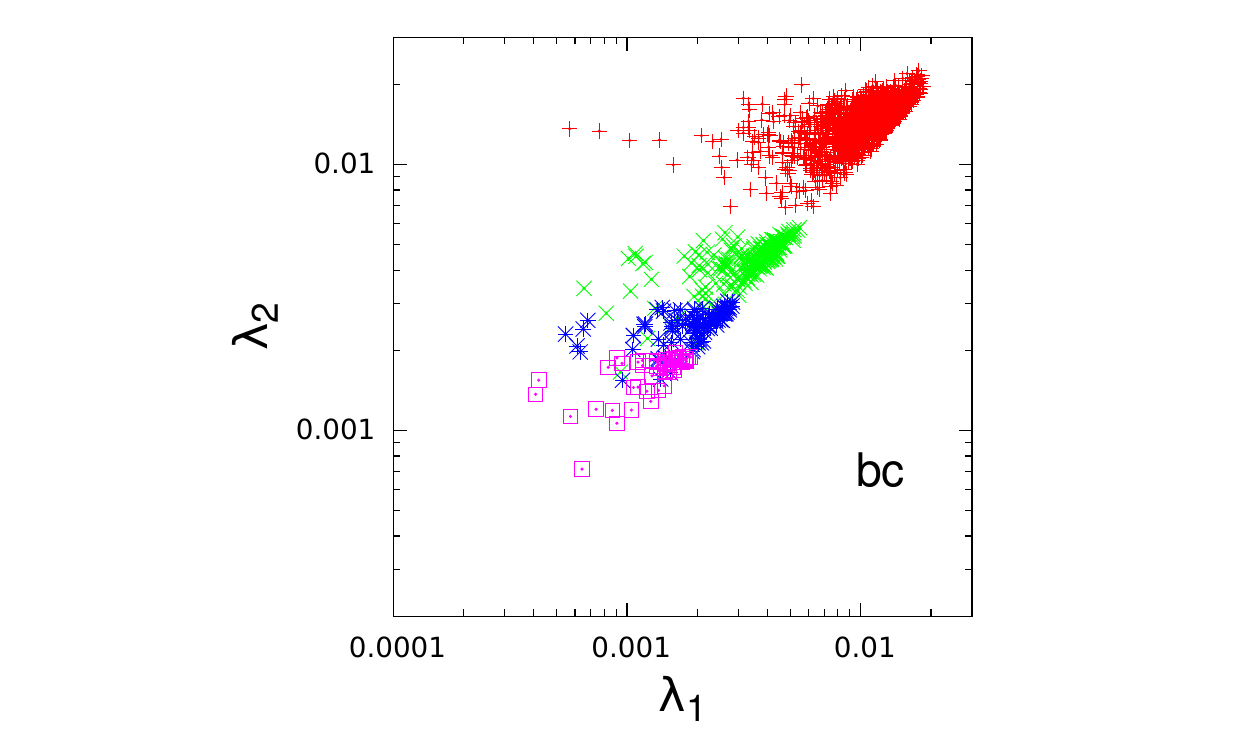}
\vspace{2mm}
\caption{\label{fig:lambdamin-lambda2} Plot of the
second smallest nonzero eigenvalue $\lambda_2$ vs.\
the smallest nonzero eigenvalue $\lambda_1$ for
the lattice volumes $V = 16^4$ (red $+$), $32^4$
(green $\times$), $48^4$ (blue $*$) and $64^4$
(magenta $\square$) and for all the configurations
used in our analysis. The quantities are in lattice
units. Four types of Gribov copies are considered
(see Section \ref{sec:num}): {\em fc} (upper left plot),
{\em sfc} (upper right plot), {\em afc} (lower left
plot) and {\em bc} (lower right plot).
Note the logarithmic scale on both axes.
}
\end{figure}

\begin{figure}
\centering
\hspace{-6.0mm}
\includegraphics[scale=1.00]{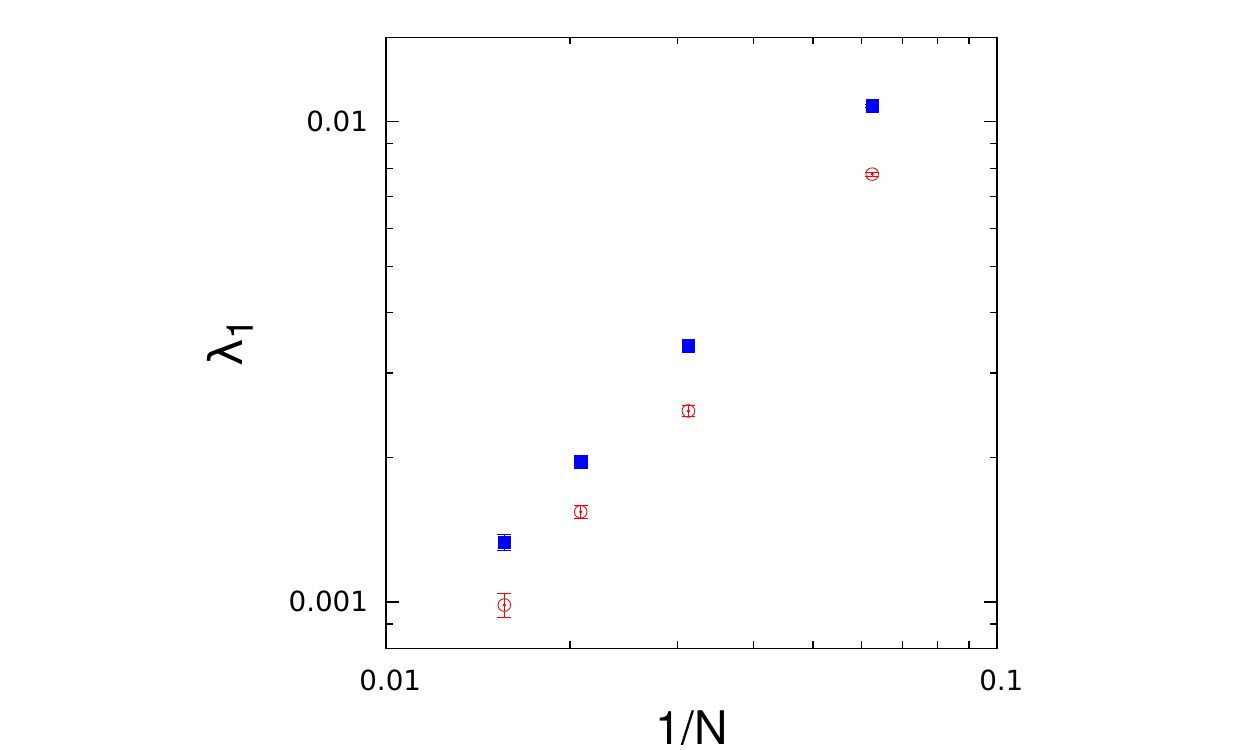}
\\
\vspace{5mm}
\includegraphics[scale=1.00]{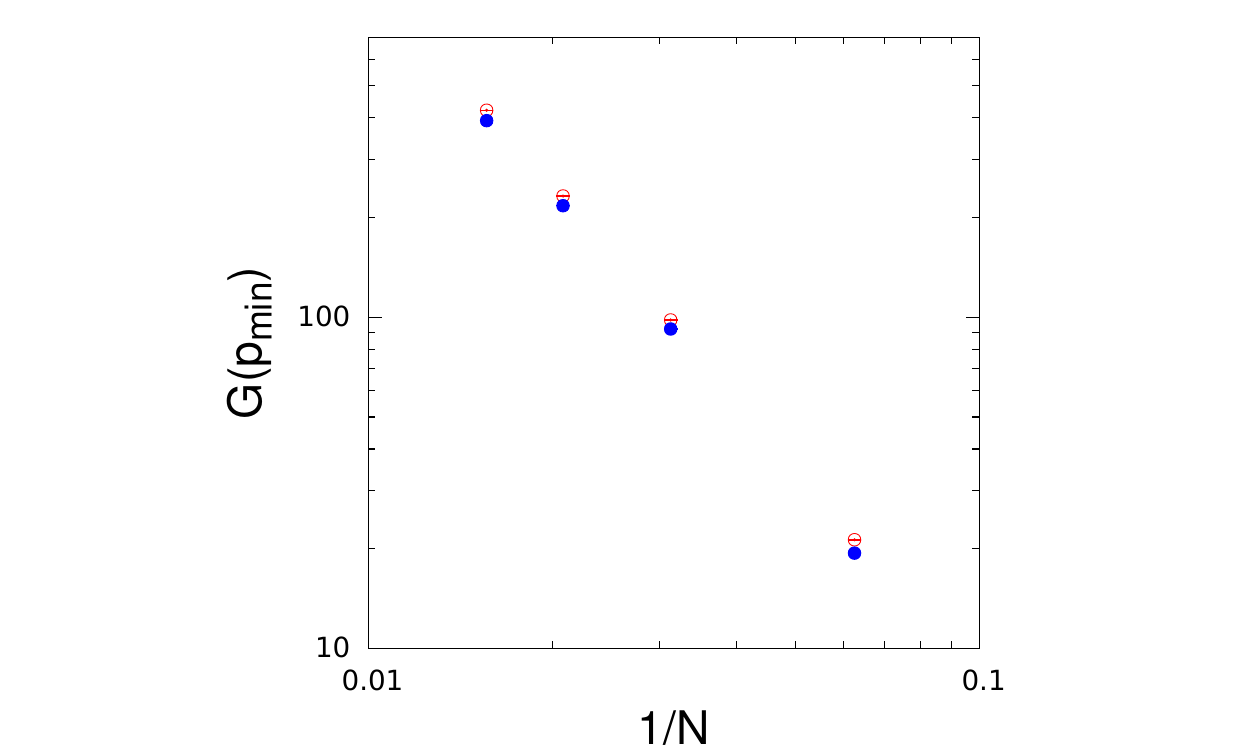}
\vspace{2mm}
\caption{\label{fig:comparison} The smallest nonzero eigenvalue
$\lambda_1$ (top plot) and the ghost propagator $G(p_{min})$
evaluated at the smallest nonzero momentum $p_{min}$
(bottom plot) for the {\em fc} data (empty circles) and for the
{\em bc} data (full circles), as a function of the inverse
lattice size $1/N$. All quantities are in lattice units.
Note the logarithmic scale on both axes.
The data points represent averages over gauge configurations, 
error bars correspond to one standard deviation.
(We consider the statistical error only.)
}
\end{figure}

In this section we present data for the volume dependence of
the ghost propagator at the smallest nonzero momentum $G(p_{min})$ 
and of the lower and upper
bounds presented in the previous section [see Eqs.\
(\ref{eq:ineq}), (\ref{eq:ineq2low}) and (\ref{eq:newupper})].
We also consider the dependence on the lattice volume $V$ for the
quantities entering the formulae of the bounds, i.e.\
the two smallest nonzero eigenvalues $\lambda_1$,
$\lambda_2$ and the (average) projections
\begin{eqnarray}
\overline{\psi}_s(p_{min}) & \equiv &
\frac{1}{N_c^2 - 1} \sum_b \;
\frac{1}{d} \sum_{\mu} \; \frac{1}{V} \, \left| \,
   \sum_x \psi_s(b,x) \, e^{- 2 \pi i x_{\mu}/N} \, \right|^2  \\[2mm]
& = & \frac{1}{N_c^2 - 1} \sum_b \;
\frac{1}{d} \sum_{\mu} \; \left| \,
{\widetilde \psi_s\left(b,\frac{2\pi}{N}\,{e_{\mu}}\right)} \, \right|^2 
\qquad s = 1, 2
\label{eq:proj}
\end{eqnarray}
of the eigenvectors $\psi_1(b,x)$ and $\psi_2(b,x)$ on the
plane waves, which are $d$-fold degenerate and correspond to the smallest
nonzero lattice momentum $p_{min} = 2 \sin{(\pi/N)}$. The data
have been generated using lattice numerical simulations for
the SU(2) gauge group at $\beta = 2.2$, corresponding (see for
example \cite{Bloch:2003sk}) to a lattice spacing $a$ of about $0.21$
fm $\approx 1.066$ GeV$^{-1}$. As discussed in Ref.\ \cite{Bloch:2003sk},
this value of $\beta$ is in the scaling region. We considered lattice sizes
$N = 16, 32, 48$ and 64 $\mbox{in}$ the four-dimensional
case\footnote{A total of 1600, 200, 100 and 50 thermalized
configurations were, respectively, generated for these four
lattice sizes.} and four types of prescriptions for the
numerical implementation of the minimal Landau
gauge.\footnote{For the gauge field and for the minimizing
functional defining the minimal Landau gauge, we consider the
usual (unimproved) lattice definitions (see for example Ref.\
\cite{Cucchieri:2010xr}).} More specifically, for each
thermalized lattice configuration we fixed the gauge by
using:
\begin{itemize}
\item a stochastic overrelaxation algorithm
   \cite{Cucchieri:1995pn,Cucchieri:1996jm} to obtain the first
   Gribov copy; we indicate this set of data below as first
   copy ({\em fc});
\item a smeared preconditioning \cite{Hetrick:1997yy}, followed
   by a standard extremization (again using the stochastic
   overrelaxation algorithm), to obtain the first Gribov copy;
   we will refer to this set of data as smeared first copy
   ({\em sfc});
\item a stochastic-overrelaxed simulated annealing
   \cite{Bali:1996dm} --- again to obtain the first Gribov copy ---
   where we alternated one extremization sweep with five
   annealing steps;\footnote{The complete annealing schedule
   consisted of 12500 steps with an increase of the inverse
   temperature by a factor of about 1.0007371 at each step.}
   this set of data will be called annealed first copy
   ({\em afc});
\item the same overrelaxed simulated annealing described in
   the above item but, this time, we considered for each configuration 
   16 Gribov copies, obtained by applying
   non-periodic Z(2) gauge transformations \cite{Bogolubsky:2005wf,
   Bogolubsky:2007bw}; among these copies we selected the one
   corresponding to the smallest value of the Landau-gauge
   minimizing functional; this set is indicated as best copy
   ({\em bc}).
\end{itemize}
The ghost propagator has been evaluated using a plane-wave source
\cite{Cucchieri:1997dx} and a point source \cite{Boucaud:2005gg},
employing a conjugate-gradient method (with even/odd
preconditioning) for the inversion of the FP matrix {\cal M}.
We have checked that the results obtained with these two
methods are in agreement within the statistical error. As for the
two smallest nonzero eigenvalues (and the corresponding eigenvectors)
of {\cal M},
they have been evaluated using the so-called power iteration method
(see for example \cite{acfd}). To this end, one needs to project
out the subspace of constant eigenvectors, when evaluating
$\lambda_1$, and the subspace spanned by constant eigenvectors
and by $\psi_1$, when evaluating $\lambda_2$. The same
eigenvalues have also been computed using a conjugate-gradient
minimization of the so-called Ritz functional (or Rayleigh quotient)
\begin{equation}
\zeta(\psi) \, = \,
  \frac{\left( \psi, {\cal M} \, \psi \right)}{
                \left(\psi,\, \psi \right)} \; ,
\end{equation}
based on the algorithm presented in Ref.\ \cite{Kalkreuter:1995mm}.
Also in this case we checked that the different methods give results
in agreement within their numerical accuracy.

\vskip 3mm
Results for the ghost propagator $G(p)$, evaluated at the
smallest nonzero momentum $p_{min}$ using a plane-wave
source, and for the two lower and the two upper bounds
as a function of the inverse lattice size $1/N$ are
reported in Figure \ref{fig:bounds}, for the four types of gauge-fixing 
prescription described above. We see that, as the
lattice size $N$ increases, the value of $G(p_{min})$ gets
closer to the two upper bounds. This result is in agreement
with the data presented in Ref.\ \cite{Cucchieri:2008fc}
and suggests that, in the infinite-volume limit, one
should find $G(p_{min}) \approx 1/\lambda_1$. At the
same time, $G(p_{min})$ is much larger than the two 
lower bounds (and this difference seems to increase with
the volume). This confirms the
results presented in Ref.\ \cite{Sternbeck:2005vs},
where it was shown that one usually needs to consider
the first 150--200 terms of Eq.\ (\ref{eq:Gp}) in order
to reproduce the value of $G(p_{min})$ within a few percent.
This observation is supported by the plots presented in
Figure \ref{fig:proj-gmin}. From them one can also see that
the relative spread of the contributions coming from the
first two terms of Eq.\ (\ref{eq:Gp}) is much larger than
the relative spread of the $G(p_{min})$ values, i.e.\ it is
indeed the sum of several (correlated) terms in Eq.\ (\ref{eq:Gp}) that
fixes the average value of $G(p_{min})$. Moreover, from Figures
\ref{fig:lambdamin-proj} and \ref{fig:lambda2-proj2}, it seems
that, for large lattice volumes, this large spread gets
complementary contributions from the projections
$\overline{\psi}_s(p_{min})$ and from the factors $1/\lambda_s$.
Indeed, for $N = 64$ and for $s = 1$ or 2, we can roughly
divide the configurations in two sets: those
characterized by a small, almost constant, value of $1/\lambda_s$
and by a large range of values for $\overline{\psi}_s(p_{min})$ and
those characterized by a small, almost constant, value of
$\overline{\psi}_s(p_{min})$ and by a large range of values for
$1/\lambda_s$. 
As a consequence, the quantity $\overline{\psi}_s(p_{min})/\lambda_s$
shows a considerable spread, as can be seen in Figure \ref{fig:distri-i}
(see also Figure \ref{fig:proj-gmin}).
At the same time, it is interesting to note that the
spread and the distribution of the points for
$\overline{\psi}_s(p_{min})/\lambda_s$, with $s=1,2$, are very
similar for our largest lattice volume (see again Figure
\ref{fig:distri-i}). This suggests that the two lowest eigenstates
give a similar contribution to the ghost propagator for increasing
volume,\footnote{One should stress that the eigenvectors $\psi_1$
and $\psi_2$ are always orthogonal when $\lambda_1 \neq \lambda_2$.}
as also discussed in the next paragraph.

In Figure \ref{fig:lambda} we show the statistical averages of
$\lambda_1$ (full circles), $\lambda_2$ (empty circles),
$\overline{\psi}_1(p_{min})$ (full squares) and 
$\overline{\psi}_2(p_{min})$ (empty squares) [see Eq.\
(\ref{eq:proj})] as a function of the inverse lattice size $1/N$.
As discussed at the end of Section \ref{sec:ineq},
for very large lattice sizes one expects $\lambda_1
\ltapprox \lambda_2$. Our data indeed suggest that, in the
infinite-volume limit, these two eigenvalues become close
to each other, in agreement with Ref.\ \cite{Sternbeck:2005vs}
(see their Figure 3). A nice visualization of this result
is also presented in Figure \ref{fig:lambdamin-lambda2} (note the
log scale along both axes).
As for the average projections $\overline{\psi}_s(p_{min})$, shown
in Figure \ref{fig:lambda}, we see that $\overline{\psi}_2(p_{min})$
has a minimum for $N$ around $32$ or $48$ (which correspond to the 
two middle data points) and a larger value for
$N = 64$ (which corresponds to the leftmost data point). 
The value of $\overline{\psi}_1(p_{min})$ seems strongly correlated 
with that of $\overline{\psi}_2(p_{min})$. On the other
hand, its qualitative behavior is not the same for the four
sets of data considered in our analysis. Indeed, in the
limit of large $N$, when considering the prescriptions {\em fc}
and {\em afc} we see for $\overline{\psi}_1(p_{min})$ a decreasing
behavior (at a lower rate than for the eigenvalues), while
for the choices {\em sfc} and {\em bc} the data suggest a
plateau value of the order of $7 \, \times \, 10^{-3}$.
Thus, in order to make a more conclusive analysis, it is
probably necessary to simulate at larger lattice
sizes\footnote{Let us note that the lattice size $N = 64$
at $\beta = 2.2$ corresponds essentially to infinite volume regarding
the study of the gluon propagator \cite{Cucchieri:2007md}.
Nevertheless, different quantities usually display
different finite-size effects. Moreover, also in the case
of the gluon propagator, simulations up to $V = 128^4$
at $\beta = 2.2$ were necessary in order to achieve
a clear description of its infrared behavior (see for
example Ref.\ \cite{Cucchieri:2011ig}).}.
At the same time, with only four
data points it is not possible to carry out reliable tests
of different fitting functions for the quantities considered 
in this work. For these reasons, we plan
to extend, in the near future, the numerical simulations
presented here.\footnote{One should also stress that, due to the
strong correlation among the data of $\overline{\psi}_1(p_{min})$
and $\overline{\psi}_2(p_{min})$, any quantitative study of
their contribution to the ghost propagator in the
infinite-volume limit would also require a full covariance
analysis.}

Finally, we note that the eigenvalues $\lambda_1$ and $\lambda_2$
(see Figures \ref{fig:lambda}, \ref{fig:lambdamin-lambda2}
and the top plot in Figure \ref{fig:comparison}) are clearly
larger for the set of data {\em bc} --- which can be viewed as
a numerical attempt of constraining the functional integration
to the fundamental modular region $\Lambda$ --- when compared
to the other three sets of data. This is in agreement with the
results presented in Refs.\ \cite{Sternbeck:2005vs,
Cucchieri:1997ns}. Accordingly, the upper and lower bounds
(see plots in Figure \ref{fig:bounds}) are always smaller for
the {\em bc} data than for the other sets. The same observation
holds for the value of $G(p_{min})$, which is smaller for the
{\em bc} set of data (see bottom plot in Figure \ref{fig:comparison}),
in agreement with Refs.\ \cite{Cucchieri:1997dx,Bakeev:2003rr,
Nakajima:2004vc,Sternbeck:2005tk,Lokhov:2005ra,Bogolubsky:2005wf,
Maas:2008ri,Bornyakov:2008yx}.
Note that throughout this section we have avoided direct comparisons of 
our data for the four different averages in the same plot. In fact,
we found that such plots were, in general, difficult to visualize. In
Figure \ref{fig:comparison}, for example, the data corresponding to
the remaining two types of averages lie somewhere between the two cases
displayed.


\section{Lower bound for $\lambda_1$}
\label{sec:lambdamin}

In Section \ref{sec:ineq} we recalled our proof of the
lower and upper bounds for the ghost propagator $G(p)$ ---
written in terms of the smallest nonzero eigenvalue $\lambda_1$
(and of the corresponding eigenvector) of the FP matrix
\cite{Cucchieri:2008fc} --- and showed how these bounds
can be improved, for example by considering the two smallest
nonzero eigenvalues $\lambda_1$ and $\lambda_2$ (and the
corresponding eigenvectors). In Section \ref{sec:num} these
results were compared to data obtained in four-dimensional
numerical simulations of the SU(2) case in minimal Landau gauge,
considering different sets of local minima. In this section we
present a simple lower bound for
$\lambda_1$, which constrains its approach to zero in the
infinite-volume limit. This result will be numerically verified
in the next section. This new bound will also help us get
a better understanding of the infinite-volume limit in minimal
Landau gauge, which in turn is discussed in Section \ref{sec:limit}.

Our proof is based on the concavity of the minimum function
(see for example Section 12.4 $\mbox{in}$ \cite{matrixalgebra})
and on three important properties of the first Gribov
region $\Omega$ in Landau gauge (see for example Section 2.1.3 $\mbox{in}$
\cite{Vandersickel:2011zc}, Section 2.2.1 $\mbox{in}$
\cite{Vandersickel:2012tz}, Appendix C in Ref.\
\cite{Zwanziger:2003cf} and Ref.\ \cite{Zwanziger:1982na}):
\begin{enumerate}
\item the trivial vacuum $A_{\mu} = 0$ belongs to $\Omega$;
\item the region $\Omega$ is convex;
\item the region $\Omega$ is bounded in every direction.
\end{enumerate}
These three properties\footnote{One should also recall that all 
gauge orbits intersect the first Gribov region \cite{Dell'Antonio:1991xt}.} 
can be proved by recalling that
the region $\Omega$ is defined as the set of gauge configurations
$A_{\mu}$ that are transverse, i.e.\ $\partial \cdot A = 0$, and
for which the FP operator
\begin{equation}
{\cal M}(b,x;c,y)[A] \, = \, - \partial \cdot D^{bc}(x,y)[A]
\end{equation}
is semi-positive definite. Here, $D^{bc}(x,y)[A]$ is the covariant
derivative. Then, the first property follows immediately, nothing that 
a null value for $A_{\mu}$ implies ${\cal M}(b,x;c,y)[0]$ equal to
(minus) the Laplacian, which is a semi-positive definite operator.

Concerning the second property listed above, the key ingredient of the
proof is that the gauge condition $\partial \cdot A = 0$ and
the operator $D^{bc}(x,y)[A]$ --- and therefore also the FP
operator ${\cal M}(b,x;c,y)[A]$ --- are linear in the gauge field
$A_{\mu}$. Indeed, if we write\footnote{Here and below we
simplify the notation and, unless necessary, we do not
explicitly show the color and space-time indices of the operators.}
\begin{equation}
{\cal M}[A] \, = \, - \partial^2 + {\cal K}[A]
\end{equation}
and we consider the linearity of the operator 
${\cal K}[A] \sim [A_{\mu},\,\partial_\mu]\,$,
we find that
\begin{eqnarray}
{\cal M}[(1-\rho) A_1 + \rho A_2] & = & - \partial^2 +
          {\cal K}[(1-\rho) A_1 + \rho A_2] 
   \, = \, - \partial^2 + (1-\rho) {\cal K}[A_1]
            + \rho \,{\cal K}[A_2] \nonumber \\[2mm]
\label{eq:Mlinearcomb-pre}
   & = & (1-\rho) \left( - \partial^2 + {\cal K}[A_1] \right)
        \, + \, \rho
            \left( - \partial^2 + {\cal K}[A_2] \right) \\[2mm]
   & = & (1-\rho) {\cal M}[A_1]  \, + \,
               \rho \,{\cal M}[A_2] \; .
\label{eq:Mlinearcomb}
\end{eqnarray}
Then, for $\rho \in [0, 1]$, it is clear that
${\cal M}[(1-\rho) A_1 + \rho A_2]$ is semi-positive
definite if ${\cal M}[A_1]$ and ${\cal M}[A_2]$ are also
semi-positive definite. At the same time, we have
\begin{equation}
(1-\rho) \, \partial \cdot A_1 \, + \, \rho
      \, \partial \cdot A_2 \, = \, 0
\end{equation}
if $\partial \cdot A_1 = \partial \cdot A_2 = 0$. Thus, the
convex combination $(1-\rho) A_1 + \rho A_2$ belongs
to $\Omega$, for any value of
$\rho \in [0, 1]$, if $A_1, A_2 \in \Omega$. This proves the
second property of $\Omega$. By combining properties
1 and 2, we can now set $A_1 = 0$, $A_2 = A$
and consider the configuration $\rho A$. In this case Eq.\
(\ref{eq:Mlinearcomb-pre}) becomes
\begin{eqnarray}
{\cal M}[\rho A] & = & - \partial^2 + {\cal K}[\rho A]
        \, = \, (1 - \rho) \, (- \partial^2) \, + \, \rho
                  \left( - \partial^2 + {\cal K}[A] \right) \\[2mm]
  & = & (1 - \rho) \, (- \partial^2) \, + \, \rho
                       \, {\cal M}[A]
\label{eq:MrhoA}
\end{eqnarray}
and, if $A$ belongs to $\Omega$, we have already proven that
$\rho A$ is also an element of $\Omega$ for any value of
$\rho \in [0, 1]$. On the other hand, we can show that, for
a sufficiently large value of $\rho > 1$, the configuration
$\rho A$ lies outside of the region $\Omega$, i.e.\ the matrix
\begin{equation}
{\cal M}[\rho A] \, = \, - \partial^2 + {\cal K}[\rho A]
            \, = \, - \partial^2 + \rho \, {\cal K}[A]
\end{equation}
is not semi-positive definite. In this case the important
observation is that the trace of the operator ${\cal K}[A]$
is zero. More specifically, since the color indices of
${\cal K}[A]$ are given by ${\cal K}^{bc}[A] \sim f^{bce}
A_{\mu}^e$ [where $f^{bce}$ are the structure constants of
the SU($N_c$) group] and since these structure constants
are completely (color) anti-symmetric, we have that all the diagonal
elements of ${\cal K}[A]$ are zero. This implies that the
sum of the eigenvalues of ${\cal K}[A]$ is also zero.
Moreover, the operator ${\cal K}[A]$ is real and symmetric
(i.e.\ invariant under simultaneous interchange of $x$ with 
$y$ and $b$ with $c$), implying that its eigenvalues are real. 
Thus, at least one of the
eigenvalues of ${\cal K}[A]$ is (real and) negative. Then,
if we indicate with $\phi_{neg}$ the corresponding eigenvector,
it is clear that for a sufficiently large (but finite) value
of $\rho > 1$ the scalar product $( \phi_{neg}, {\cal M}[\rho A]
\phi_{neg} )$ must be negative, so that ${\cal M}[\rho A]$
is not semi-positive definite and the configuration
$\rho A$ does not belong to $\Omega$.

\vskip 3mm
Using Eq.\ (\ref{eq:MrhoA}) above we can now find
a lower bound for the smallest nonzero eigenvalue. To this
end we consider a configuration $A'$ belonging to the boundary
$\partial \Omega$ of $\Omega$ and we write\footnote{Here and below,
we indicate with $\lambda_1\left[{\cal M}\right]$ the smallest nonzero
eigenvalue of the matrix ${\cal M}$.}
\begin{equation}
\lambda_1\left[ \, {\cal M}[\rho A'] \, \right] \, = \,
\lambda_1\left[(1 - \rho) \, (- \partial^2)
      \, + \, \rho \, {\cal M}[A']\right]
\end{equation}
for $\rho \in \Re$.
From the second property of $\Omega$ we know that $\rho A'
\in \Omega$ for $\rho \in [0,1]$. At the same time, we have
\begin{equation}
\lambda_1\left[(1 - \rho) \, (- \partial^2)
  \, + \, \rho \, {\cal M}[A']\right]
\, = \,
       \min_{\chi \neq \mbox{constant}} \,
      \left( \chi \, , \left[ \, (1 - \rho) \, (- \partial^2)
 \, + \, \rho \, {\cal M}[A'] \, \right] \, \chi \right) \; ,
\end{equation}
where the vectors $\chi(b,x)$ are assumed to be normalized
as $( \chi \, , \chi) = 1$. Then, using the concavity
of the minimum function \cite{matrixalgebra} we obtain
\begin{eqnarray}
\label{eq:newboundlambda-ini}
\lambda_1\left[ \, {\cal M}[\rho A'] \, \right] & = &
   \min_{\chi \neq \mbox{constant}}
  \, \left( \chi \, , \left[ \, (1 - \rho) \, (- \partial^2)
 \, + \, \rho \, {\cal M}[A'] \, \right] \, \chi \right) \\[2mm]
  & \geq & (1 - \rho) \min_{\chi \neq \mbox{constant}}
  \, \left( \chi \, , (- \partial^2) \, \chi \right) 
  \, + \,
   \rho \min_{\chi \neq \mbox{constant}} \, \left( \chi \, ,
             {\cal M}[A'] \, \chi \right) \nonumber \\[2mm]
  & = & (1 - \rho) \, \lambda_1\left[ \, - \partial^2
           \right] \, + \, \rho \, \lambda_1
                 \left[ \, {\cal M}[A'] \, \right] \nonumber \\[2mm]
  & = & (1 - \rho) \, p^2_{min} \; ,
\label{eq:newboundlambda}
\end{eqnarray}
where in the last step we used the fact that $A' \in \partial
\Omega$, i.e.\ the smallest non-trivial eigenvalue of the FP
matrix ${\cal M}[A']$ is null, and that the smallest non-trivial
eigenvalue of (minus) the Laplacian $ - \partial^2$ is the
magnitude squared of the smallest nonzero momentum $p^2_{min}$.
(Let us recall that this eigenvalue is $d$-fold degenerate.)
Therefore, as the lattice size $N$ goes to infinity, we find that
$\lambda_1\left[ \, {\cal M}[\rho A'] \, \right]$ cannot
go to zero faster than $(1 - \rho) \, p^2_{min}$. At the same
time, since $p^2_{min}$ behaves as $1/N^2$ at large $N$, we
have that $\lambda_1$ behaves as $\, N^{-2-\alpha} \,$ in
the same limit, with $\alpha > 0$, only if the quantity $1 - \rho$
goes to zero at least as fast as $N^{-\alpha}$. It is also
interesting to note that in the Abelian case one has
${\cal M} = - \partial^2$ and therefore $\lambda_1 = p^2_{min}$.
Thus, all the non-Abelian effects of the theory are essentially
included in the $(1 - \rho)$ factor.\footnote{For a numerical verification
of the inequality, see Section \ref{sec:limit}.}
It is furthermore important to stress that the above result applies to
any Gribov copy belonging to $\Omega$. (Clearly, different Gribov copies 
$A^{(g)}$ will have different values of $\rho$.)
Finally, note that the above inequality may equivalently be written as
\begin{equation}
\lambda_1\left[ \, {\cal M}[A] \, \right]
\;\geq\; \left( 1 - \rho \right) \, p^2_{min} \;,
\label{eq:newboundlambda2}
\end{equation}
where $\rho\leq 1$ measures the distance of a generic configuration
$A\in \Omega$ from the ``origin'' $A=0$ (in such a way that
$\rho^{-1} A$ will lie on $\partial \Omega$).

\vskip 3mm
A fortuitous feature of the above bound is
that it is written in terms of the smallest nonzero lattice
momentum. Then, if we combine this lower bound for
$\lambda_1$ with the upper bound in Eq.\ (\ref{eq:ineq}),
we find
\begin{equation}
   G(p) \, \leq \, \frac{1}{\lambda_1} \, \leq \,
       \frac{1}{[1 - \rho] \, p^2_{min}} \; ,
\label{eq:inequpp}
\end{equation}
where we have stressed that all quantities have been
evaluated for the gauge configuration $A$. Thus, in order
to have a ghost propagator $G(p)$ enhanced in the
infrared limit (compared to the tree-level behavior
$1/p^2$), a necessary condition is that the quantity
$1 - \rho$ go to zero sufficiently fast in the
infinite-volume limit. Moreover, for $p = p_{min}$
and by writing the ghost propagator as
\begin{equation}
G(p_{min}) \, = \, \frac{1}{p^2_{min}} \,
              \frac{1}{1 - \sigma(p_{min})} \; ,
\label{eq:sigma0}
\end{equation}
where $\sigma(p)$ is the Gribov ghost form-factor, we have
the inequality
\begin{equation}
\sigma(p_{min}) \, \leq \, \rho
\label{eq:sigma}
\end{equation}
[with $\rho \leq 1$]. Let us recall that the
restriction of the physical configuration space to the
region $\Omega$ is usually done by imposing the so-called
no-pole condition (see for example \cite{Cucchieri:2012cb,
Capri:2012wx} and references therein)
\begin{equation}
\sigma(p) < 1 \qquad \mbox{for} \qquad p^2 > 0
\label{eq:nopole}
\end{equation}
and that $\sigma(p)$ is monotonically decreasing as
$p^2$ increases (see Section 2.D in Ref.\ \cite{Cucchieri:2012cb}).
Here we have proven that, for a given lattice configuration
$A \in \Omega$, the Gribov ghost form-factor satisfies the
stronger inequality (\ref{eq:sigma}), which is related to
a simple geometrical characterization of the configuration
$A$.

Another simple consequence of Eq.\ (\ref{eq:inequpp}) is an upper
bound for the so-called $b$ parameter, used to characterize different
Gribov copies in Refs.\ \cite{Maas:2009se,Maas:2011ba}.
Note that, on the 
lattice, the $b$ factor is proportional to the ghost propagator
$G(p)$ evaluated at the smallest nonzero momentum $p_{min}$.
Thus, if we indicate this proportionality constant with $\nu^2$,
we can write
\begin{equation}
\max \, b[A] \, \leq \, \frac{\nu^2}{p^2_{min}}
   \, \max_{g} \frac{1}{1 - \rho\left[A^{(g)}\right]} \,
   \; ,
\label{eq:maxb}
\end{equation}
where the configurations $\left\{ A \right\}$ and
$\left\{ A^{(g)} \right\}$ are related by the gauge
transformation $g$ and the maximum is taken over all Gribov 
copies in $\Omega$. We see that the
volume dependence of the upper bound for the $b$
parameter is given by the approach to 1 of the
quantity $\max_{g} \rho\left[A^{(g)}\right]$ towards the
infinite-volume limit (and by how well the upper
bound is saturated). This could explain the
over-scaling observed in Ref.\ \cite{Maas:2009ph}
using simulations at $\beta = 0$, apparently in violation
of the uniqueness result \cite{Fischer:2009tn} for the
scaling solution \cite{von Smekal:1997vx,
Zwanziger:2001kw,Lerche:2002ep,Fischer:2006ub} of the
Dyson-Schwinger equations for gluon and ghost propagators.
Indeed, in the strong-coupling regime, most Gribov copies should be
just lattice artifacts not related to the continuum
theory, such as the lattice configurations inducing
confinement in the compact $U(1)$ case. Therefore, it is conceivable
that an exhaustive search of Gribov copies might generate
a gauge-fixed configuration $A^{(g)}$ 
characterized by a value of $\rho\left[A^{(g)}\right]$ 
very close to 1 and almost saturation of the bound (\ref{eq:maxb}).
In this way, strong-coupling lattice artifacts would produce an overly 
enhanced ghost propagator in the infrared limit.

Finally, following the analysis in Section 3 of Ref.\
\cite{Zwanziger:1991ac}, we can consider (in a finite lattice
volume) the nonzero eigenvalues ${\overline \lambda} = p^2(k)$ 
(for $k\neq 0$) of the operator ${\cal M}(b,x;c,y)[0]
= - \delta^{bc} \, \delta^d(x - y) \, \partial^2$, which are
characterized by a degree of degeneracy $g({\overline \lambda})$.
[For example, for the smallest nonzero eigenvalue $p^2_{min}$
one has $g({\overline \lambda}) = d (N_c^2 - 1)$.] This degeneracy is
in general lifted for the eigenvalues $\lambda$ of the operator
${\cal M}(b,x;c,y)[A]$, where $A$ is a generic gauge configuration. 
More precisely, to a given eigenvalue ${\overline \lambda}(p) = p^2$,
with degeneracy $g$, one can associate the eigenvalues
$\lambda_i(p)$ of ${\cal M}(b,x;c,y)[A]$, with $i = 1, 2,
\ldots, g$ and $\lambda_i(0,p) = p^2$. By rescaling these
eigenvalues by ${\overline \lambda}(p)$, i.e.\ by considering the ratios
\begin{equation}
L_i(p) \, = \, \frac{\lambda_i(p)}{p^2} \; ,
\end{equation}
and after defining the center of gravity
\begin{equation}
m(p) \, = \, \frac{1}{g} \, \sum_{i=1}^g \, L_i(p)\,,
\end{equation}
it was proven in Appendix C of Ref.\ \cite{Zwanziger:1991ac} that,
for very large volumes, the following equality holds
\begin{equation}
m(p) \, = \, 1 \, - \, \frac{H}{d V (N_c^2 - 1)} \; ,
\label{eq:theorem}
\end{equation}
where $H$ is the so-called horizon function, defined
in Eq.\ (3.10) of \cite{Zwanziger:1991ac}. Since the
r.h.s.\ of the above result is independent of the momentum $p$,
we can evaluate it at the smallest nonzero 
eigenvalue ${\overline \lambda} = p^2_{min}$ of ${\cal M}(b,x;c,y)[0]$. 
At the same time we have that
$\lambda_i(p_{min}) \geq \lambda_1$, where
$\lambda_1$ is the smallest nonzero eigenvalue
of ${\cal M}(b,x;c,y)[A]$, implying $L_i(p_{min})
\geq \lambda_1/p^2_{min}$. As a consequence, using
the inequality (\ref{eq:newboundlambda2}), we obtain
\begin{equation}
m(p_{min}) \, \geq \, \frac{\lambda_1}{p^2_{min}}
\, \geq \, 1 \, - \, \rho
\end{equation}
and the theorem in Eq.\ (\ref{eq:theorem}) yields 
\begin{equation}
\frac{H}{d V (N_c^2 - 1)} \, \equiv \,  h \, \leq \, \rho \; .
\label{eq:hbound}
\end{equation}
One immediately notices that this result is very similar
to the bound in (\ref{eq:sigma}). This is not
surprising since it has been recently proven \cite{Capri:2012wx}
that, in the continuum and to all orders in the gauge
coupling, the Gribov ghost form-factor $\sigma(0)$
at zero momentum is indeed equal to the normalized
horizon function $h$.

We can conclude this section by stressing that our
new bounds [see Eqs.\ (\ref{eq:newboundlambda2}),
(\ref{eq:inequpp}), (\ref{eq:sigma}) and (\ref{eq:hbound})]
suggest all non-perturbative features of a minimal-Landau-gauge 
configuration $A \in \Omega$ to be related to its
normalized distance $\rho$ from the ``origin'' $A=0$
(or, equivalently, to its normalized distance $1-\rho$
from the boundary $\partial \Omega$). The same formulae
also represent a clear mathematical description\footnote{One
can compare, for example, the above result (\ref{eq:hbound})
to the qualitative discussion presented in the Conclusions
of Ref.\ \cite{Zwanziger:1992qr}.} of the crucial role of
the boundary $\partial \Omega$ in the Gribov-Zwanziger
approach.


\section{Lower bound for $\lambda_1$: numerical results}
\label{sec:num2}

One can verify numerically some of the analytic
results presented in the previous section. In particular,
one can check the third property of the region $\Omega$,
presented at the beginning of the section, and the new bound
(\ref{eq:newboundlambda2}) --- or, equivalently, the sequence
of bounds in Eq.\ (\ref{eq:inequpp}) --- as well as the bounds
(\ref{eq:sigma}) and (\ref{eq:hbound}). To this end, we have
generated 70 new configurations for $V=16^4, 24^4, 32^4$ and
$40^4$ at $\beta = 2.2$ and 50 new configurations for $V=48^4,
56^4, 64^4, 72^4$ and $80^4$ at the same $\beta$ value. For
each of these thermalized configurations\footnote{For the
gauge fixing we used a stochastic overrelaxation algorithm
for the first Gribov copy, i.e.\ these data correspond to
the statistical average of the gauge-fixing prescription
indicated above with {\em fc}. Of course,
the same numerical analysis can also be carried out for the
other three types of statistical averages considered in Section
\ref{sec:num}.} we have evaluated the ghost propagator $G(p)$,
the two smallest nonzero eigenvalues and the respective eigenvectors. 
We then studied the geometrical properties discussed in the previous
section, by applying scale transformations\footnote{One should
stress that this procedure is a simple way of ``simulating'' the
mathematical proofs presented in the previous section. On the
other hand, by rescaling the gauge field we withdraw the unitarity
of the link variables, thus loosing the connection with the usual
Monte Carlo simulations. Nevertheless, as it will be shown below,
this approach gives us useful insights into the properties
of the Faddeev-Popov matrix and of the first Gribov region.}
to the gauge configuration $A$.
More precisely, we multiply the gauge-fixed gauge field $A_{\mu}(x)$
by a constant factor $\tau_1$, slightly larger than 1. Clearly,
the new gauge field $A_{\mu}^{(1)}(x) = \tau_1 A_{\mu}(x)$
still satisfies the Landau gauge condition $\partial_{\mu}
A_{\mu}^{(1)}(x) = 0$. Also, if $\tau_1$ is not too
large, it can be verified that the smallest nonzero
eigenvalue of ${\cal M}[A^{(1)}]$ remains positive, i.e.\
the configuration $A^{(1)}$ belongs to the first
Gribov region $\Omega$. 
By iterating this procedure, we consider a sequence of
values $\tau_2$, $\tau_3$, $\ldots$ 
for the scale transformation\footnote{For the factor $\tau_i$ we 
\label{foot:steps}
used the following prescription:
$\tau_0 = 1$, $\tau_i = \delta \, \tau_{i-1}$,
$\delta = 1.001$ if $\lambda_1 \geq 5 \, \times \, 10^{-3}$,
$\delta = 1.0005$ if $\lambda_1 \in [ 5 \, \times \, 10^{-4}, \,
5 \, \times \, 10^{-3})$ and $\delta = 1.0001$ if $\lambda_1 < 5 \,
\times \, 10^{-4}$, with $\lambda_1$ evaluated at the step $i-1$.}
of $A$. After $n$ steps, we end up with a modified gauge field
$A_{\mu}^{(n)}(x) = \tau_n A_{\mu}(x)$ that
does not belong anymore to the region $\Omega$, i.e.\ the
eigenvalue $\lambda_1$ of ${\cal M}[A^{(n)}]$
is negative (while $\lambda_2$ is still positive). In Table \ref{tab:cross} 
we report, for each lattice size $N$, the average
number of steps $\langle n \rangle$ necessary to find a modified gauge
field $\tau_n A_{\mu}(x) \notin \Omega$.  We also show, for
each $N$, the largest and the smallest values of $n$. This
represents a direct verification of the third property
described in the previous section. It is also interesting
to note that the number of steps $n$ necessary to ``bring''
the gauge configuration outside the region $\Omega$ decreases
with $N$, confirming that configurations with larger physical
volume are (on average) closer to the boundary $\partial \Omega$.

\setlength{\tabcolsep}{4pt}
\vspace{1cm}
\begin{table}[t]
\centering
\begin{tabular}{|c|c|c|c|c|c|} \hline
$N$   &   $\max(n)$  &   $\min(n)$   &   $\langle n \rangle$   &   $R_{\rm before}$   &   $R_{\rm after}$ \\ \hline
16    &   30         &         6          & 17.2   &  15(3)    &  -30(12) \\
24    &   27         &         4          & 15.1   &  20(7)    &  -26(6) \\
32    &   19         &         5          & 11.7   &  26(9)    &  -51(20) \\
40    &   18         &         4          &  9.4   &  155(143) &  -21(6) \\
48    &   13         &         2          &  7.8   &  21(5)    &  -21(5) \\
56    &   12         &         3          &  7.6   &  16(4)    &  -21(7) \\
64    &   11         &         2          &  6.8   &  20(7)    &  -42(18) \\
72    &   11         &         2          &  6.1   &  129(96)  &  -42(13) \\
80    &   12         &         3          &  6.1   &  15(4)    &  -24(4) \\ \hline
\end{tabular}
\vspace{2mm}
\caption{The maximum, minimum and average number of steps $n$,
necessary to ``cross the Gribov horizon'' along the direction
$A_{\mu}^b(x)$, as a function of the lattice size $N$. We also
show the ratio $R$ [see Eq.\ (\ref{eq:ratio})], divided by 1000,
for the modified gauge fields $\tau_{n-1} A_{\mu}^b(x)$ and
$\tau_n A_{\mu}^b(x)$, i.e.\ for the configurations immediately
before and after crossing $\partial \Omega$.
Here we used the gauge-fixing prescription {\em fc} (see Section \ref{sec:num}).
The data in the last two columns represent averages over gauge 
configurations, errors correspond to one standard deviation.
(We consider the statistical error only.)
\label{tab:cross}}
\end{table}
\begin{figure}
\centering
\hspace{-6.0mm}
\includegraphics[scale=1.00]{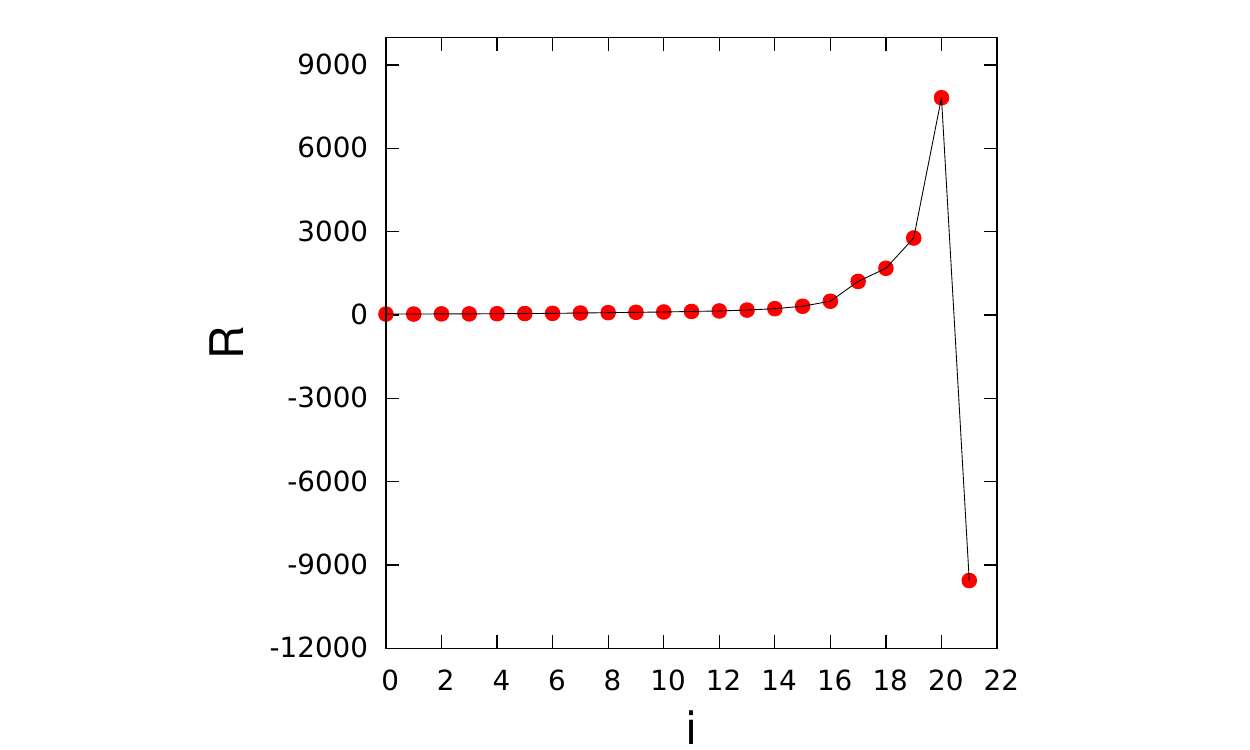}
\\
\vspace{5mm}
\includegraphics[scale=1.00]{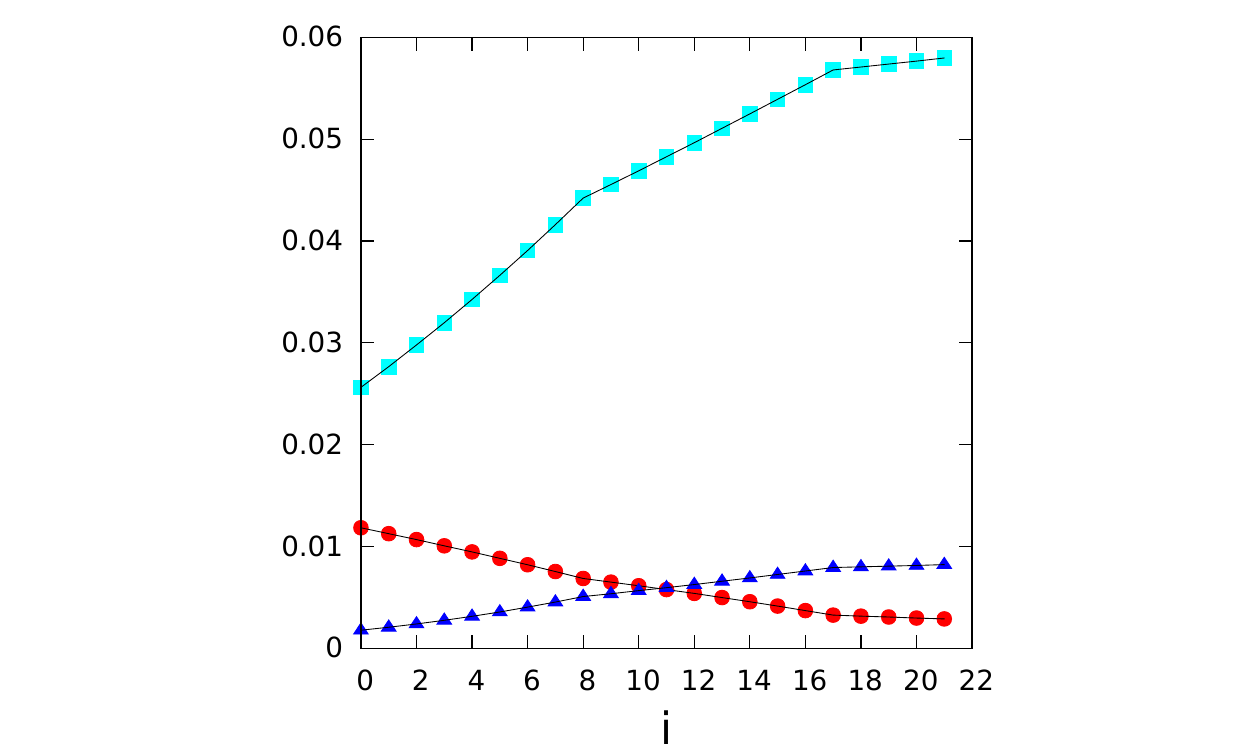}
\vspace{2mm}
\caption{\label{fig:ratio} Top plot: the ratio $R$ [see Eq.\
(\ref{eq:ratio})], as a function of the iteration step $i$,
for a typical configuration with lattice volume $16^4$.
Bottom plot: the second smallest (non-trivial) eigenvalue
$\lambda_2$ (full circles), the absolute value of the
third derivative of the minimizing function $| {\cal E}''' \, |$
(full squares) and the fourth derivative of the
minimizing function $ {\cal E}'''' \, $ (full triangles)
as a function of the iteration step $i$, for the same
configuration considered in the top plot. One can
easily recognize the three different values for the $\delta$ step
used in the iteration process (see footnote \ref{foot:steps}).
Here we used the gauge-fixing prescription {\em fc} (see Section \ref{sec:num}).
In all cases the solid line is only to guide the eye.
}
\end{figure}
\begin{figure}
\centering
\includegraphics[scale=1.00]{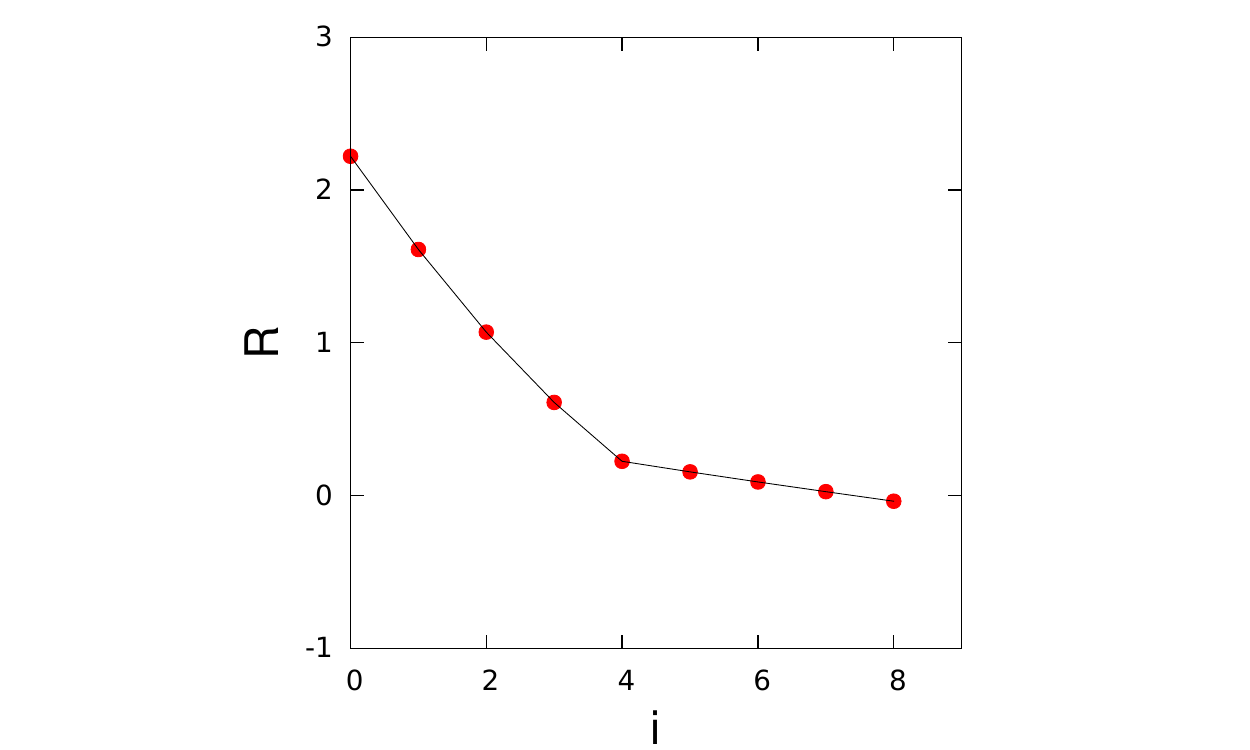}
\\
\vspace{5mm}
\includegraphics[scale=1.00]{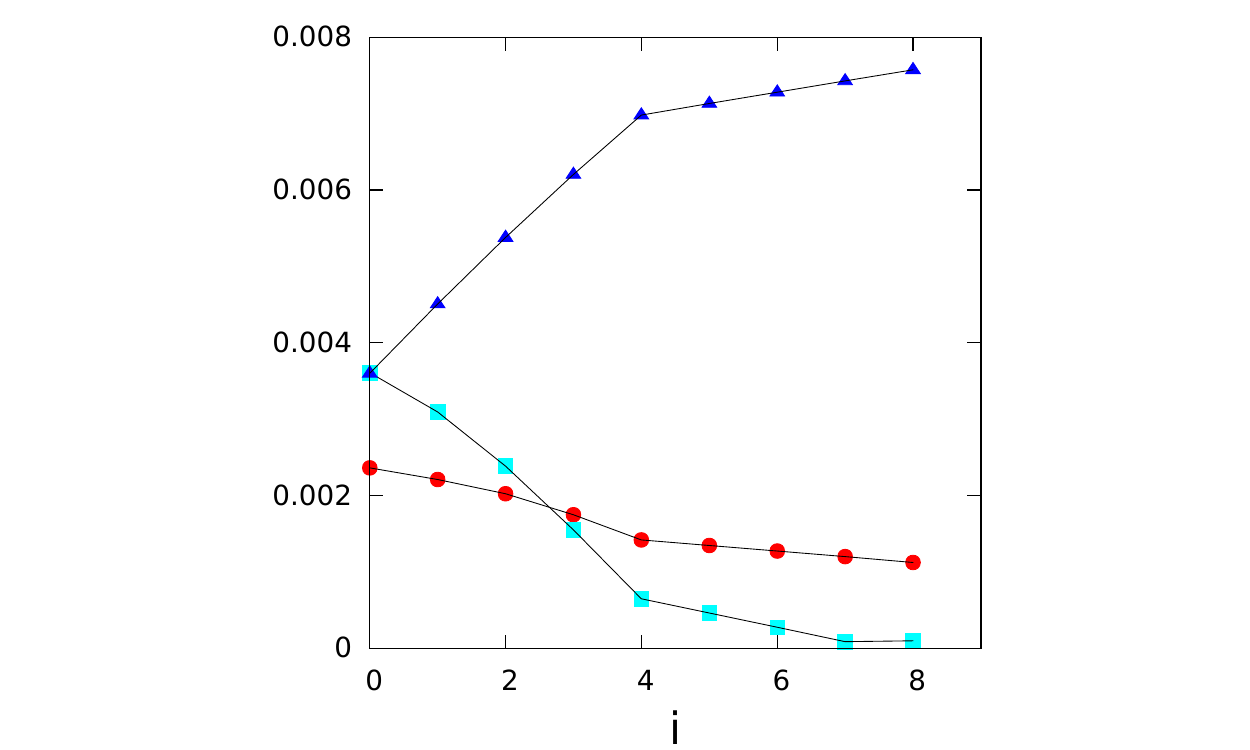}
\vspace{2mm}
\caption{\label{fig:ratio2} Top plot: the ratio $R$ [see Eq.\
(\ref{eq:ratio})], as a function of the iteration step $i$,
for a configuration, with lattice volume $48^4$, that is
a possible candidate for an element of the common boundary
$\partial\Omega \, \cap \, \partial\Lambda$.
(Note the small range of values
on the $y$ axis when compared to the
corresponding plot in Fig.\ \ref{fig:ratio}.)
Bottom plot: the second smallest (non-trivial) eigenvalue
$\lambda_2$ (full circles), the absolute value of the
third derivative of the minimizing function $| {\cal E}''' \, |$
(full squares) and the fourth derivative of the
minimizing function $ {\cal E}'''' \, $ (full triangles)
as a function of the iteration step $i$, for the same
configuration considered in the top plot.
Here we used the gauge-fixing prescription {\em fc} (see Section \ref{sec:num}).
In all cases the solid line is only to guide the eye. 
}
\end{figure}

\begin{figure}
\centering
\hspace{-6.0mm}
\includegraphics[scale=1.00]{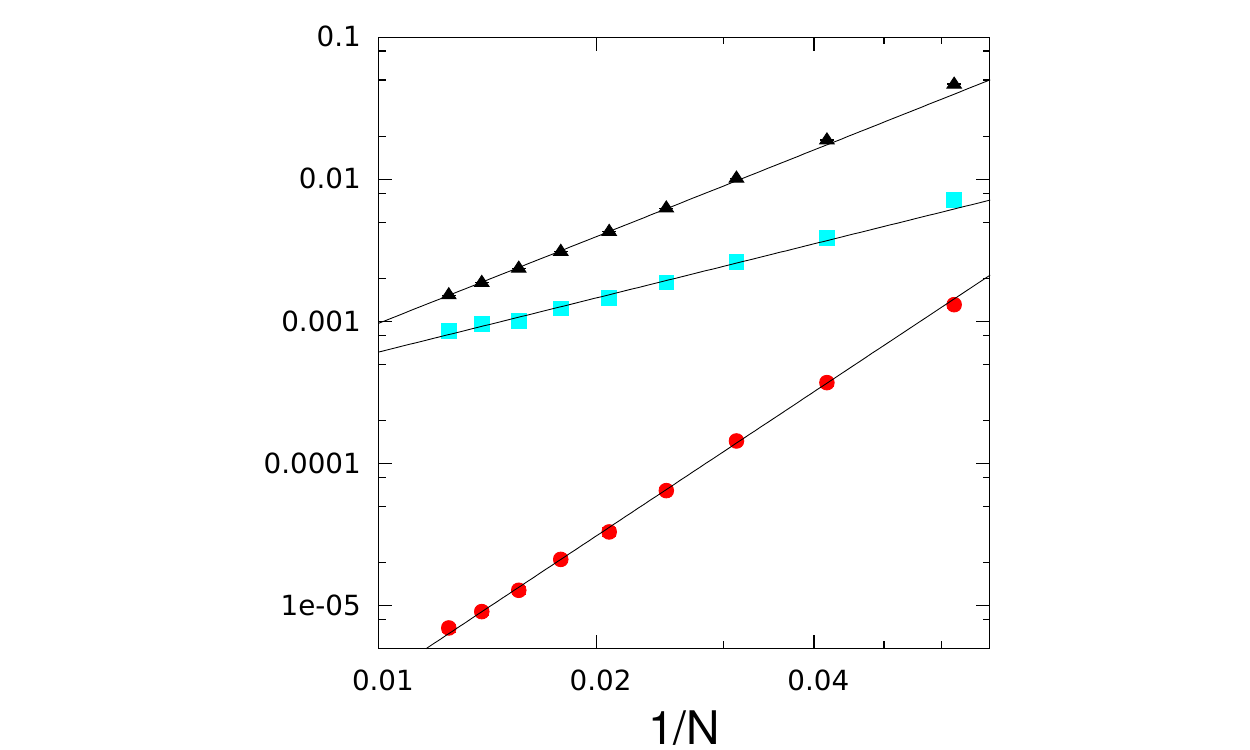}
\vspace{2mm}
\caption{\label{fig:newbounds-fit} The inverse ghost
propagator $1/G(p_{min})$ (full triangles), evaluated at the
smallest nonzero momentum $p_{min}$, the smallest nonzero
eigenvalue $\lambda_1$ (full squares) and the lower-bound
estimate in Eq.\ (\ref{eq:newboundlambda2}) (full circles)
as a function of the inverse lattice size $1/N$. We also
show, for each quantity, the fit to the function $c/N^{\eta}$
(see discussion in the text).
Here we used the gauge-fixing prescription {\em fc} (see Section \ref{sec:num}).
All quantities are in lattice units. 
Note the logarithmic scale on both axes.
The data points represent averages over gauge configurations, 
error bars correspond to one standard deviation.
(We consider the statistical error only.)
}
\end{figure}

\begin{figure}
\centering
\hspace{-6.0mm}
\includegraphics[scale=1.00]{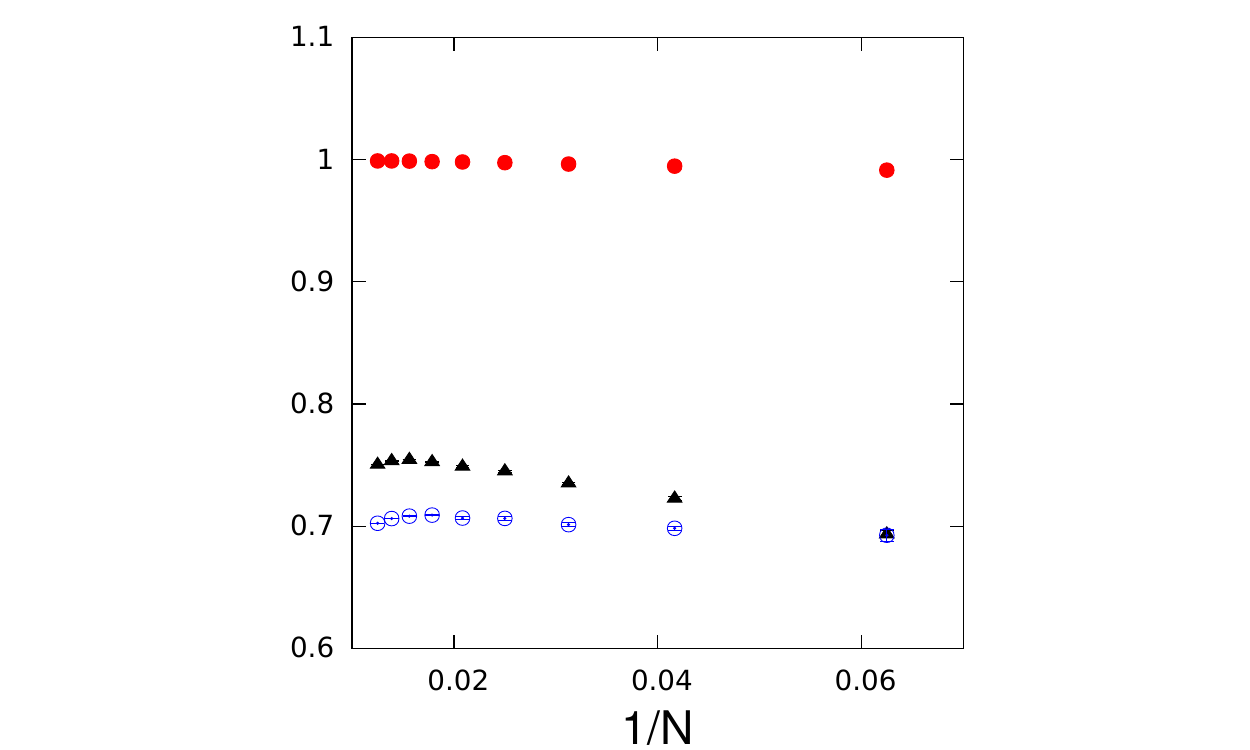}
\vspace{2mm}
\caption{\label{fig:newbounds}
The normalized horizon function $h$ (empty circles),
the Gribov ghost form-factor $\sigma$ evaluated at the
smallest nonzero momentum $p_{min}$ (full triangles) and
the upper bound $\rho$ (full circles) as a function of the
inverse lattice size $1/N$.
Here we used the gauge-fixing prescription {\em fc} (see Section \ref{sec:num}).
All quantities are in lattice units.
The data points represent averages over gauge configurations, 
error bars correspond to one standard deviation.
(We consider the statistical error only.)
}
\end{figure}

For each configuration we also show, in the same table,
the average value of the ratio
\begin{equation}
R \, = \, \frac{ \left( \, {\cal E}''' \, \right)^2 }{ 
       {\cal E}'' \; {\cal E}'''' }
\label{eq:ratio}
\end{equation}
just before and just after crossing $\partial\Omega$.
Here ${\cal E}'', \, {\cal E}'''$ and ${\cal E}''''$
are, respectively, the second, third and fourth
derivatives of the minimizing functional ${\cal E}[A]$
[see Eq.\ (\ref{eq:min})]
evaluated along the direction defined by the eigenvector
$\psi_1(b,x)$.\footnote{One should note that, then,
${\cal E}'' = \lambda_1$.} As shown in Ref.\
\cite{Cucchieri:1997ns}, this ratio characterizes the
shape of the minimizing functional ${\cal E}$, around
the local minimum considered, when one applies to
${\cal E}$ a fourth-order Taylor expansion (see in
particular Figure 2 of the same reference).
We find that this ratio is the only
quantity that shows an abrupt jump across the boundary
$\partial \Omega$ (see the top plot in Fig.\ \ref{fig:ratio}).  
Indeed, we checked that the second,
third and fourth derivatives, as well as the various
terms contributing to them and defined in Eqs.\ (11),
(13) and (14) of Ref.\ \cite{Cucchieri:1997ns}, have a
slow and continuous dependence on the factors $\tau_i$
(see for example the bottom plot in Fig.\ \ref{fig:ratio}
and in Fig.\ \ref{fig:ratio2}).
On the other hand, since ${\cal E}''$ decreases as $\tau_i$
increases, we find that the ratio $R$ usually increases
with $\tau_i$ and that $R_{n-1} \approx - R_n$, due to
the change in sign of ${\cal E}''$ as the first Gribov
horizon is crossed (see the fourth and the fifth
columns in Table \ref{tab:cross}). This behavior can
also be seen in the top plot of Figure \ref{fig:ratio},
where the value of $R$ is shown (as a function
of the number of steps $i$) for one of the typical
configurations generated for the volume $16^4$.
At the same time one can check (see the bottom plot
in Figure \ref{fig:ratio}) that the second smallest
(non-trivial) eigenvalue stays positive, i.e.\ the final
configuration $\tau_{n} A_{\mu}(x)$ belongs to the
second Gribov region.

On the other hand, for a few configurations\footnote{More
exactly, for eleven configurations (out of a total of 530)
the value of $|R|$ was smaller than 10 immediately before
crossing the Gribov horizon and
immediately after the crossing.} we found a very
small value for the ratio $R$, also when the configuration
is very close to $\partial \Omega$ (see Figure
\ref{fig:ratio2}). In all these cases the absolute value of the third
derivative ${\cal E}'''$ of $A_{\mu}(x)$ is much smaller (typically by
a factor of order of 20) compared to the average (absolute)
value obtained for the other configurations with the same
lattice volume, thus forcing the value of $R$ to stay small 
along the path $\tau_i A_{\mu}(x)$. Let us recall
that a null value for ${\cal E}'''$ is indicative
\cite{Zwanziger:1991ac,Zwanziger:1993dh} of a configuration
belonging to the common boundary\footnote{This observation
clarifies why it is important to consider at least a
fourth-order Taylor expansion in the analysis of the
minimizing function ${\cal E}$ (see for example Appendix
A.1 in Ref.\ \cite{Zwanziger:2003cf}). Indeed, the first derivative
of ${\cal E}$ is null for all the configurations in
$\Omega$. At the same time, its second derivative is zero
for configurations belonging to $\partial \Omega$ and the
third derivative is also zero when one considers the common
boundary of $\Omega$ and $\Lambda$. More precisely, one can show
(see Refs.\ \cite{sefr,vanBaal:1991zw})
that if ${\cal E}'''[A] \neq 0$ for $A \in \partial \Omega$,
then the configuration $(1 - s) A \in \Omega$, with
$0 \leq s $ and $s$ sufficiently small,
cannot be an absolute minimum, i.e.\ $(1 - s) A \notin \Lambda$.} 
of the first Gribov
region $\Omega$ and of the fundamental modular region
$\Lambda$. These configurations, characterized by a
small value of $R$ for all factors $\tau_i$,
are therefore good candidates to belong to
$\partial\Omega \, \cap \, \partial\Lambda$. 

\vskip 3mm
Using the above results we can now easily verify the new
bound (\ref{eq:newboundlambda2}). Indeed, for each
configuration $A_{\mu}(x)$, we can consider\footnote{Since
$\tau_{n} = \delta \tau_{n-1}$ and $\delta = 1 + \epsilon$
with $\epsilon \ll 1$, the definition ${\widetilde \tau} \equiv
(\tau_{n-1} + \tau_{n})/2$ is numerically equivalent (up to
order $\epsilon$) to the definition ${\widetilde \tau}
\equiv \sqrt{\tau_{n-1} \, \tau_{n}}$.}
\begin{equation}
A' _{\mu}(x) \, = \,
  {\widetilde \tau} \, A_{\mu}(x) \, \equiv \,
     \frac{ \tau_{n-1} + \tau_{n} }{2} \, A_{\mu}(x)
\end{equation}
as a candidate for a configuration on the boundary
of $\Omega$, since $\tau_{n-1} A_{\mu}(x) \in
\Omega$ and $\tau_{n} A_{\mu}(x) \notin \Omega$.
Equivalently, with the notation of the previous section,
we can write
\begin{equation}
A_{\mu}(x) \, = \,
  \rho \, A' _{\mu}(x) \; ,
\end{equation}
with $\rho = 1/ {\widetilde \tau} < 1$, and the bound
(\ref{eq:newboundlambda2}) becomes
\begin{equation}
\lambda_1\left[ \, {\cal M}[A] \, \right]
\, = \, \lambda_1\left[ \, {\cal M}[\rho A'] \, \right]
    \; \geq \;
  (1 - \rho) \, p^2_{min} \, = \,
  \left( 1 - \frac{1}{{\widetilde \tau}} \right)
  \, p^2_{min} \; .
\label{eq:ineq-tau}
\end{equation}
This last result is checked numerically in
Figure \ref{fig:newbounds-fit}.
In the same figure, we show the inverse
of the ghost propagator evaluated ad the smallest nonzero
momentum $p_{min}$. As before, we see that $\lambda_1$ 
approaches $1/G(p_{min})$ at large volumes. On the other
hand, the lower bound $(1 - \rho) \, p^2_{min}$ seems to
decrease faster than $1/G(p_{min})$. Indeed, if one tries
to fit the data (see again Figure \ref{fig:newbounds-fit})
using the fitting function $c/N^{\eta}$, the value for the exponent
$\eta$ is given\footnote{For these fits we considered only
the data with $N > 16$.} by $\eta\approx 2.0$ for the inverse
ghost propagator, $\eta \approx 3.4$ for the lower bound
$(1 - \rho) \, p^2_{min}$ and by $\eta\approx 1.3$ for
$\lambda_1$. This last result is in disagreement with
the results presented in \cite{Sternbeck:2005vs}, where
$\lambda_1$ displayed an exponent slightly larger than two
(using significantly smaller lattice volumes and the
so-called {\em best copy} average), and only in qualitative
agreement with Ref.\ \cite{Cucchieri:2008fc}, where we found
$\lambda_1 \sim L^{-1.53}$ (using, however, mostly data
in the strong-coupling regime). [Here $L = a N$ is the
physical size of the lattice.] One should also note that,
since the above fits give $1/G(p_{min}) < \lambda_1$ for
large values of $N$, simulations at larger
lattice volumes are necessary in order to see how the
inequality $1/G(p_{min}) \geq \lambda_1$ is realized
in the infinite-volume limit. 

As for the inequality (\ref{eq:newboundlambda2}), one should
stress that it becomes
an equality only when the eigenvectors corresponding to the
smallest nonzero eigenvalues of the two matrices on the
r.h.s.\ of Eq.\ (\ref{eq:newboundlambda-ini}) coincide.
Thus, the fact that the new bound is very far from being
saturated tells us that the eigenvector $\psi_1$ is
very different from the plane waves corresponding to the
smallest eigenvalue of the (lattice) Laplacian. That this
is indeed the case can also be seen from the very small
values obtained (for all configurations and Gribov copies)
for the average projection $\overline{\psi}_1(p_{min})$
(see Figure \ref{fig:lambdamin-proj}).

Finally, in Figure \ref{fig:newbounds} we check the
inequalities (\ref{eq:sigma}) and (\ref{eq:hbound}), i.e.\
\begin{equation}
\sigma(p_{min}), \, h \, \leq \, \rho \; . 
\label{eq:sigmaall}
\end{equation}
To this end, the normalized horizon function $h$ has been evaluated
using Eq.\ (3.12.b) in Ref.\ \cite{Zwanziger:1993dh} (see Ref.\
\cite{Cucchieri:1997ns} for details of the numerical evaluation
of $h$). One clearly sees that these inequalities are verified,
for all values of the lattice size $N$. Moreover, $\sigma(p_{min})$
and $h$ appear to stay far away from their upper bound
$\rho \approx 1$. At the same time, one sees from the plot
that $\sigma(p_{min}) \approx h$ at small volume.
On the other hand, in the infinite-volume limit, the value of
$\sigma(0)$ does not seem to agree with the value of $h$. This
small disagreement, in contradiction with the result of
Ref.\ \cite{Capri:2012wx}, could be related to different
finite-lattice-spacing effects,\footnote{We include in
these effects the dependence on the lattice spacing $a$
introduced by the renormalization conditions for the considered
quantities.} in which case the observed difference
should disappear in the continuum limit, and/or to possible
non-perturbative contributions to these two quantities,
since the proof in Ref.\ \cite{Capri:2012wx} is valid at
the perturbative level.


\section{The infinite-volume limit}
\label{sec:limit}

In order to interpret the new bound (\ref{eq:newboundlambda2})
we recall that a distance function $\delta_l$ on SU($N_c$)
lattice configurations can be defined (see Appendix A in
Ref.\ \cite{Zwanziger:1993dh}) as
\begin{equation}
\delta^2_l(U,W) \, = \, \frac{1}{d \, V \, N_c}
     \sum_{\mu, x} \, \Tr \, \left\{ 
       \left[ U_{\mu}(x) - W_{\mu}(x) \right]^{\dagger}
    \, \left[ U_{\mu}(x) - W_{\mu}(x) \right] \right\} \; ,
\end{equation}
where $U_{\mu}(x)$ and $W_{\mu}(x)$ are elements of the gauge
group and $^{\dagger}$ indicates the conjugate transpose
matrix. Similarly, in the continuum, we can write
\begin{equation}  
\delta^2_c(A,B) \, = \,
    \frac{1}{d \, V \, ( N_c^2 - 1 )}
       \sum_{\mu, x} \, \Tr \, \left\{ 
     \left[ A_{\mu}(x) - B_{\mu}(x) \right]^{\dagger}
  \, \left[ A_{\mu}(x) - B_{\mu}(x) \right] \right\} \; ,  
\label{eq:deltadef}
\end{equation}  
where $A_{\mu}(x)$ and $B_{\mu}(x)$ are now elements of the
algebra generating the gauge group. Both definitions are
invariant with respect to global and to local gauge
transformations,\footnote{If the configurations $U_{\mu}(x),
W_{\mu}(x)$ [or $A_{\mu}(x), B_{\mu}(x)$] belong to the
same gauge orbit, this invariance indicates that the
distance between the two configurations is independent
of the ``choice of the origin'' for the given orbit.}
i.e.\ $\delta_l^2(U^g, W^g) = \delta_l^2(U, W)$ for a general 
gauge transformation $g$ [and similarly for $\delta_c^2(A, B)$].
Then, if $B = \rho A$ and $\rho \in \Re$, we have
\begin{equation}
\delta^2_c(A,B) \, = \,
  \frac{\left( 1 - \rho \right)^2}{d \, V \, ( N_c^2 - 1 )}
            \sum_{\mu, x} \, \Tr \, \left\{
                   A_{\mu}(x)^{\dagger} \,
                   A_{\mu}(x) \right\} \, = \,
       \left( 1 - \rho \right)^2 \, \| A \|^2 \; ,
\end{equation}
where
\begin{equation}
\| A \|^2 \, \equiv \, \frac{1}{d \, V \, ( N_c^2 - 1 )}
            \sum_{\mu, x} \, \Tr \, \left\{
                   A_{\mu}(x)^{\dagger} \,
                   A_{\mu}(x) \right\} \, = \,
      \frac{1}{d \, V \, ( N_c^2 - 1 )}
            \sum_{\mu, x, b} \, | \,
                   A_{\mu}^b(x) |^2 
\end{equation}
is the (average) norm square of the components of the
gauge field. Therefore, following the notation of the
previous section, i.e.\ considering a configuration
$A' \in \partial \Omega$ and a configuration $B =
\rho A' \in \Omega$, with $\rho \in [0,1]$, we have 
from (\ref{eq:deltadef}) that
\begin{equation}
\rho \, = \, \sqrt{ \frac{\delta^2_c(0,\rho A')}{
           \| A' \|^2} } \; .
\end{equation}
This means that the factor $\rho$ can be viewed as a
normalized distance of the configuration $B = \rho A'$
from the ``origin'' $A=0$, along a direction parallel
to $A'$. Equivalently, the factor $1 - \rho$ in
Eq.\ (\ref{eq:newboundlambda}) is a normalized
distance of the configuration $\rho A'$ from the
boundary of the region $\Omega$ (along the same
direction).\footnote{Let us note that the dependence
of the FP eigenvalues on the distance of the
configuration from the first Gribov horizon has been
considered in Ref.\ \cite{Greensite:2010hn} within
a perturbative study of the FP spectrum.} The
inequality (\ref{eq:newboundlambda2}) is therefore a
simple mathematical realization of the intuitive statement
that the behavior of $\lambda_1$ in the infinite-volume
limit is controlled by the speed at which an average
thermalized and gauge-fixed configuration approaches
the boundary $\partial \Omega$ (see discussion in the
Introduction).

\setlength{\tabcolsep}{3pt}
\vspace{1cm}
\begin{table}[t]
\centering
\begin{tabular}{|c|c|c|c|c|c|c|c|c|} \hline
$N$ & \multicolumn{2}{|c|}{\mbox{\em fc}} & \multicolumn{2}{|c|}{\mbox{\em sfc}}
    & \multicolumn{2}{|c|}{\mbox{\em afc}} & \multicolumn{2}{|c|}{\mbox{\em bc}} \\ \hline
16  & 0.0248(8) & 5.6 \% & 0.0245(8) & 4.6 \% & 0.0240(8) & 5.6 \% & 0.029(1) & 5.4 \% \\
32  & 0.023(2)  & 5.5 \% & 0.021(2)  & 7.5 \% & 0.022(2)  & 7.0 \% & 0.020(2) & 7.0 \% \\
48  & 0.026(4)  & 10.0\% & 0.023(3)  & 13.0\% & 0.019(2)  & 9.0 \% & 0.029(5) & 14.0\% \\
64  & 0.019(4)  & 20.0\% & 0.018(4)  & 20.0\% & 0.029(5)  & 20.0\% & 0.035(8) & 10.0\% \\ \hline
\end{tabular}
\vspace{2mm}
\caption{The average value of $| {\cal E}''' \, |$ and the
percentage of configurations that are ``very close'' to the common boundary
$\partial\Omega \, \cap \, \partial\Lambda$, i.e.\ with
a value of $| {\cal E}''' \, |$ smaller than 1/20 of the
corresponding average value, for each lattice volume and
for the four types of gauge-fixing prescription considered
(see Section \ref{sec:num}).
The data represent averages over gauge configurations, 
errors correspond to one standard deviation.
(We consider the statistical error only.)
\label{tab:ratio}}
\end{table}

\begin{figure}
\centering
\includegraphics[trim=80 0 40 0, clip, scale=1.00]{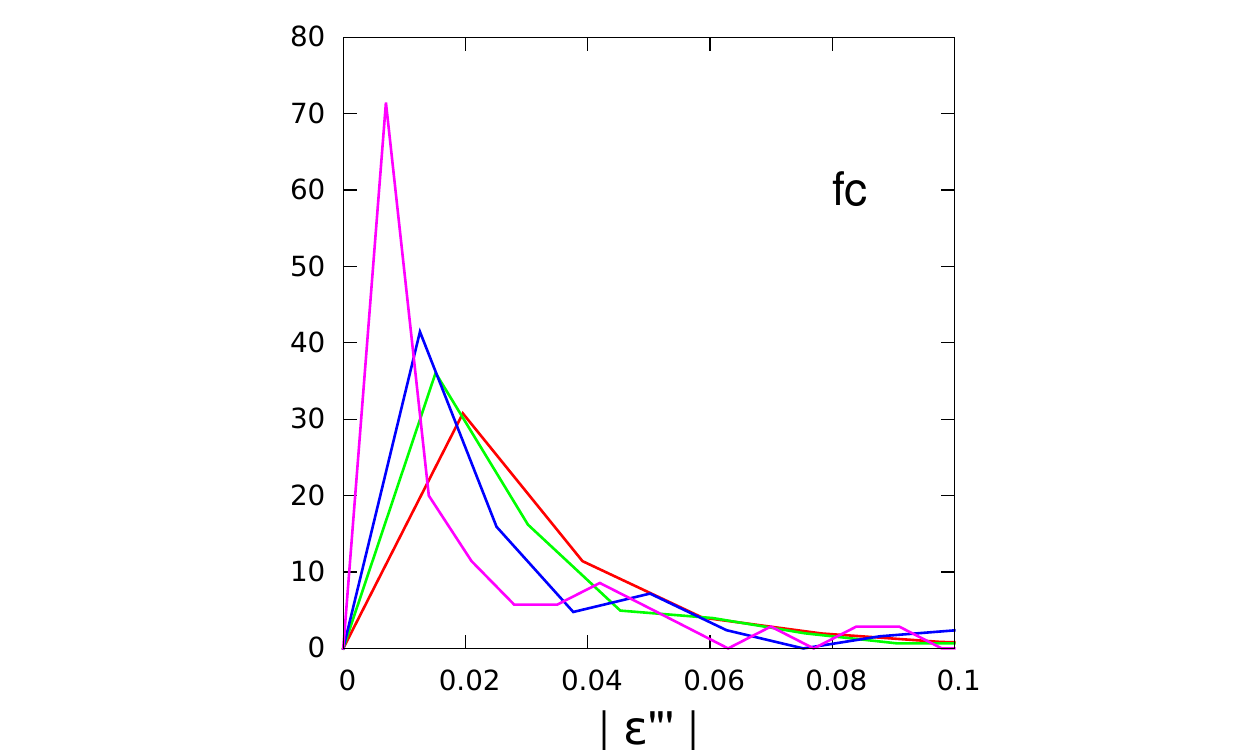}
\hspace{-5mm}
\includegraphics[trim=80 0 40 0, clip, scale=1.00]{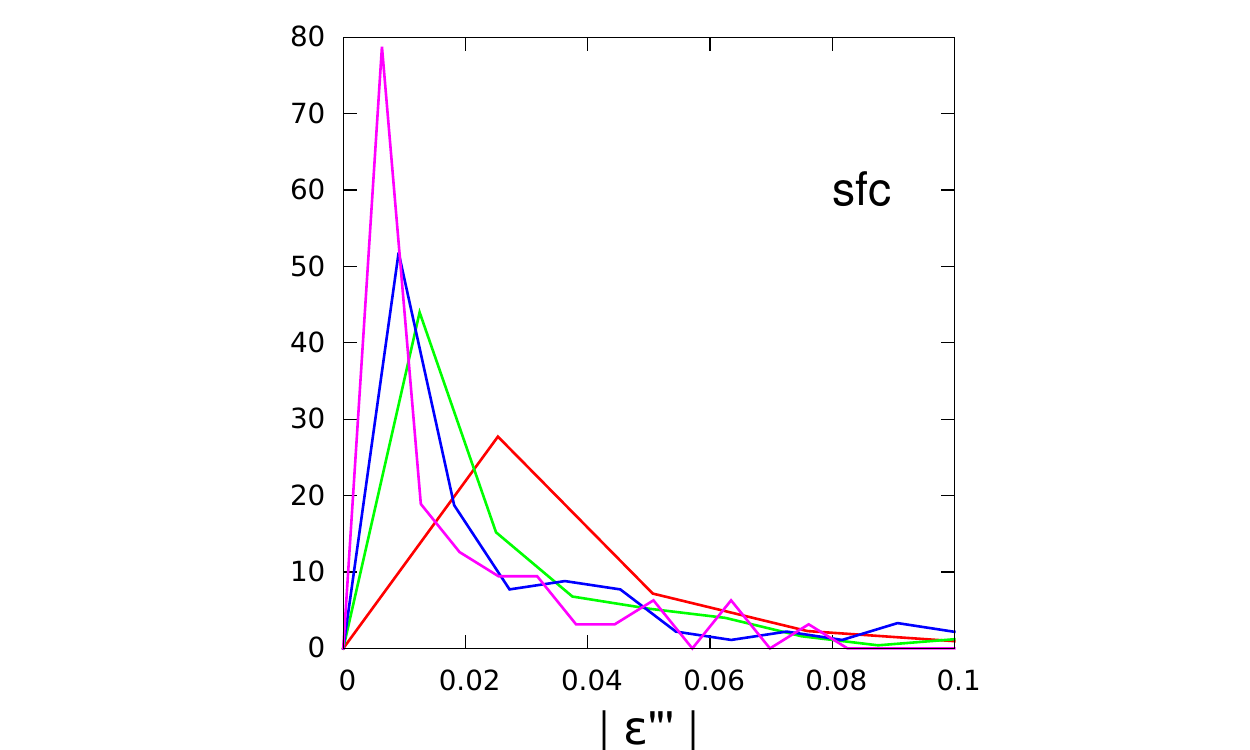}
\\
\vspace{5mm}
\includegraphics[trim=80 0 40 0, clip, scale=1.00]{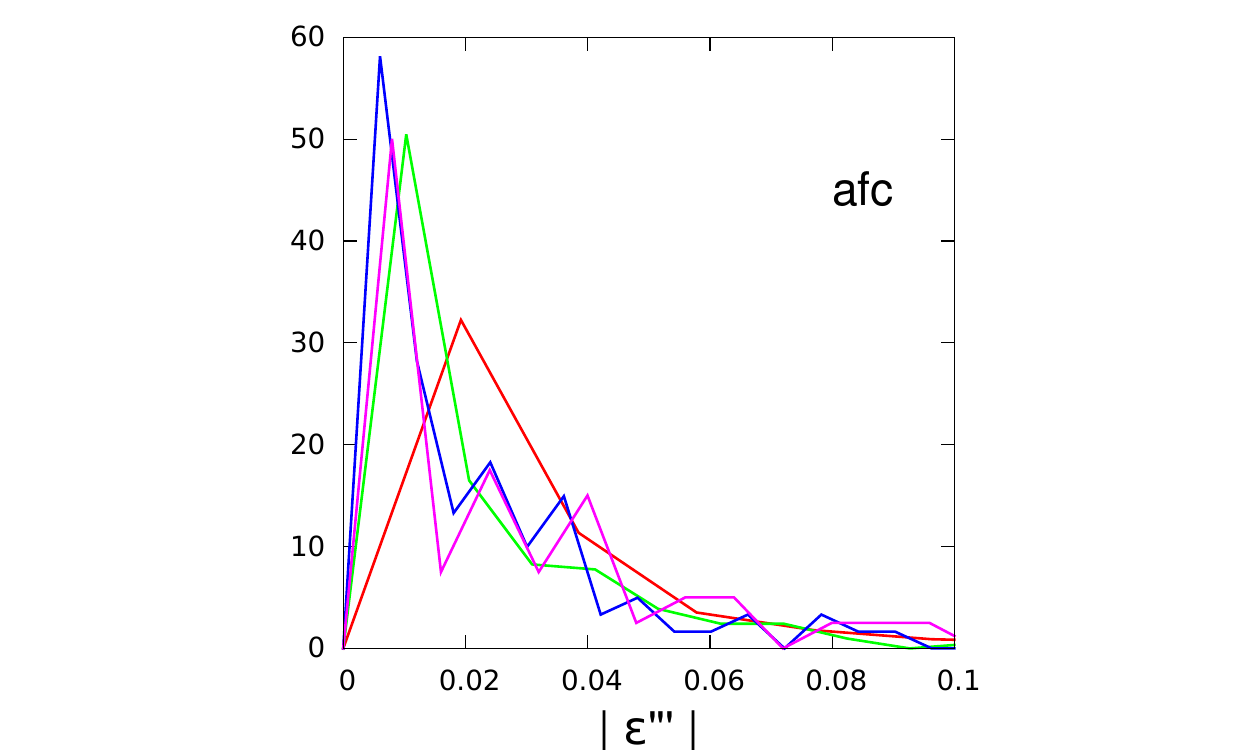}
\hspace{-5mm}
\includegraphics[trim=80 0 40 0, clip, scale=1.00]{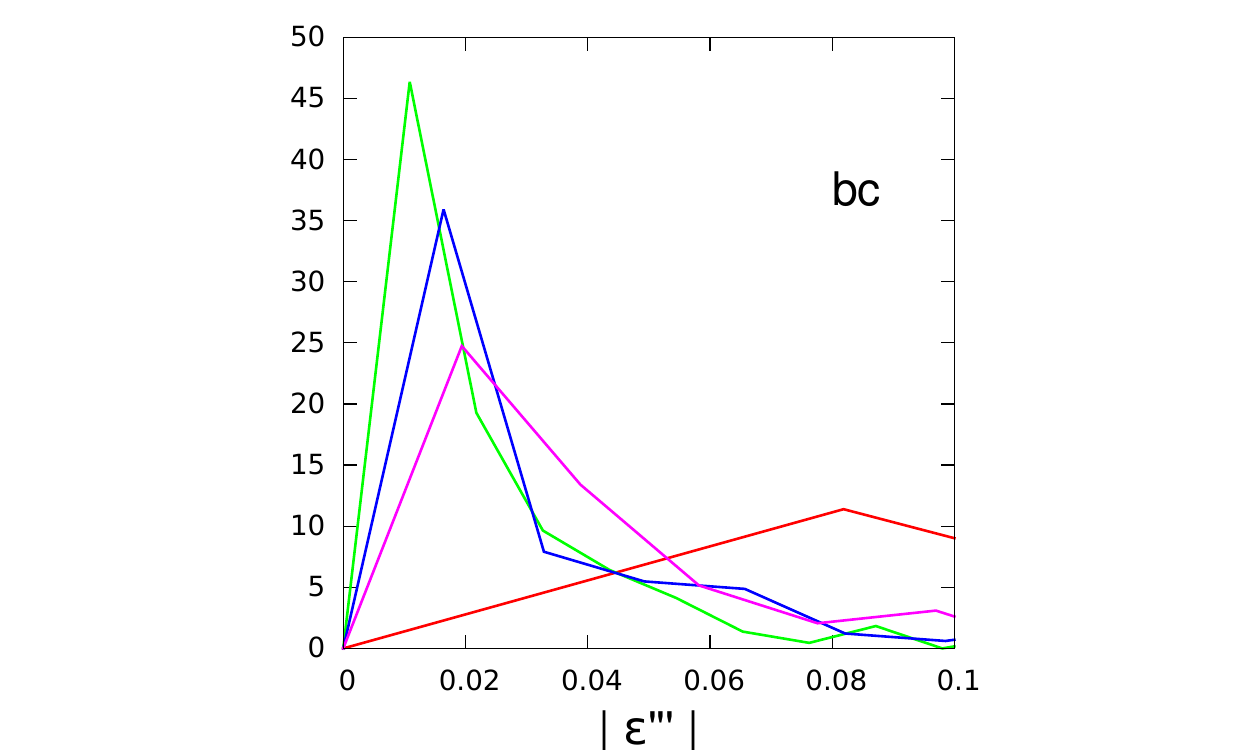}
\vspace{2mm}
\caption{\label{fig:histod3} Histogram distribution of
$ | {\cal E}''' \, | $ for the lattice sizes
$N = 16$ (red line), 32 (green line), 48 (blue line) and 64
(magenta line) using 20 bins. The area of the histograms is
normalized to 1. In order to reveal the distribution of the
data with a small value of $ | {\cal E}''' \, | $, we show
only the interval $[0,0.1]$. Four types of gauge-fixing prescription
are considered (see
Section \ref{sec:num}): {\em fc} (upper left plot), {\em sfc}
(upper right plot), {\em afc} (lower left plot) and
{\em bc} (lower right plot).
}
\end{figure}

\begin{figure}
\centering
\includegraphics[trim=80 0 40 0, clip, scale=1.00]{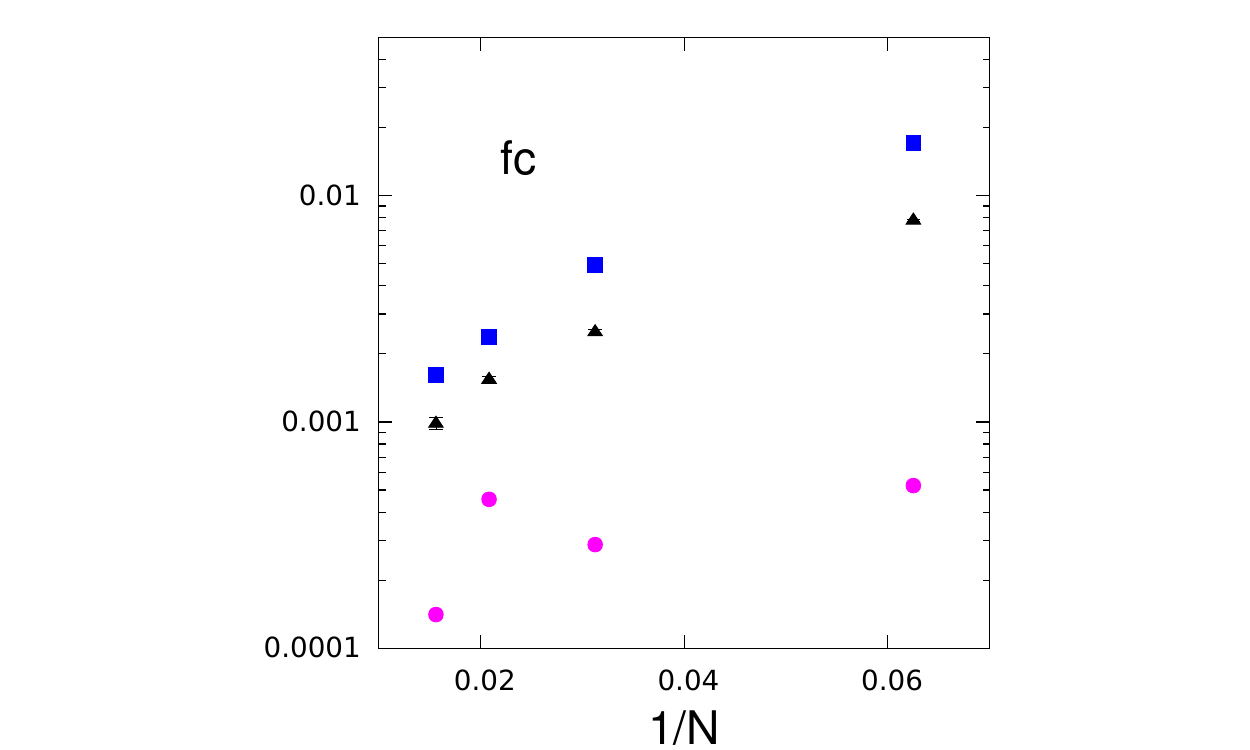}
\hspace{-5mm}
\includegraphics[trim=80 0 40 0, clip, scale=1.00]{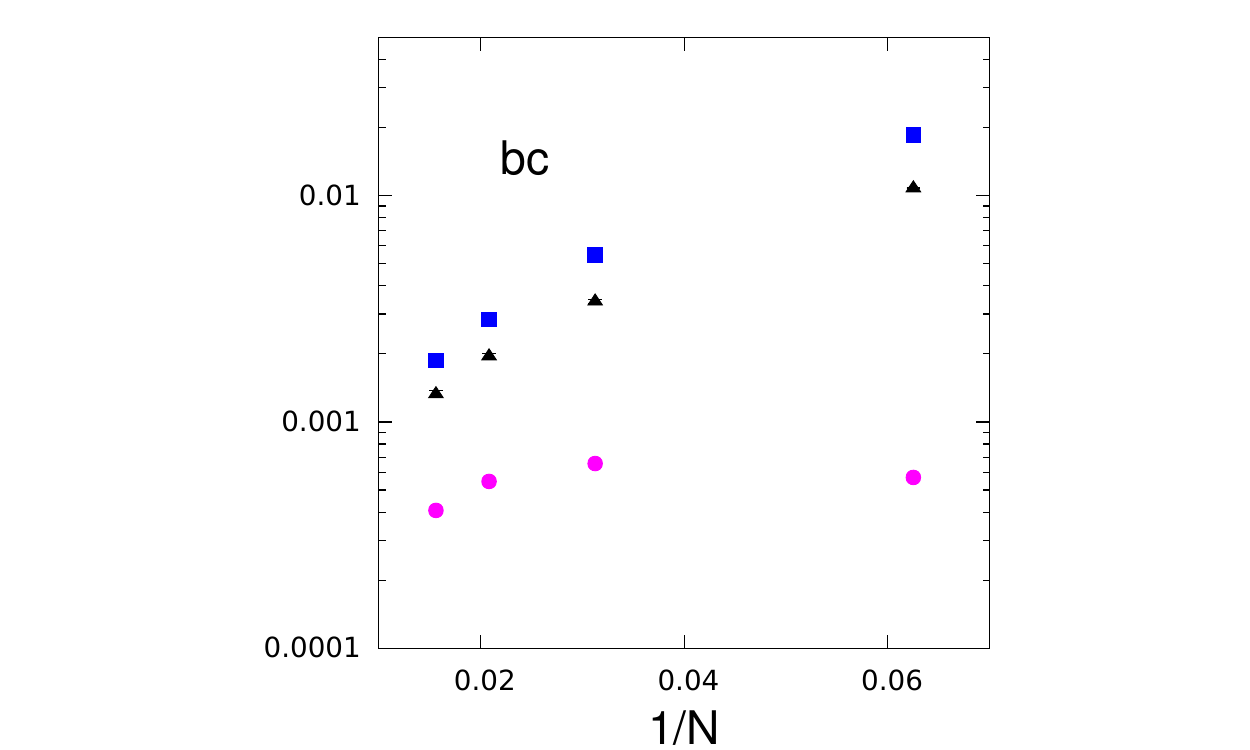}
\vspace{2mm}
\caption{\label{fig:distri} The largest (full squares),
the smallest (full circles) and the average value (full
triangles) of the smallest nonzero eigenvalue $\lambda_1$ as a
function of the inverse lattice size $1/N$. All quantities
are in lattice units.
Two types of Gribov copies are
considered (see Section \ref{sec:num}): {\em fc}
(left plot) and {\em bc} (right plot). Similar
results have been obtained for the sets of
Gribov copies indicated with {\em sfc} and {\em afc}.
Note the logarithmic scale on the $y$ axis.
Error bars around the average value of $\lambda_1$ correspond 
to one standard deviation. (We consider the statistical error only.)
}
\end{figure}

\begin{figure}
\centering
\includegraphics[trim=80 0 40 0, clip, scale=1.00]{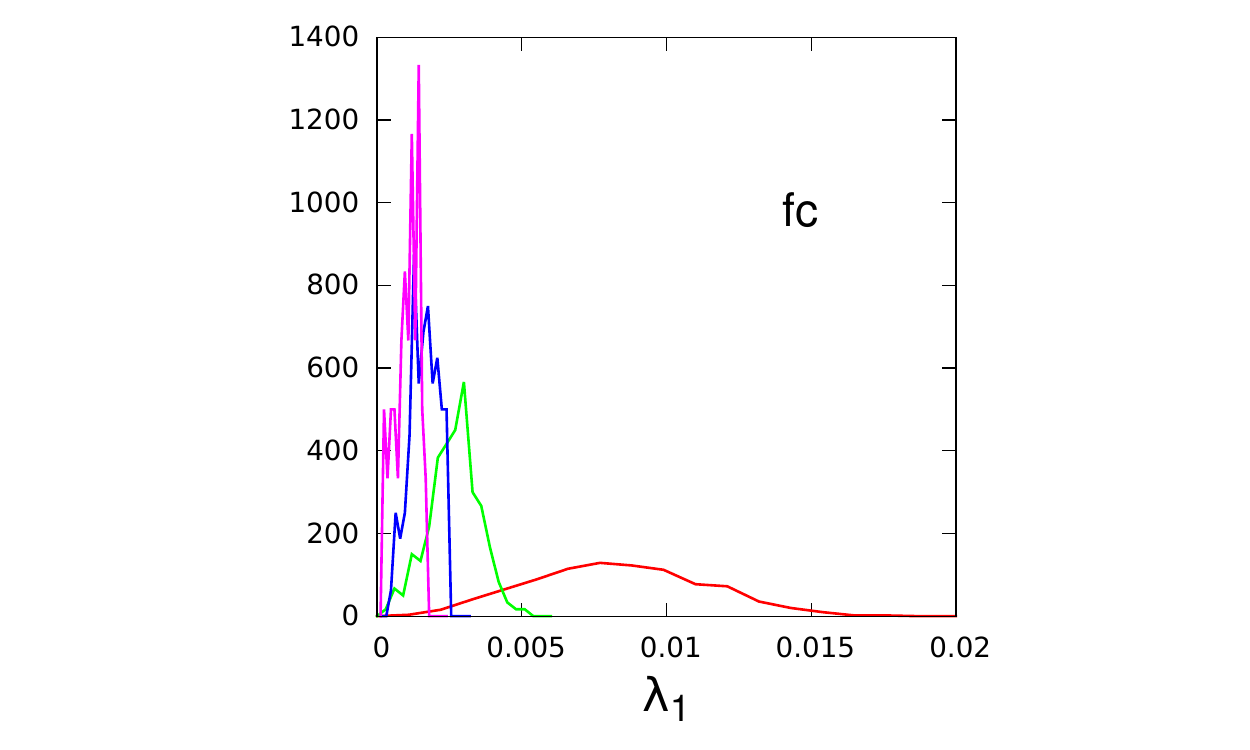}
\hspace{-5mm}
\includegraphics[trim=80 0 40 0, clip, scale=1.00]{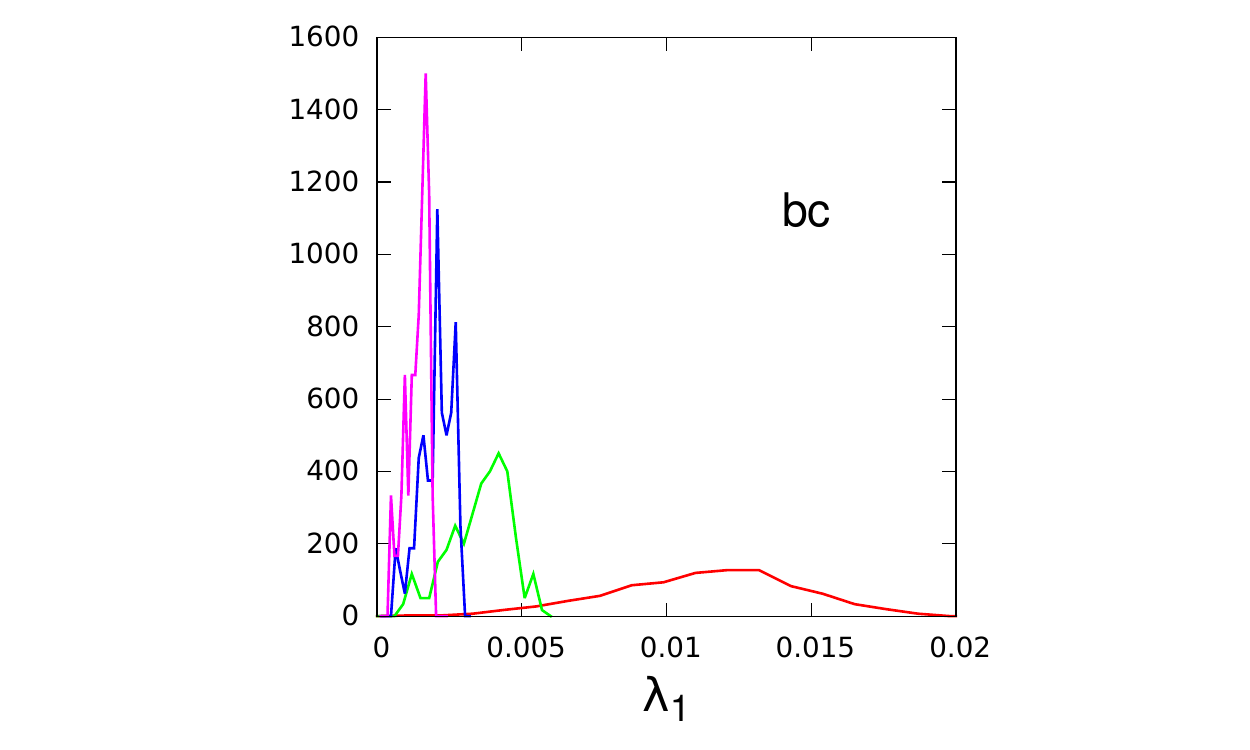}
\vspace{2mm}
\caption{\label{fig:histoc} Histogram distribution of the
smallest nonzero eigenvalue $\lambda_1$ for the lattice sizes
$N = 16$ (red line), 32 (green line), 48 (blue line) and 64
(magenta line) using 20 bins. The area of the histograms is
normalized to 1.
Two types of Gribov copies are
considered (see Section \ref{sec:num}): {\em fc}
(left plot) and {\em bc} (right plot). Similar
results have been obtained for the sets of
Gribov copies indicated with {\em sfc} and {\em afc}.
}
\end{figure}

\begin{figure}
\centering
\includegraphics[trim=80 0 40 0, clip, scale=1.00]{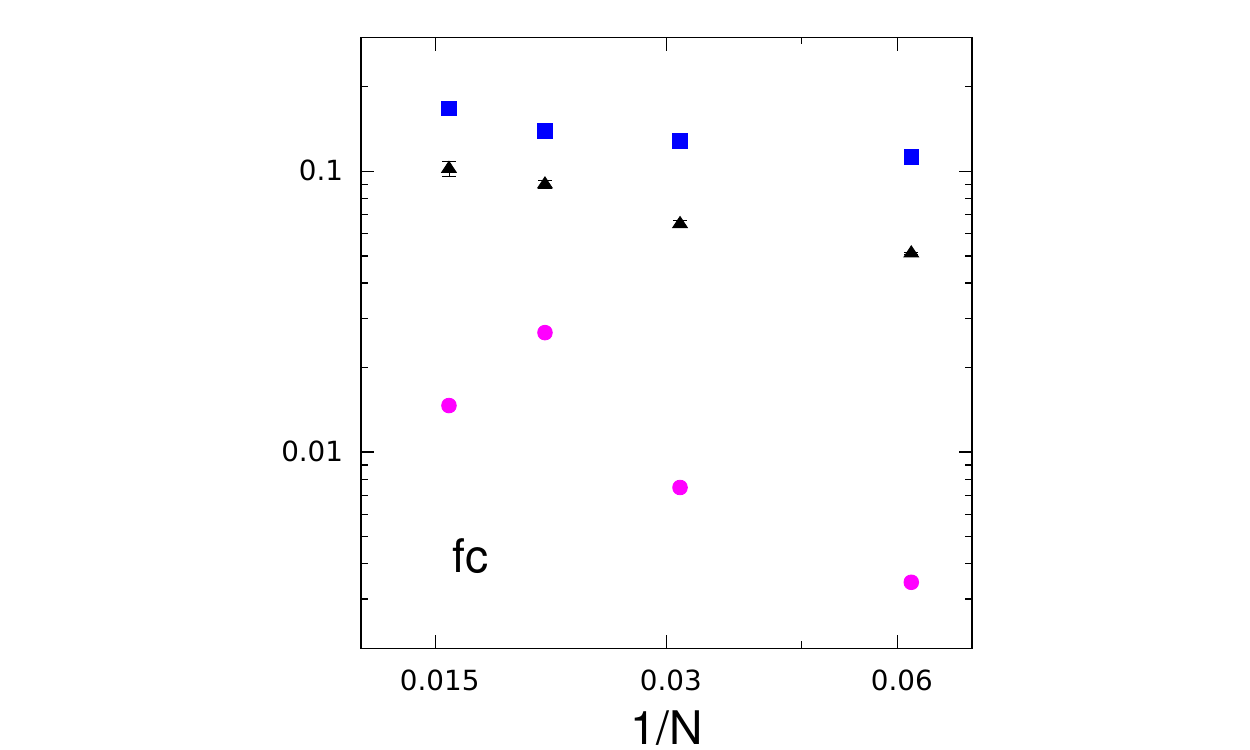}
\hspace{-5mm}
\includegraphics[trim=80 0 40 0, clip, scale=1.00]{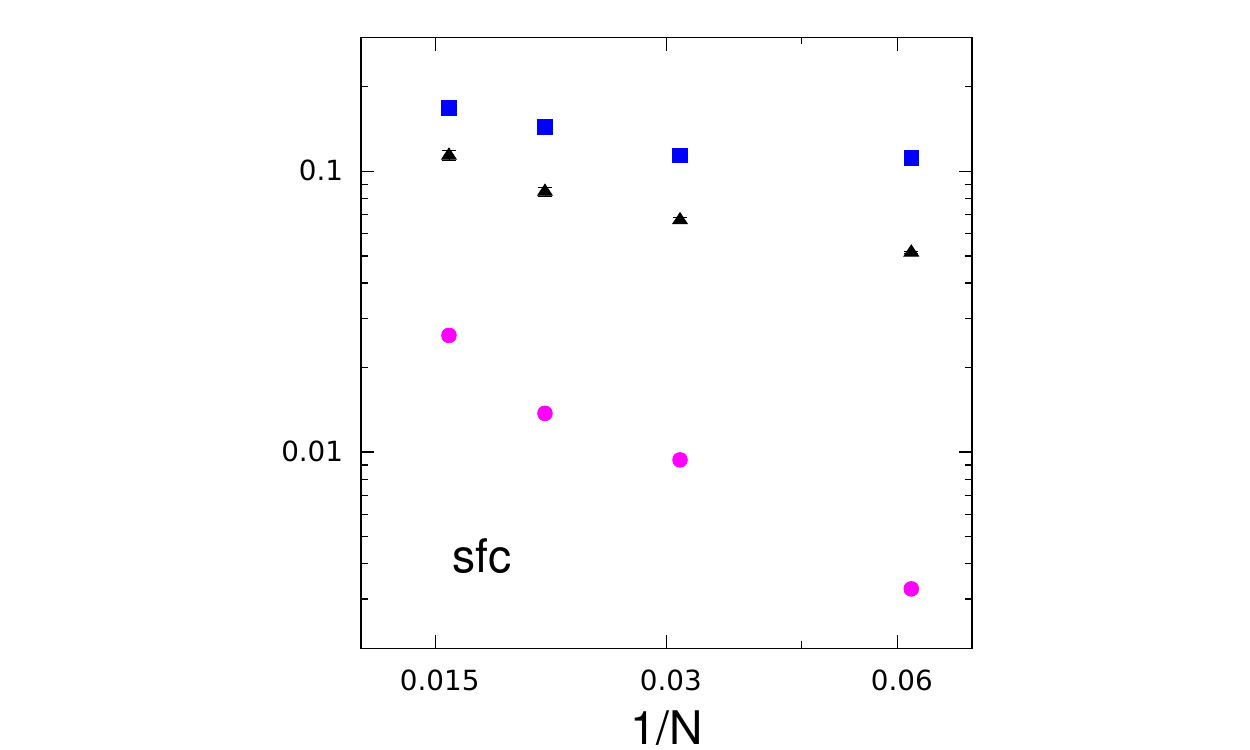}
\\
\vspace{5mm}
\includegraphics[trim=80 0 40 0, clip, scale=1.00]{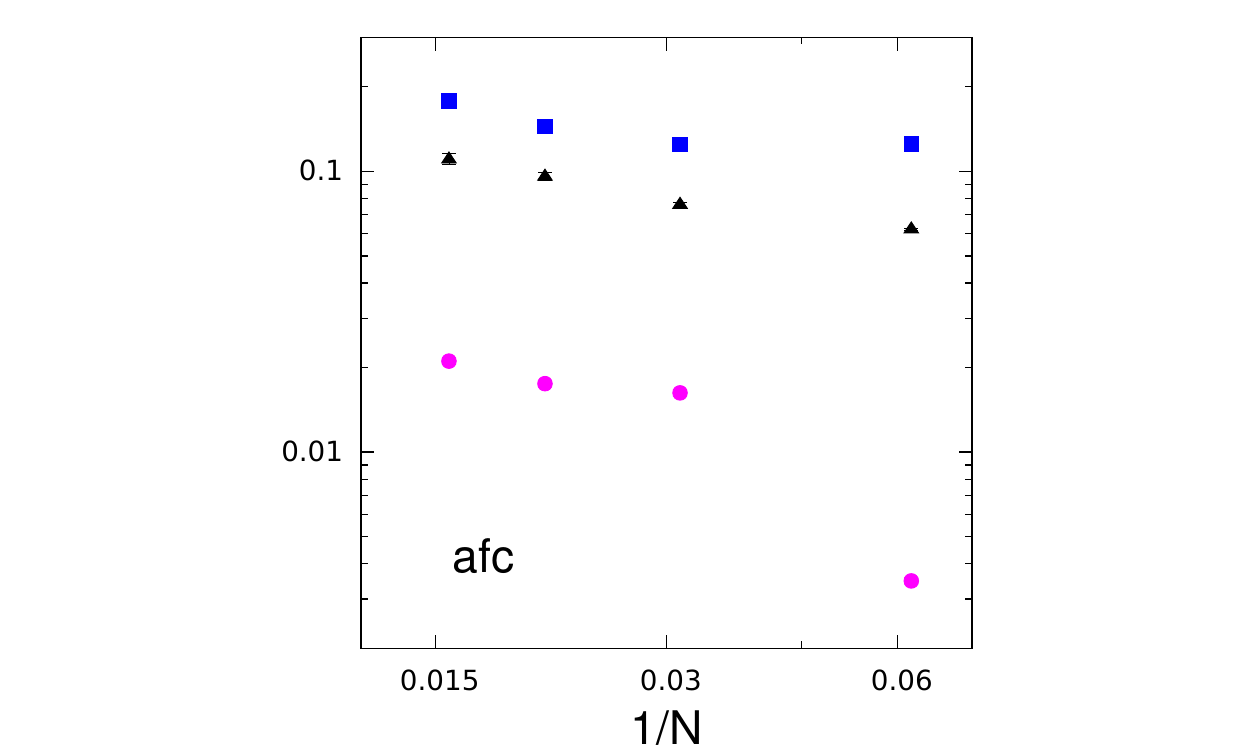}
\hspace{-5mm}
\includegraphics[trim=80 0 40 0, clip, scale=1.00]{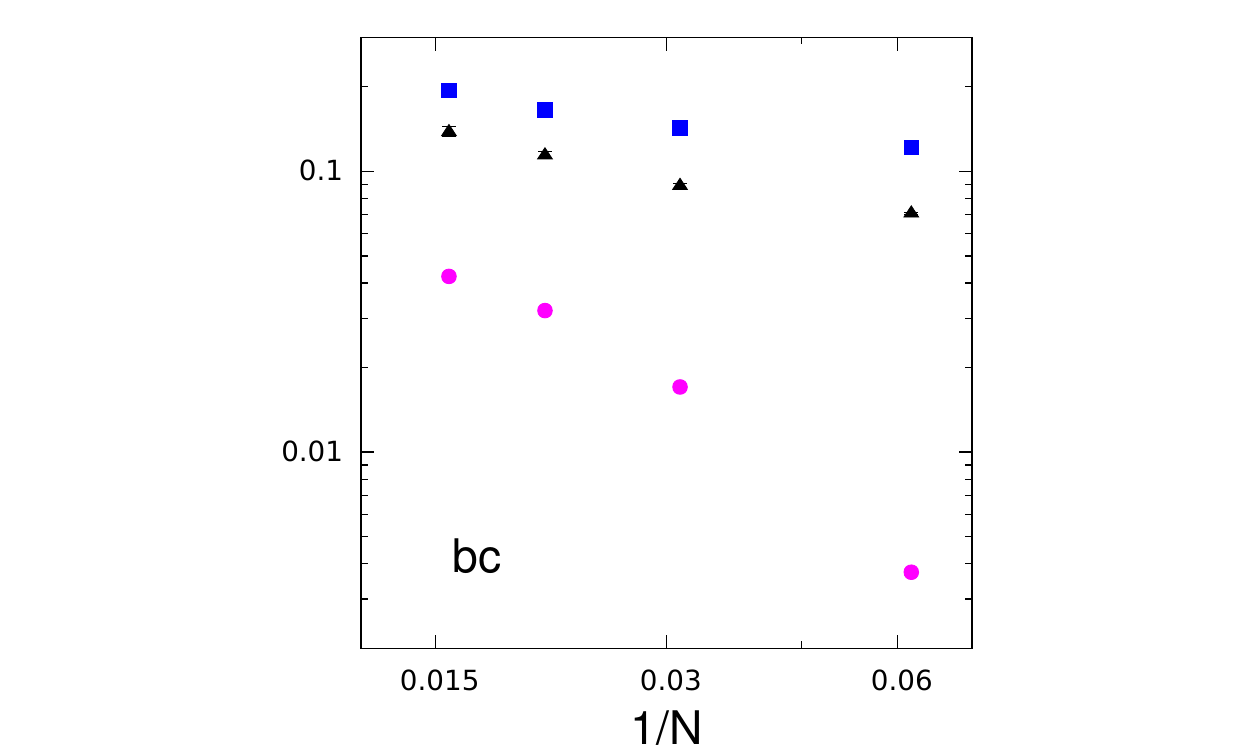}
\vspace{2mm}
\caption{\label{fig:distrimod} The largest (full squares),
the smallest (full circles) and the average value (full
triangles) of the smallest nonzero eigenvalue $\lambda_1$, all divided by the
smallest nonzero momentum squared $p_{min}^2$, as a function
of the inverse lattice size $1/N$. All quantities are in
lattice units. Four types of gauge-fixing prescription are considered (see
Section \ref{sec:num}): {\em fc} (upper left plot),
{\em sfc} (upper right plot), {\em afc} (lower left plot)
and {\em bc} (lower right plot). Note the logarithmic scale
on both axes.
Error bars around the average value of $\lambda_1/p_{min}^2$ 
correspond to one standard deviation. (We consider the statistical error only.)
}
\end{figure}


The new bounds give us useful and quantitative insights
into the realization of the infinite-volume limit in
the ghost sector. In particular, one can easily prove
some generally accepted facts about the Gribov-Zwanziger
confinement scenario. For example, a simple consequence of
the inequality (\ref{eq:newboundlambda2}) is that if,
in the infinite-volume limit, the functional integration
gets concentrated in a region (strictly) inside $\Omega$,
then we have $1 - \rho > 0$ and the ghost propagator will
display a tree-level behavior $\sim 1/p^2$ in the
infrared region.\footnote{See also the discussion in
Sec.\ 4.1 of Ref.\ \cite{Zwanziger:1991ac}.} This
observation is relevant when the theory is defined in
the fundamental modular region $\Lambda$
(what is sometimes called the absolute Landau gauge
\cite{Maas:2008ri,Maas:2009se}). Indeed, we know that
$\Lambda \subset \Omega$ \cite{Dell'Antonio:1991xt}
and that the boundaries $\partial \Lambda$ and $\partial
\Omega$ of these two regions have common points
\cite{vanBaal:1991zw}. Then, we can immediately conclude,
for the absolute Landau gauge and in the limit
of infinite volume, that a measure concentrated on the
common boundary of $\Lambda$ and $\Omega$ is a necessary
condition for an infrared-enhanced ghost propagator.
At the same time, one can infer some interesting implications
from published data. Indeed, the same line of reasoning
and the fact that the lower bound of the so-called
$b$-corridor displayed in Figure 3 of Ref.\ \cite{Maas:2009se}
is essentially flat tell us that (at least for relatively
small lattice volumes) it is always possible to find
Gribov copies very far away from $\partial \Omega$.

\vskip 3mm
Using the results presented in the previous section, i.e.\
looking at the data for the third derivative ${\cal E}'''$
of the minimizing function,\footnote{Here and below
we used for the analysis the same set of configurations
considered in Section \ref{sec:num}.} one can also try
to estimate the percentage of configurations that are
``very close'' to the common boundary $\partial\Omega \,
\cap \, \partial\Lambda$. To this end we
evaluated the average value of $| {\cal E}''' \, |$.
One notices (see Table \ref{tab:ratio}) that the results
obtained are more or less independent of the lattice size
$N$ and of the type of Gribov copy considered, with values
displaying small fluctuations around 0.024. We also checked
how many configurations are characterized by a value of
$| {\cal E}''' \, |$ smaller than one twentieth of the
corresponding average value. As one can see from the same
Table, as the lattice size increases there is a
noticeable increase in the percentage of configurations
that are good candidates for the common boundary
$\partial\Omega \, \cap \, \partial\Lambda$. At the same
time, by looking at the distribution of the values of
$| {\cal E}''' \, |$ (see Figure \ref{fig:histod3}) we
find that, for the {\em fc} and {\em sfc} gauge-fixing
choices, the distribution gets strongly concentrated
below the value 0.01 as the lattice size $N$ increases.
On the other hand, for the {\em afc} and {\em bc} gauge-fixing
choices, i.e.\ for configurations that are ``closer''
to the fundamental modular region $\Lambda$, the behavior of
the peak of the distribution is not monotonically decreasing
with $N$. Of course, in the case of $N= 48$ and $64$, 
the number (only 100 or 50) of data points 
may be insufficient to sample correctly
the distribution of $| {\cal E}''' \, |$ and to produce
a reliable 20-bin histogram. Thus, these results should be
verified by considering larger sets of data and (possibly)
larger lattice volumes.

\vskip 3mm
In order to study the infinite-volume limit in more
detail, we can now follow Refs.\ \cite{Sternbeck:2005vs,
Sternbeck:2012mf} and consider the distribution of
the smallest nonzero eigenvalue $\lambda_1$ as a
function of the lattice size (see Figures \ref{fig:distri}
and \ref{fig:histoc}). In particular,
from Figure \ref{fig:distri} it is clear that the
average value of $\lambda_1$, as well as its
largest value, are monotonically decreasing as the
lattice size increases, for the four types of statistics
considered here. Conversely, its smallest value
shows in some cases a non-monotonic behavior. At the
same time, from Figure \ref{fig:histoc}, one sees that
the distribution of $\lambda_1$ changes
considerably when going from $N=16$ to $N=32$. Indeed,
in the former case, one has a very broad distribution
for the smallest nonzero eigenvalues, with long tails
(compared to the central value), while in the latter
case we already see a more concentrated 
distribution\footnote{These results are in qualitative agreement
with Figure 1 of Ref.\ \cite{Sternbeck:2005vs}.} 
for the values of $\lambda_1$. The same
relatively small spread for the values of $\lambda_1$
is seen for $N=48$ and $64$. Thus, results obtained at small lattice volumes
are strongly affected by how well (or badly)
the tails of the $\lambda_1$-distribution is sampled.
Indeed, for small volumes we expect the thermalized
and gauge-fixed configuration to belong to the center
of the distribution in most of the cases. On the other
hand, occasionally, one can obtain a configuration
in the long tails of the distribution and, in particular,
one that strongly feels the effects of $\partial \Omega$ or, equivalently, 
a configuration for which the inequality
(\ref{eq:newboundlambda2}) approaches an equality. 
This observation supports and explains results obtained by various groups
for the so-called exceptional configurations, found in the ghost sector
of Landau gauge \cite{Bakeev:2003rr,Furui:2004bq,
Sternbeck:2005tk,Sternbeck:2005vs,Furui:2006nm,
Cucchieri:2006tf,Maas:2008ri} and of Coulomb gauge
\cite{Greensite:2009zz,Greensite:2010tm,Burgio:2012bk,
Greensite:2011np}. Indeed, it seems clear that
the presence of these exceptional configurations
\begin{itemize}
\item increases with the coupling $\beta$ \cite{Bakeev:2003rr},
i.e.\ when the physical lattice volume is typically smaller;
\item decreases when the statistical average is constrained
to the fundamental modular region \cite{Sternbeck:2005tk},
i.e.\ when the values of $\lambda_1$ are typically larger
(compare the {\em bc} average to the {\em fc} average in Figures
\ref{fig:distri} and \ref{fig:histoc});
\item is not correlated to anomalous values of the
Polyakov loop and to ``toron'' excitations \cite{Bakeev:2003rr};
\item is correlated to very small values of the smallest
nonzero eigenvalue of the FP matrix \cite{Greensite:2009zz,
Greensite:2010tm,Burgio:2012bk,Greensite:2011np,Sternbeck:2012mf}
or, equivalently, to large contributions to the
ghost propagator \cite{Sternbeck:2005vs,Sternbeck:2005tk};
\item depends on the ``direction'' of the momentum $p$
used to evaluate the ghost propagator --- i.e.\ rotational
invariance is broken \cite{Sternbeck:2005tk} --- since
the smallest eigenvalue of the (lattice) Laplacian,
entering the new bound (\ref{eq:newboundlambda2}), is
$d$-fold degenerate (in the $d$-dimensional case);
\item is correlated to very long minimization processes
of the numerical gauge fixing \cite{Cucchieri:2006tf,Maas:2008ri,
Burgio:2012bk}, since the structure
of the gauge orbit becomes more complicated 
near $\partial \Omega$ (see for
example Figures 1 and 2 in \cite{vanBaal:1991zw} and Figure 2
in \cite{Cucchieri:1997ns}), with the presence of various stationary
points and flat directions.
\end{itemize}
Let us note that different approaches have been considered in
the literature for dealing with these anomalous configurations.
For example, in Coulomb gauge, the evaluation of the Coulomb
potential \cite{Greensite:2009zz,Greensite:2010tm,
Greensite:2011np} --- which involves two powers of the
inverse FP matrix --- and of the ghost propagator
\cite{Burgio:2012bk} is usually done by taking these
configurations out of the statistical average. On the
other hand, since the anomalous behavior in the ghost sector is not
rotationally invariant, the authors of Ref.\ \cite{Sternbeck:2005tk}
suggested to average the data over different realizations of the same
momentum $p$, in order to reduce the systematic
effects due to these exceptional configurations.
Finally, in Refs.\ \cite{Furui:2004bq,Furui:2006nm}, the
authors observed that using only the exceptional configurations 
can give results in a better agreement with the scaling solution 
\cite{von Smekal:1997vx, Zwanziger:2001kw,Lerche:2002ep,Fischer:2006ub} 
of the Dyson-Schwinger equations. 

More recently, exceptional configurations
have been viewed \cite{Maas:2009se,Sternbeck:2012mf,Maas:2013vd} as 
representatives in the implementation of minimal Landau gauge with 
additional conditions. Such gauge choices are introduced
mainly as a possible lattice realization\footnote{See
Ref.\ \cite{Cucchieri:2011um} for various possible criticisms
to this interpretation.} of the scaling solution and of
the one-parameter family \cite{Boucaud:2008ji,Boucaud:2008ky,
Fischer:2008uz,RodriguezQuintero:2010wy} of the so-called
massive solutions \cite{Aguilar:2002tc,Aguilar:2004sw,
Aguilar:2006gr,Binosi:2009qm,Aguilar:2009ke} (see also
Refs.\ \cite{Frasca:2007uz,Dudal:2008sp,Kondo:2009ug,
Tissier:2010ts,Tissier:2011ey,Weber:2011nw,Pennington:2011xs,
Cucchieri:2012cb,LlanesEstrada:2012my} for other approaches
and points of view on the scaling and/or the massive
solutions).

\vskip 3mm
Considering again the inequality (\ref{eq:newboundlambda2}),
we also see that, from the numerical point of view, the ratio
$\lambda_1[ \, {\cal M}[A] \,] / p^2_{min}$ can be used
as an upper-bound estimate of the distance
$1 - \rho$ of a configuration $A$ from the boundary
$\partial \Omega$. It is therefore interesting to rescale
the data presented in Figure \ref{fig:distri} for
the smallest nonzero eigenvalue $\lambda_1$ by
$1/p^2_{min}$. The results are shown in Figure \ref{fig:distrimod}.
It seems that, for the four types of statistics
considered in this work, this upper bound increases in the
large-volume limit, i.e.\ $\lambda_1$ goes to zero more slowly
than $p^2_{min} \sim 1/N^2$. Of course, since the ratio
$\lambda_1[ \, {\cal M}[A] \,] / p^2_{min}$ is an
upper bound for $1 - \rho$, this result
does not imply that the quantity $\rho$ does not go to 1
in the infinite-volume limit, as indeed seems to be the case
(at least for the {\em fc} average, see the plot in
Figure \ref{fig:newbounds}).

We plan to extend our numerical simulations in the near future.
In particular, since we did not do an extensive search of Gribov copies
as in Refs.\ \cite{Maas:2009se,Sternbeck:2012mf,Maas:2013vd},
our results apply, for the moment, only to the four types
of statistics considered here. On the other hand, one should
recall that a systematic search of lattice Gribov copies is
numerically very difficult, already for very small lattice volumes
\cite{Hughes:2012hg}, and that some of the lattice copies are
just lattice artifacts, as in the compact $U(1)$ case.


\section{Conclusions}
\label{sec:con}

In this section we sum up the main results presented in this work. 
Considering the various bounds proven in the text and 
their numerical verification, we can say that:
\begin{enumerate}
\item The bound (\ref{eq:newboundlambda2}) allows a simple
      mathematical realization of the assumptions considered in the
      Introduction. It connects non-Abelian aspects of
      Yang-Mills theory in Landau gauge to a simple geometrical 
      parameter $\rho$, which can be interpreted
      as a normalized distance of a configuration
      $A$ in the Gribov region $\Omega$ from the origin $A=0$.
      (Equivalently, $\rho^{-1} A$ is a configuration on the
      boundary $\partial \Omega$ of $\Omega$.)
\item In the infinite-volume limit one finds $\lambda_1 \approx
      \lambda_2$, as expected. On the other hand, this equality
      is approached quite slowly by the lattice data. Indeed,
      when going from $N=16$ to $N=64$ (for $\beta = 2.2$)
      we see that the ratio $\lambda_1/\lambda_2$ changes from
      about 0.7 to about 0.78, for the four types of
      statistics considered here.
\item As the lattice size increases, it seems easier to
      find candidate configurations for the common boundary 
      $\partial \Omega \, \cap \, \partial \Lambda$.
      These are anomalous lattice configurations characterized by 
      a very small absolute value for the third derivative of the minimizing
      functional along the direction of $\psi_1$, the eigenvector
      corresponding to the smallest nonzero eigenvalue $\lambda_1$.
\item For large lattice sizes $N$ the relations
\begin{equation}
G(p_{min}) \, \approx \, \frac{1}{\lambda_1} \, < \,
\frac{1}{p_{min}^2 \, \left( 1 - \rho \right)}
\end{equation}
      are satisfied by lattice data in minimal Landau gauge.
\item Since the two upper bounds (\ref{eq:ineq}) and (\ref{eq:newupper})
      are almost saturated by our data, we conclude that
      $\lambda_i \approx \lambda_1$ [see comment below Eq.\
      (\ref{eq:ineq2})]. This constitutes a numerical
      verification of the statement that ``all horizons are one horizon'',
      discussed in Section 3 of Ref.\ \cite{Zwanziger:1991ac}.
\item Since the lower bounds (\ref{eq:ineq}), (\ref{eq:ineq2low}) and
      (\ref{eq:newboundlambda2}) are far from being saturated by
      the lattice data, we can say that, for a generic configuration,
      the eigenvector $\psi_1$ is very different from the
      plane waves corresponding to the eigenvalue $p^2_{min}$ of (minus)
      the lattice Laplacian.
\item The smallest nonzero eigenvalue $\lambda_1$ of the FP matrix
      goes to zero more slowly than $1/N^2$ for the four types of
      statistics considered here, supporting the so-called
      massive solution of the Dyson-Schwinger equations.
\item The lower bound $p_{min}^2 \, \left( 1 - \rho \right)$
      goes to zero as $1/N^{\eta}$ with $\eta$ reasonably larger
      than 2, at least for the {\em fc} gauge-fixing prescription.
      Moreover, even for small lattice volumes, the
      value of $\rho$ is very close to 1, i.e.\ most lattice
      configurations are indeed very close to the first
      Gribov horizon $\partial \Omega$. 
\end{enumerate}

Observations 4, 5 and 6 above explain why
it is ``difficult to find'' a scaling solution on
the lattice. Indeed, our data suggest that configurations
producing an infrared-enhanced ghost propagator should almost
saturate the new bound (\ref{eq:newboundlambda2}), i.e.\ their
eigenvector $\psi_1$ should have a large projection on
at least one of the plane waves corresponding to $p^2_{min}$.
In this case, $\lambda_1$ would go to zero faster than
$1/N^2$ and one would also find the value of the
ghost propagator $G(p_{min})$ to be mostly given by the first term
of the sum (\ref{eq:Gp}), i.e.\ the lower bound (\ref{eq:ineq})
would be almost saturated. On the other hand, this would contradict
the intuitive picture that, in the infinite-volume limit: i) the spectrum 
of the FP matrix should become continuous (see Observation 2 above)
and ii) the eigenvalue {\em density} around $\lambda \sim 0$ should
be relevant for the IR behavior of $G(p)$. Moreover, this would imply
that nonperturbative effects, such as color confinement,
be driven by configurations whose FP matrix ${\cal M}$ is 
``dominated'' by an eigenvector $\psi_1$ very similar
to the corresponding eigenvector of 
${\cal M} = - \partial_{\mu} \partial^{\mu} \,$, i.e.\ to
$\psi_1$ of the free case! In this sense we agree with
the observation made in the Conclusions of Ref.\ \cite{LlanesEstrada:2012my}
that it is very difficult (if not impossible) to find a scaling
solution on the lattice.

The above considerations also answer the question posed in
Ref.\ \cite{LlanesEstrada:2012my}: which are
the ``appropriate boundary conditions'' for a lattice configuration
in order to find a scaling solution in Monte Carlo
simulations of Yang-Mills theory? Indeed, we believe that
the above discussion clarifies the properties of a ``would-be
typical lattice-scaling configuration'', which, up to now, was
defined only operationally \cite{Maas:2009se,Maas:2013vd}.
At the same time, Observation 8 seems to disprove the
conjecture of Ref.\ \cite{LlanesEstrada:2012my} that
``the scaling solution is related to the formation of the Gribov
horizon''. In fact, since essentially all lattice configurations,
even for $N = 16$, are very close to $\partial \Omega$, it seems
to us that it is not enough to have a Gribov horizon in order to
produce a scaling solution. On the contrary, the key ingredient
is {\em how} this boundary is encoded in the FP matrix.

From the analytic point of view, if one desires to isolate configurations
displaying a scaling behavior, the simplest candidates for which $\psi_1$ 
is almost a plane wave are probably Abelian configurations, since in
this case it might be easier to minimize the contribution of the operator 
$K[A]$ to the eigenvalue $\lambda_1$. On the other hand, one should note 
that it is not enough to find specific examples of lattice configurations 
whose FP matrix satisfies the properties discussed above.
One should also show that this type of minima exists for all
(or almost all) gauge orbits, so that their contribution to
the functional integration (or, equivalently, to a Monte Carlo
sampling) can be relevant. 


\section*{Acknowledgments}

We thank M. Chiapparini, M.S. Guimar\~aes and
S.P. Sorella for useful discussions.
We acknowledge partial support from FAPESP {\bf grant \# 2009/50180-0}
and from CNPq.
We would like to acknowledge computing time provided on the Blue Gene/P
supercomputer supported by the Research Computing Support Group (Rice
University) and Laborat\'orio de Computa\c c\~ao Cient\'\i fica Avan\c cada
(Universidade de S\~ao Paulo).



\end{document}